\documentclass[]{aa}  
\usepackage{graphicx}
\usepackage[integrals]{wasysym}
\usepackage{siunitx}
\usepackage{txfonts}
\usepackage{hyperref}
\hypersetup{colorlinks,citecolor=blue, linkcolor=black, urlcolor=blue}
\newcounter{magicrownumbers}
\newcommand\rownumber{\stepcounter{magicrownumbers}\arabic{magicrownumbers}}

\usepackage{cellspace, makecell} 

\newcommand{\lf}{\left}
\newcommand{\rg}{\right}
\newcommand{\tonda}[1]{\!\left ( #1 \right )}
\newcommand{\quadra}[1]{\left [ #1 \right ]}

\newcommand{\sistemai}{\lf\{\begin{aligned}}
\newcommand{\sistemaf}{\end{aligned}\rg.}
\newcommand{\sopra}[1]{\overline{#1}}

\newcommand{\eq}[1]{\begin{equation} #1 \end{equation}}

\newcommand{\asun}{\astrosun}

\newcommand{\Mbh}{M_\text{BH}}

\begin{document}

   \title{The ALMA view of the high-redshift relation between supermassive black holes and their host galaxies}


   \author{A. Pensabene
          \inst{\ref{difa},\ref{inaf-bo},\ref{dip}}
          \thanks{\email{antonio.pensabene2@unibo.it}}
          \and
          S. Carniani
          \inst{\ref{sns}}
          \and
          M. Perna
          \inst{\ref{arcetri},\ref{CSIC}}
          \and
          G. Cresci
          \inst{\ref{arcetri}}
          \and
          R. Decarli
          \inst{\ref{inaf-bo}}
          \and
          R. Maiolino
          \inst{\ref{cav},\ref{kavli}}
          \and
          A. Marconi
          \inst{\ref{dip},\ref{arcetri}}
          }

   \institute{Dipartimento di Fisica e Astronomia, Alma Mater Studiorum, Universit\`a di Bologna, Via Gobetti 93/2, I-40129 Bologna, Italy\label{difa}
        \and
        INAF-Osservatorio di Astrofisica e Scienza dello Spazio, via Gobetti 93/3, I-40129 Bologna, Italy\label{inaf-bo}
        \and
        Dipartimento di Fisica e Astronomia, Universit\`a degli Studi di Firenze, Via G. Sansone 1, I-50019 Sesto Fiorentino (Firenze), Italy\label{dip}
         \and
         Scuola Normale Superiore, Piazza dei Cavalieri 7, I-56126 Pisa, Italy\label{sns}    
         \and
         INAF-Osservatorio Astrofisico di Arcetri, Largo E. Fermi 2, I-50125 Firenze, Italy\label{arcetri}
         \and
         Centro de Astrobiolog\'ia (CSIC-INTA), Torrej\'on de Ardoz, 28850 Madrid, Spain\label{CSIC}    
         \and
         Cavendish Laboratory, University of Cambridge, 19 J. J. Thomson Ave., Cambridge CB3 0HE, UK\label{cav}
         \and
        Kavli Institute for Cosmology, University of Cambridge, Madingley Road, Cambridge CB3 0HA, UK\label{kavli}
             }

        \date{Received XXX; accepted YYY}
 
  \abstract
  {The existence of tight correlations between supermassive black holes (BHs) and their host galaxies' properties in the local Universe suggests a closely linked evolution. Investigating these relations up to the high redshifts ($z\apprge6$) is crucial in order to understand the interplay between star formation and BH growth across the cosmic time and to set constraints on galaxy formation and evolution models. In this work, we focus on the relation between BH mass ($\Mbh$) and the dynamical mass ($M_{dyn}$) of the host galaxy.}
  {Previous works suggest an evolution of the $\Mbh-M_{dyn}$ relation with redshift indicating that BH growth precedes the galaxy mass assembly during their co-evolution at $z>3$. However, dynamical galaxy masses at high redshift are often estimated through the virial theorem, thus introducing significant uncertainties. Within the scope of this work, our aim is to study the $\Mbh-M_{dyn}$ relation of a sample of $2<z<7$ quasars by constraining their galaxy masses through a full kinematical modelling of the cold gas kinematics, thus avoiding all possible biases and effects introduced by the rough estimates usually adopted so far.}
  {For this purpose, we retrieved public observations of 72 quasar host galaxies observed in [CII]$_{158\mu m}$ or CO transitions with the Atacama Large Millimeter Array (ALMA). We then selected those quasars whose line emission is spatially resolved, and performed a kinematic analysis on ALMA observations. We estimated the dynamical mass of the systems by modelling the gas kinematics with a rotating disc, taking into account geometrical and instrumental effects. Our dynamical mass estimates, combined with $\Mbh$ obtained from literature and our own new CIV$\lambda1550$ observations allowed us to investigate the $\Mbh/M_{dyn}$ \mbox{in the early Universe.}}
  {Overall, we obtained a sample of ten quasars at $z\sim2-7,$ in which line emission is detected with high SNR ($\apprge 5-10$) and the gas kinematics are spatially resolved and dominated by ordered rotation. The estimated dynamical masses place six out of ten quasars above the local relation yielding to $\Mbh/M_{dyn}$ ratios $\sim10\times$ higher than those estimated in low-$z$ galaxies. On the other hand, we found that four quasars at $z\sim 4-6$ have dynamical-to-BH-mass ratios consistent with what is observed in early-type galaxies in the local Universe.}
   {}

               
   \titlerunning{The ALMA view of the BH-galaxy relation at high-$z$}
   \authorrunning{Pensabene et al.}
   \maketitle

%

\section{Introduction}
\label{sect:introduction}
Supermassive black holes (BHs; $\Mbh\sim 10^6-10^{10} \rm{M}_{\astrosun}$) are believed to reside at the centre of all nearby galaxies and are likely the relics of a past quasar (QSO) activity \citep[e.g.][]{Soltan+1982,Hopkins+2008}. Such BHs have likely played a key role in shaping galaxies during their assembly at early epochs, with the implication that BH growth and galaxy formation are closely linked \citep{Heckman+2014}.

The discovery of the strong correlations (in the local Universe) between the mass of the central black hole ($\Mbh$) and the physical properties of host galaxies (e.g. stellar velocity dispersion of the bulge stars, mass of the bulge, etc.; \citealt{Tremaine+2002, Haring-Rix+2004}; see also \citealt{Kormendy+2013} for an extensive review and references therein) has been one of the most significant breakthroughs of the past decades and represents a key building block for our understanding of galaxy formation and evolution across the cosmic time. In the framework of co-evolution between BHs and their host galaxies, the observed local relations are believed to arise from the balance between the energy released by the active galactic nucleus (AGN), which generates galactic-scale outflows expelling gas from the galaxy, and the gravitational potential that keeps the galactic system bound. According to current galaxy evolution models \citep[see e.g.][]{Lamastra+2010, Sijacki+2015}, AGN are able to regulate the star formation activity in the host and constrain both the final stellar mass and dynamical properties of the galaxy \citep[e.g.][]{DiMatteo+2005, Menci+2008, Hopkins+2008, Kormendy+2013}. Therefore, investigating the onset of BH-galaxy relations at high redshift is fundamental to exploring the interplay between BH accretion and star formation activity in the host galaxies, and to constrain, accordingly, galaxy formation and evolution models.

In this work, we focus on the relation between BH mass and that of the host galaxy ($\Mbh-M_{gal}$ relation). The latter has been widely sampled for active and quiescent galaxies in the local Universe ($z<1$), indicating that BH mass is a defined fraction of the bulge stellar mass ($\Mbh\sim10^{-3}M_{gal}$, e.g. \citealt{MarconiHunt2003,Haring-Rix+2004}). More recently, several groups \citep{Treu+2004, Treu+2007, Walter+2004, Peng+2006, Peng+2006b, Shields+2006, Woo+2006, Woo+2008, Ho+2007, Decarli+2010, Merloni+2010, Wang+2010, Bennert+2011, Canalizo+2012, Targett+2012, Bongiorno+2014} attempted to sample this relation beyond the local Universe, showing that there are indications for a possible evolution with redshift. In particular, these works suggest a parameterisation of the ratio $\Gamma=\Mbh/M_{gal}$ as a function of redshift, $\Gamma\propto(1+z)^{\beta}$. The published values of $\beta$ span the range $0.7-2$ \citep{McLure+2006, Bennert+2010, Bennert+2011, Decarli+2010, Merloni+2010} with the implication that, at higher redshifts, galaxies host black holes that are more massive than the local counterparts (e.g. a factor of  $\sim7$ at $z\sim3$; \citealt{Decarli+2010}). Therefore, during the competitive accretion of matter from the galactic halo that occurred at early epochs, black hole growth possibly must have preceded that of the host galaxy \citep[e.g.][]{Decarli+2010, Bongiorno+2014, Wang+2016}. 

However, the aforementioned results are affected by observational biases and instrumental limits. The selection of host galaxies revealed at high redshift ($z>3$), is driven by AGN luminosity, so more massive black holes are preferably selected \citep{Lauer+2007,Vestergaard+2008,Volonteri+2011,Portinari+2012,Schulze+2014,Volonteri+2016}. Then, in these sources, the luminosity of the central region overwhelms the emission from the host galaxy, and the disentangling of the two components is challenging even with high-resolution observations. Since the galaxy stellar mass estimates used to derive the $\Mbh - M_{gal}$ relation are based upon photometric methods, they are significantly contaminated by light from the central non-stellar source, and are thus very uncertain. Finally, since the average gas fraction of galaxies increases with the redshift \citep[e.g.][]{Magdis+2017, Tacconi+2018}, primordial galaxies may not yet have converted a large fraction of their gas into stars, therefore their stellar mass content may not be a reliable tracer of the total mass \citep[but there is also evidence of luminous QSOs with low gas fractions possibly related to the effect of an AGN-driven feedback mechanism, see e.g.][]{Carniani+2017,Kakkad+2017,Brusa+18,CresciMaiolino2018,Perna+18}. The galaxy's capability of retaining its gas under the influence of AGN activity, is indeed determined by the gravitational potential of the whole galaxy traced by the total (dynamical) mass.

The recent advent of ALMA (Atacama Large Millimeter and Sub-Millimeter Array) opened a new era of cold gas observations. Thanks to its unparalleled capability in terms of sensitivity, signal-to-noise ratio (SNR), and angular resolution, it is now possible to spatially resolve the gas kinematics in quasar host galaxies up to the higher redshifts targeting the brightest line emission of the cold gas, such as [CII]$_{158\mu m}$ or CO rotational line transitions with sub-mm spectroscopic observations \citep[see][for a comprehensive review]{CarilliWalter+2013, Gallerani+2017}. In fact, the emission of radio-quiet AGN in the sub-mm band is dominated by the cold gas mass and the dust continuum in their hosts, thus allowing observations that are not affected by the non-stellar emission of the central source. Therefore, thanks to the efforts of many groups, ALMA has made it possible to trace the BH-galaxy relation at very high redshift using dynamical mass estimations of host galaxies \citep[e.g.][]{Wang+2013, Wang+2016, Willott+2013, Willott+2015mbh, Venemans+2012, Venemans+2016, Venemans+2017, Decarli+2017, Trakhtenbrot+2017, Feruglio+2018}. The dynamical masses provided in these works are estimated assuming rotating disc geometry and by simply combining the full width half maximum (FWHM) of the observed line emission, the observed size of the emitting region, and the inclination angle of the galaxy disc with respect to the sky plane. However, it is hard to test the basic assumption that the cold atomic/molecular gas of the galaxy is a rotating disc. Furthermore, the disc inclination is calculated from the observed morphology by using the axial ratio of the flux map, and is thus affected by significant uncertainties.

In this work, we studied a large sample ($\sim 70$) of $2 < z < 7$ quasars observed by ALMA targeting the [CII]$_{158\mu m}$ atomic fine-structure line or the CO rotational line emission, which we exploited in order to trace the morphology and kinematics of quasar host galaxies. Overall, we identified ordered rotational motion in a sample of ten quasars (for which high SNR allowed a spatially resolved analysis). By carefully modelling the kinematics with rotating discs, we were able to measure their host galaxy dynamical mass, at variance with previous work where rough estimates are usually adopted. Our dynamical mass measurements, combined with $\Mbh$ estimates obtained from the literature allowed us to trace the evolution of the $\Mbh-M_{gal}$ relation and to study the trend of $\Gamma=\Mbh/M_{gal}$ across the cosmic time.

The paper is organised as follows: %
in Sect.~\ref{sect:data}, we outline our starting sample and the data reduction performed on the raw data. In Sect.~\ref{sect:methods}, we illustrate the methods of data analysis to retrieve the information on the morphology and kinematics of the host galaxies. In Sect.~\ref{sect:kinematical_model}, we present the kinematical model used to measure the galaxy dynamical mass. In Sect.~\ref{sect:BH_masses}, we obtain the BH masses from the literature and from LBT data. In Sect.~\ref{sect:subselections}, we recap the different sub-selections of the starting sample that occurred during this work. In Sect.~\ref{sect:comparison}, we compare our dynamical mass estimates with previous similar studies and discuss the uncertainties on our measurements. Then, we investigate limits of validity of the assumptions. In Sect.~\ref{sect:BH-galaxy_relation}, the $\Mbh-M_{dyn}$ relation and the trend of $\Mbh/M_{dyn}$ ratio across cosmic time are presented. Then, in Sect.~\ref{sect:discussion}, we discuss our results and compare them with previous works. We also examine how possible additional uncertainties and biases could affect the results both from observational and theoretical points of view. Then, we compute the virial masses of our final sample, and we compare them with our dynamical mass estimates. Finally, in Sect.~\ref{sect:conclusions}, we draw our conclusions.

Throughout the paper, we assume a standard $\Lambda\rm{CDM}$ cosmology with $H_0=69.3\,\si{km\,s^{-1}Mpc^{-1}}$, $\Omega_{m}=0.287$, $\Omega_{\Lambda}=1-\Omega_{m}$ from \citet{Hinshaw+2013}.
\begin{figure}[!t]
        \centering
        \includegraphics[width=\hsize]{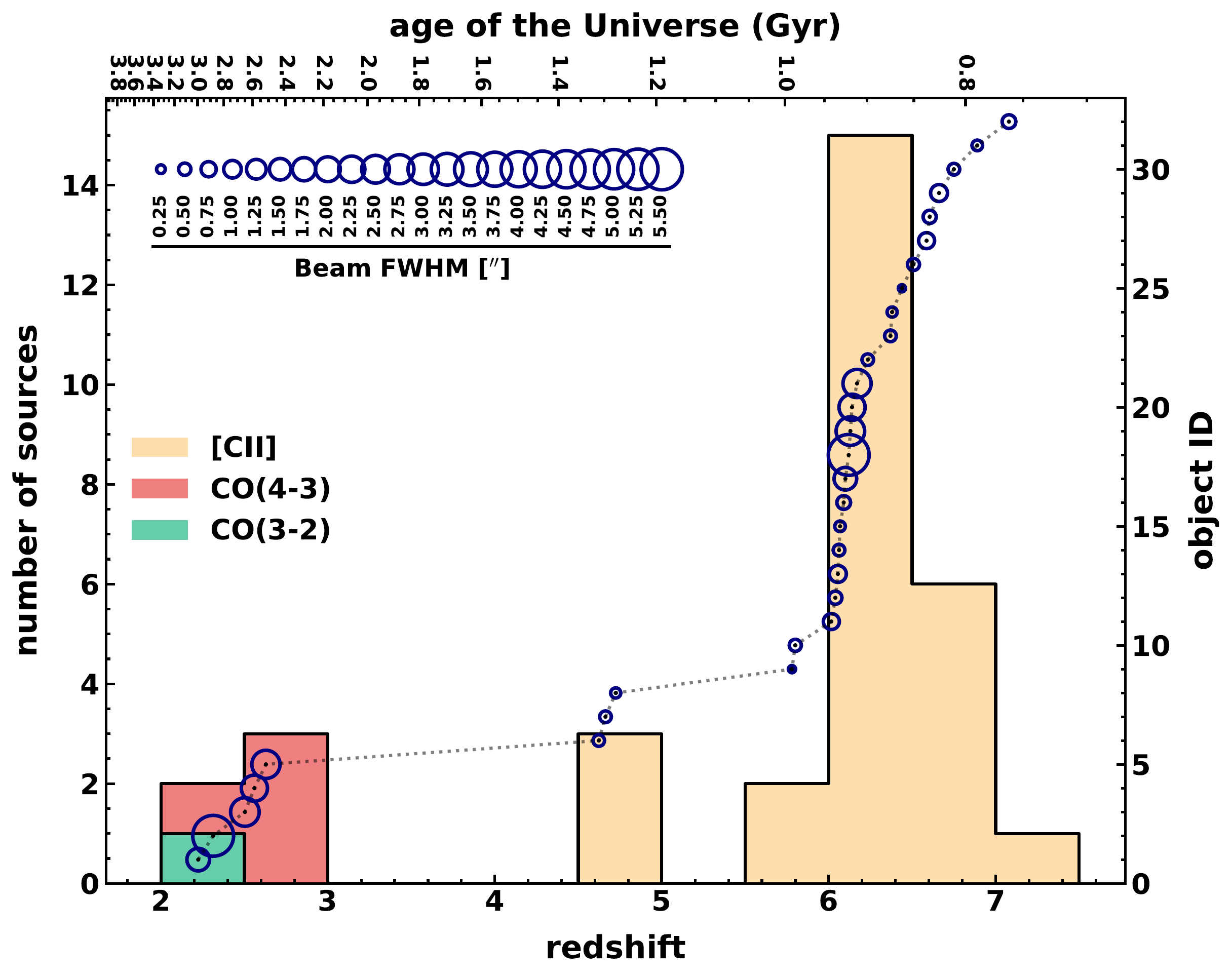}
         \caption{Redshift distribution of 32 QSOs listed in Table~\ref{tbl:sample}. The histogram colours indicate the emission line detected in the ALMA data. The figure also shows the sample cumulative redshift distribution (dotted line and right axis). Individual sources are marked with blue circles (sizes are proportional to the ALMA synthesised beam FWHM).}
         \label{fig:redshift-distr}
\end{figure}

\begin{table*}
\caption{List of 32 QSOs revealed with emission line detection significant at $\apprge 3\sigma$ and related information about the ALMA observing project.}             
\label{tbl:sample}      
\centering
\begin{minipage}{1.0\textwidth} 
\centering 
\begin{tabular}{l l c c c c c}
\hline\hline       
No.             &  Object ID                                    & R.A. (J2000)                         &  DEC. (J2000)                                 &  $z_\text{cat}$ $^a$ & Observed line    & Flags [rot, $\beta$, $\Mbh$] $^b$\\
\hline
\rownumber & \object{CXOCDFS J0332-2746}                 & $03^{h} 32^{m} 31^{s}.46      $ & $-27\si{\degree} 46\si{\arcmin} 23\si{\arcsecond}.18          $ & $2.2234     $ & CO(4-3) & r, u, -- \\
\rownumber & \object{VHS J2101-5943}                     & $21^{h} 01^{m} 19^{s}.5         $ & $-59\si{\degree} 43\si{\arcmin} 45\si{\arcsecond}         $ & $2.313          $ & CO(3-2) & r, b, a \\
\rownumber & \object{ULAS J1234+0907}                    & $12^{h} 34^{m} 27^{s}.52      $ & $+09\si{\degree} 07\si{\arcmin} 54\si{\arcsecond}.2     $ & $2.503             $ & CO(3-2) & u, --, -- \\
\rownumber & \object{ULAS J2315+0143}                    & $23^{h} 15^{m} 56^{s}.23      $ & $+01\si{\degree} 43\si{\arcmin} 50\si{\arcsecond}.38   $ & $2.560     $ & CO(3-2) & r, u, -- \\
\rownumber & \object{ULAS J0123+1525}                    & $01^{h} 23^{m} 12^{s}.52         $ & $+15\si{\degree} 25\si{\arcmin} 22\si{\arcsecond}.52   $ & $2.629   $ & CO(3-2) & u, --, -- \\
\rownumber & \object{SDSS J1328-0224}                    & $13^{h} 28^{m} 53^{s}.65      $ & $-02\si{\degree} 24\si{\arcmin} 41\si{\arcsecond}.79    $ & $4.62271  $ & [CII] & r, c, e \\
\rownumber & \object{SDSS J0923+0247}                    & $09^{h} 23^{m} 03^{s}.52         $ & $+02\si{\degree} 47\si{\arcmin} 39\si{\arcsecond}.68   $ & $4.66307         $ & [CII] & r, c, e \\
\rownumber & \object{SDSS J0331-0741}                    & $03^{h} 31^{m} 19^{s}.66      $ & $ -07\si{\degree} 41\si{\arcmin} 43\si{\arcsecond}.16   $ & $4.72426  $ & [CII] & r, u, -- \\
\rownumber & \object{SDSS J0129-0035}                    & $01^{h} 29^{m} 58^{s}.5         $ & $-00\si{\degree} 35\si{\arcmin} 39\si{\arcsecond}         $ & $5.780          $ & [CII] & r, l, a \\
\rownumber & \object{SDSS J1044-0125}                    & $10^{h} 44^{m} 33^{s}.04         $ & $-01\si{\degree} 25\si{\arcmin} 02\si{\arcsecond}.07    $ & $5.800         $ & [CII] & r, c, a \\
\rownumber & \object{SDSS J1306+0356}                    & $13^{h} 06^{m} 08^{s}.25         $ & $+03\si{\degree} 56\si{\arcmin} 26\si{\arcsecond}.33   $ & $6.016         $ & [CII] & r, c, e \\
\rownumber & \object{SDSS J2310+1855}                    & $23^{h} 10^{m} 38^{s}.88         $ & $+18\si{\degree} 55\si{\arcmin} 19\si{\arcsecond}.72   $ & $6.040         $ & [CII] & r, c, e \\
\rownumber & \object{SDSS J0842+1218}                    & $08^{h} 42^{m} 29^{s}.43         $ & $+12\si{\degree} 18\si{\arcmin} 50\si{\arcsecond}.48   $ & $6.055         $ & [CII] & u, --, -- \\
\rownumber & \object{SDSS J2054-0005}                    & $20^{h} 54^{m} 06^{s}.49         $ & $-00\si{\degree} 05\si{\arcmin} 14\si{\arcsecond}.57    $ & $6.062         $ & [CII] & r, c, a \\
\rownumber & \object{[WMH2013] 05}                               & $02^{h} 26^{m} 27^{s}.03       $ & $-04\si{\degree} 52\si{\arcmin} 38\si{\arcsecond}.3      $ & $6.068          $ & [CII] & u, --, -- \\
\rownumber & \object{CFHQS J2100-1715}                   & $21^{h} 00^{m} 54^{s}.62         $ & $-17\si{\degree} 15\si{\arcmin} 22\si{\arcsecond}.5      $ & $6.09           $ & [CII] & u, --, -- \\
\rownumber & \object{DES J0454-4448}                     & $04^{h} 54^{m} 01^{s}.79         $ & $-44\si{\degree} 48\si{\arcmin} 31\si{\arcsecond}.1      $ & $6.100          $ & [CII] & u, --, -- \\
\rownumber & \object{CFHQS J1509-1749}                   & $15^{h} 09^{m} 41^{s}.8         $ & $-17\si{\degree} 49\si{\arcmin} 27\si{\arcsecond}         $ & $6.120          $ & [CII] & u, --, -- \\
\rownumber & \object{ULAS J1319+0950}                    & $13^{h} 19^{m} 11^{s}.29         $ & $+09\si{\degree} 50\si{\arcmin} 51\si{\arcsecond}.34   $ & $6.130         $ & [CII] & r, l, e \\
\rownumber & \object{PSO J065-26}                                & $04^{h}21^{m}38^{s}.05                $ & $ -26\si{\degree}57\si{\arcmin}15\si{\arcsecond}.60   $ & $6.14               $ & [CII] & u, --, -- \\
\rownumber & \object{[CLM2003] J0228-04161}              & $02^{h} 28^{m} 02^{s}.97      $ & $-04\si{\degree}16\si{\arcmin}18\si{\arcsecond}.3     $ & $6.17       $ & [CII] & u, --, -- \\
\rownumber & \object{PSO J308-21}                                & $20^{h} 32^{m} 10^{s}.00       $ & $-21\si{\degree} 14\si{\arcmin} 02\si{\arcsecond}.4      $ & $6.2342         $ & [CII] & r, b, u \\
\rownumber & \object{VIKING J1152+0055}                  & $11^{h} 52^{m} 21^{s}.27         $ & $+00\si{\degree} 55\si{\arcmin} 36\si{\arcsecond}.6 $ & $6.37               $ & [CII] & u, --, -- \\
\rownumber & \object{PSO J159-02}                                & $10^{h}36^{m}54^{s}.19         $ & $-02\si{\degree}32\si{\arcmin}37\si{\arcsecond}.94  $ & $6.38               $ & [CII] & u, --, -- \\
\rownumber & \object{PSO J183+05}                                & $12^{h}12^{h}26^{s}.98                $ & $+05\si{\degree}05\si{\arcmin} 33\si{\arcsecond}.49    $ & $6.4386    $ & [CII] & r, l, u \\
\rownumber & \object{PSO J167-13}                                & $11^{h} 10^{m} 33^{s}.98       $ & $-13\si{\degree} 29\si{\arcmin} 45\si{\arcsecond}.6      $ & $6.508          $ & [CII] & r, c, e \\
\rownumber & \object{PSO J231-20}                                & $15^{h}26^{m} 37^{s}.84       $ & $ -20\si{\degree}50\si{\arcmin}00\si{\arcsecond}.8    $ & $6.58651    $ & [CII] & r, u, -- \\
\rownumber & \object{VIKING J0305-3150}                  & $03^{h} 05^{m} 16^{s}.92         $ & $-31\si{\degree} 50\si{\arcmin} 56\si{\arcsecond}.0      $ & $6.605        $ & [CII] & r, c, e \\
\rownumber & \object{VIKING J1048-0109}                  & $10^{h}48^{m}19^{s}.08         $ & $-01\si{\degree}09\si{\arcmin}40\si{\arcsecond}.29  $ & $6.661      $ & [CII] & u, --, -- \\
\rownumber & \object{VIKING J0109-3047}                  & $01^{h} 09^{m} 53^{s}.13         $ & $-30\si{\degree} 47\si{\arcmin} 26\si{\arcsecond}.3      $ & $6.750        $ & [CII] & r, u, -- \\
\rownumber & \object{VIKING J2348-3054}                  & $23^{h} 48^{m} 33^{s}.34         $ & $-30\si{\degree} 54\si{\arcmin} 10\si{\arcsecond}.0      $ & $6.890        $ & [CII] & u, --, -- \\
\rownumber & \object{ULAS J1120+0641}                    & $11^{h} 20^{m} 01^{s}.48         $ & $+06\si{\degree} 41\si{\arcmin} 24\si{\arcsecond}.3     $ & $7.080         $ & [CII] & u, --, -- \\
\hline
\end{tabular}
\end{minipage}\hfill
\setcounter{magicrownumbers}{0}
\begin{minipage}{0.5\hsize}
\vspace{0.2cm}
\resizebox{\hsize}{!}{%
\begin{tabular}{l c c c c c |}
\hline\hline
No.$^{c}$               &ALMA ID               &P.I.$^{d}$      &$\sigma_{line}$ $^{e}$                  &$\theta_{beam}$ $^{f}$         &Ref.$^{g}$\\
                   &                           &                &(mJy/beam)                                     &$(\si{\arcsecond})$                    &\\
\hline
\rownumber &2015.1.00228.S &GP          &0.60                                           &1.66                                   &P17\\  
\rownumber &2015.1.01247.S &MB          &0.68                                           &5.39                                   &B17\\  
\rownumber &2015.1.01247.S &MB          &1.20                                           &2.65                                   &B17\\  
\rownumber &2015.1.01247.S &MB          &0.73                                           &2.22                                   &B17\\  
\rownumber &2015.1.01247.S &MB          &0.74                                           &2.58                                   &B17\\  
\rownumber &2013.1.01153.S &PL          &0.71                                           &0.44                                   &T17\\  
\rownumber &2013.1.01153.S &PL          &0.69                                           &0.45                                   &T17\\  
\rownumber &2013.1.01153.S &PL          &1.60                                           &0.34                                   &T17\\  
\rownumber &2012.1.00240.S &RW          &0.66                                           &0.17                                   &--\\           
\rownumber &2011.0.00206.S &RW          &1.84                                           &0.48                                   &W13\\  
\rownumber &2015.1.01115.S &FW          &1.35                                           &0.83                                   &D18\\  
\rownumber &2011.0.00206.S &RW          &1.39                                           &0.57                                   &W13\\  
\rownumber &2015.1.01115.S &FW          &1.35                                           &0.93                                   &D17\\  
\rownumber &2011.0.00206.S &RW          &1.81                                           &0.45                                   &W13\\  
\rownumber &2013.1.00815.S &CW          &0.58                                           &0.37                                   &W15\\  
\rownumber &2015.1.01115.S &FW          &1.30                                           &0.62                                   &D17\\  

\hline
\end{tabular}
}
\end{minipage}\hfill
\begin{minipage}{0.5\hsize}
\vspace{0.2cm}
\resizebox{\hsize}{!}{%
\begin{tabular}{| l c c c c c}
\hline\hline
No.$^{c}$               &ALMA ID               &P.I.$^{d}$      &$\sigma_{line}$ $^{e}$  &$\theta_{beam}$ $^{f}$         &Ref.$^{g}$\\
                   &                           &                        &(mJy/beam)                     &$(\si{\arcsecond})$            &       \\
\hline
\rownumber &2015.1.01115.S &FW          &1.25                           &1.24                                   &V18\\          
\rownumber &2015.1.01115.S &FW                  &1.10                           &1.04                                   &D18\\  
\rownumber &2012.1.00240.S &RW          &0.62                           &0.22                                   &J17\\  
\rownumber &2015.1.01115.S &FW          &1.28                           &0.82                                   &D18\\  
\rownumber &2013.1.00815.S &CW          &0.60                           &0.39                                   &W15\\  
\rownumber &2016.A.00018.S &RD          &0.54                           &0.27                                   &D17\\  
\rownumber &2015.1.01115.S &FW          &1.21                           &0.92                                   &D18\\  
\rownumber &2015.1.01115.S &FW          &1.20                           &0.89                                   &D18\\  
\rownumber &2016.1.00544.S &EB          &0.64                           &0.25                                   &--\\           
\rownumber &2015.1.00606.S &CW           &0.58                          &0.63                                   &W17\\  
\rownumber &2015.1.01115.S &FW           &1.37                          &0.97                                   &D17\\  
\rownumber &2015.1.00399.S &BV          &0.51                           &0.19                                   &--\\           
\rownumber &2015.1.01115.S &FW          &1.01                           &0.98                                   &D18\\  
\rownumber &2012.1.00882.S &BV          &1.45                           &0.36                                   &V16\\  
\rownumber &2012.1.00882.S &BV          &1.88                           &0.42                                   &V16\\  
\rownumber &2012.1.00882.S &BV          &0.67                           &0.20                                   &V17\\  
\hline
\end{tabular}
}
\end{minipage}
\tablefoot{\textit{Top table:} $^{a}$Redshift values retrieved from SIMBAD Astronomical Database (\url{http://simbad.u-strasbg.fr/simbad/}). $^b$The flags provide information on kinematics (rot), disc inclination angle ($\beta$) as resulting from our kinematical modelling, and black hole mass ($\Mbh$) as follows: [rot]: unresolved (u) or resolved (r) kinematics. [$\beta$]: unconstrained (u) or constrained (c) disc inclination angle, lower limit (l), bimodal distribution (b). [$\Mbh$]: unavailable in the literature (u), single-epoch virial mass estimated in this work (e) or value retrieved the literature (a). \textit{Bottom table:} $^{c}$Identification numbers (No.) refer to those of the top table.  $^{d}$Principal investigator of the project: BV (Venemans, B.), CW (Willott, C.), EB (Ba\~{n}ados, E.), FW (Walter, F.), GP (Popping, G.), MB (Banerji, M.), PL (Lira, P.), RD (Decarli, R.), RW (Wang, R.). $^{e}$Line sensitivity over $10\,\si{km\,s^{-1}}$. $^{f}$Angular resolution. $^{g}$References: V16, V17, V18 \citep{Venemans+2016, Venemans+2017, Venemans+2018}, B17 \citep{Banerji+2017}, D17, D18 \citep{Decarli+2017,Decarli+2018}, J17 \citep{Jones+2017} P17 \citep{Popping+2017}, T17 \citep{Trakhtenbrot+2017}, W13 \citep{Wang+2013}, W15, W17 \citep{Willott+2015, Willott+2017}.}
\end{table*}

\section{ALMA data selection and reduction}
\label{sect:data}
We started by collecting all [CII]$_{158\mu m}$ and CO($J$$\rightarrow$$J$-$1$) (rotational quantum number $J=4,3$) observations of $z>1.5$ QSOs on the ALMA data archive public as of June 2017 for a total of $72$ QSOs in the redshift range $1.5<z<7.1$. Different ALMA bands were involved according to the atomic/molecular transition targeted and the redshift of the sources. The collected data were calibrated using the ALMA pipeline in the Common Astronomy Software Applications, \textsc{CASA} \citep{McMullin}, by executing the appropriate ALMA calibration scripts corresponding to each specific observation. Continuum images were produced for each quasar from the calibrated visibilities, by combining the line-free channels from all spectral windows in multi-frequency synthesis mode using the \textsc{CASA} task \texttt{tclean} and \texttt{briggs} weighting scheme (with robustness parameter $R=0.5$) to maximise both the signal-to-noise ratio and angular resolution. The line-free channels were determined by inspecting the visibilities in all the frequency sidebands. For those quasars in which the FIR line was not detected, we selected the line-free channel by adopting a line width of $300\,{\rm km\,s^{-1}}$ and the redshift from literature.

These same channels were also used to produce a UV plane model by fitting the continuum emission with a zeroth order polynomial\footnote{For a typical SNR$\sim60-100$ over a bandwidth of 4 GHz in ALMA band 3, the continuum emission is well-described by a zeroth-order polynomial within the uncertainties.} that was then subtracted from the spectral windows containing the line using the \textsc{CASA} task \texttt{uvcontsub}. The continuum-subtracted line visibilities were then imaged using \texttt{tclean}. In order to recover all the information within the resolution element, the pixel size was commonly set to $\sim B_{min}/7$, where $B_{min}$ is the minor FWHM of ALMA's synthesised beam. Therefore, we obtained cubes with a typical pixel size of $0.025\si{\arcsecond}-0.05\si{\arcsecond}$ and with a spectral bin width set to $40-70$ km/s. Self-calibration was attempted but showed no additional improvement for almost all observations and was not used for the final cubes.
Finally, both continuum images and the line cubes were corrected for the primary beam response.

Among these observations, we selected the cubes in which the line detection was significant ($\apprge 3\sigma$). This first selection reduced the sample to 32 QSOs at $2.2<z<7.1$ on which we performed all of the analyses described in the following sections. Different sub-selections occurred at each step of the analysis (see Sect.~\ref{sect:subselections}) and the final sample is composed of only ten sources, for which we obtained constraints on the host galaxy's dynamical mass. We thus picked deeper observations from the archive for this final sub-sample of sources that became public while the work was in progress (by the end of February 2018). In Table~\ref{tbl:sample}, we list the starting sample of 32 objects and the characteristics of the observations, including the aforementioned deeper observations for the final sub-sample. The distribution of redshifts of our quasars is illustrated in Fig.~\ref{fig:redshift-distr}.

\section{Methods of data analysis}
\label{sect:methods}
Our goal is to measure the dynamical mass of our sample of host galaxies (listed in Table~\ref{tbl:sample}) by modelling the gas kinematics as traced by [CII]$_{158\mu m}$ or CO line emission with rotating discs. Therefore, in order to obtain the kinematical maps, we performed a spaxel-by-spaxel fit of the emission line profile by adopting a single Gaussian model with three free parameters: the amplitude $A$, the central frequency $\nu_{\it obs,}$ and the standard deviation $\sigma$. For this purpose, we designed an algorithm to achieve a robust residual minimisation in each pixel. Since the beam smearing affects the observed emission, we expect the spatial shape of the line to change smoothly from one pixel to the adjacent one, and the signal-to-noise ratio to decrease as a function of the distance from the centre of the galaxy. The underlying idea of the procedure is the subsequent performance of the line fit in all spaxels starting from the central pixel and moving away following a spiral-like path. The basic operations are: %
(1) performing a 2D Gaussian fit on the continuum image and defining the central spaxel; (2) extracting the spectrum from the central spaxel and computing a 1D Gaussian line fit in which the starting points are properly chosen by inspecting the line shape; (3) following a spiral-like path to select the next spaxel, extracting the spectrum, and performing a 1D Gaussian line fit by using the best-fitting results from the neighbour spaxels as starting points for the spectral fit; (4) continuously repeating step 3 for the consecutive spaxel until the end of the spiral-like path. We used the minimum chi-square method to estimate the best-fitting parameters.

The result of the fit in each pixel is accepted or rejected on the basis of criteria illustrated in Sect.~\ref{ssect:maps}, while the stopping criteria to break the entire fitting procedure can be fixed by setting the dimension $d$ of the spiral path, that is the distance from the central pixel. In the case of our datacubes, a typical value of $d\sim20-25$ pixels ($\sim 0.5\si{\arcsecond}-1.25\si{\arcsecond}$ depending on the pixel size, see Sect.~\ref{sect:data}) turned out to be adequate to fit the line throughout the emitting region with a total of $\sim1200-2000$ pixels analysed for each source. This fitting strategy enables a more robust minimisation compared with using a unique set of initial guess parameters for all the pixels, thus avoiding numerical problems arising from incorrectly chosen starting-points. Finally, we retrieved the information regarding the line together with the uncertainties on each spectral channel of the cube by measuring the $rms$ of the noise ($r_{\nu}$) over a wide spatial region where no emission is detected. 

\begin{figure*}[!htbp]
        \centering
        \includegraphics[width=\textwidth]{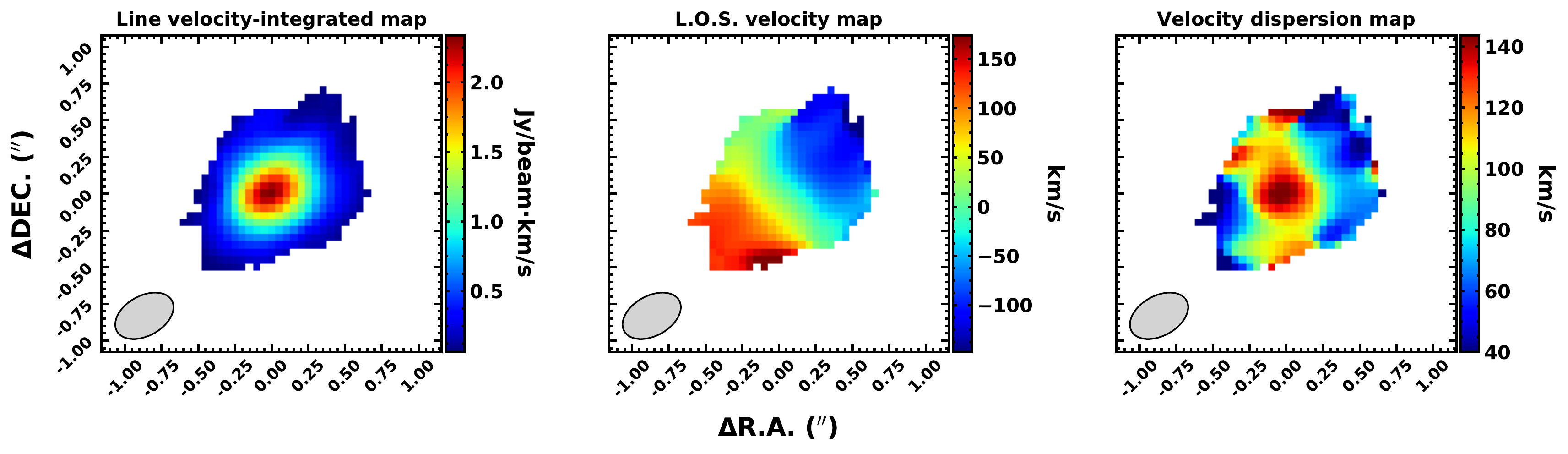}
         \caption{Observed maps of SDSS J0923+0247. From left panel to right we report, respectively, flux map, velocity map, and velocity-dispersion map along the line of sight. At the bottom-left corner of each panel, we report the ALMA synthesised beam FWHM. The coordinates indicate offsets with respect to the map centre.}
         \label{fig:3_maps}
\end{figure*}
\begin{figure}[!htbp]
        \centering
        \includegraphics[width=\hsize]{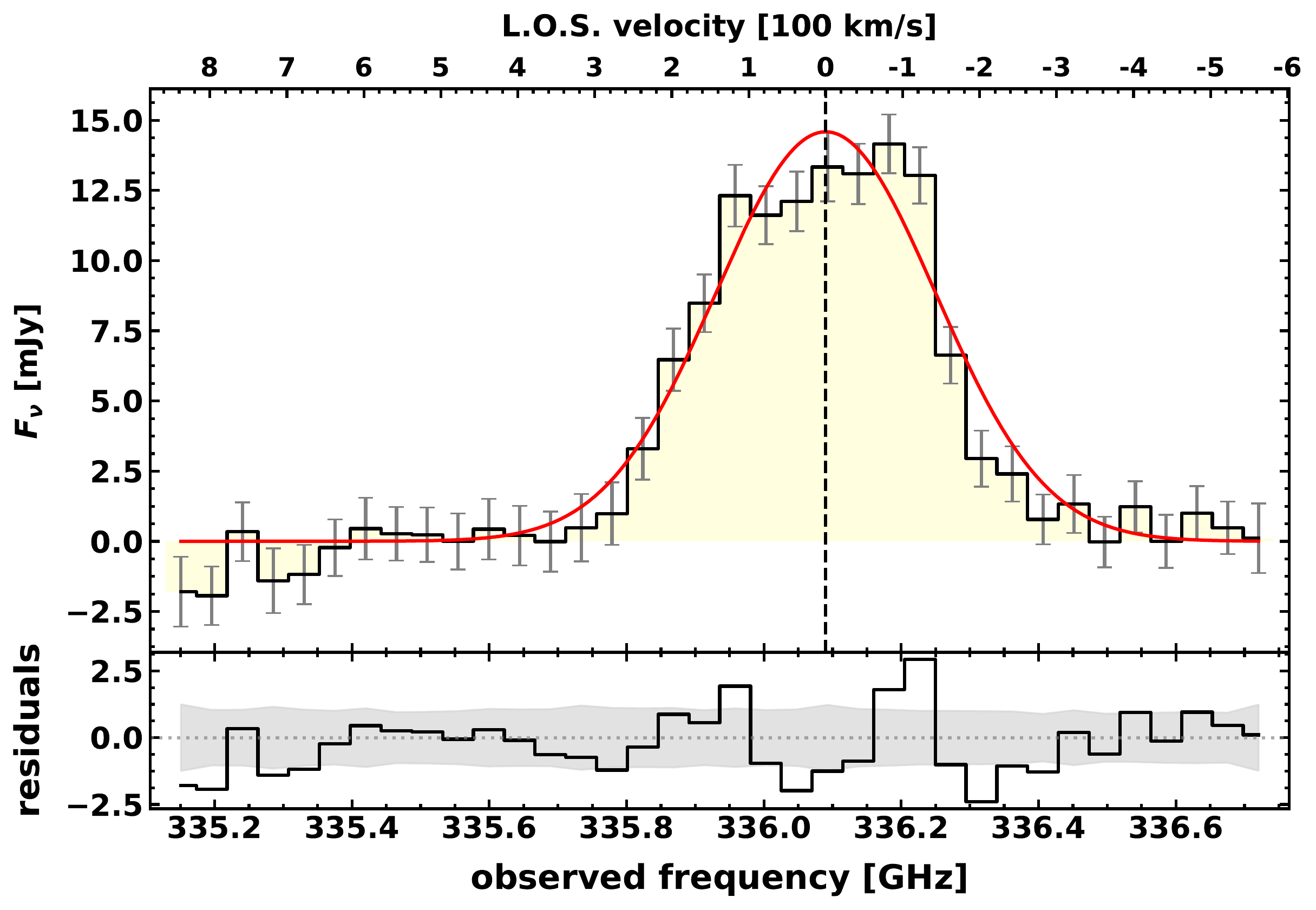}
        \caption{Integrated spectrum of source SDSS J0923+0247. \textit{Top panel}: the observed data is shown in light yellow with error bars in grey ($rms$ in each channel). The red solid curve is the best-fit Gaussian model. The velocity scale (top axes) has as its referecence %
        the central frequency of the best fit. \textit{Bottom panel}: the fit residuals (\textit{model-data}), the grey filled area shows the $rms$ along the spectral axis.}
        \label{fig:tot_flux}
\end{figure}

\subsection{Integrated spectra and derived quantities}
\label{ssec:integrated_spectrum}
We obtained the integrated spectra of all the sources by adding all the fitted spectra in spaxels selected based on criteria illustrated in Sect.~\ref{ssect:maps} (e.g. Fig.~\ref{fig:tot_flux}, see also Appendix~\ref{sect:all_maps}). Then, the resulting integrated spectrum was fitted using a Monte Carlo method in order to estimate the redshift uncertainty. Firstly, we collected a large number (e.g. 2000) of different integrated spectra obtained by adding a random value extracted from a normal distribution defined by a zero mean, and a standard deviation equal to the corresponding $rms$ in that channel ($r_\nu$) to each channel of the original spectrum. Then, we performed the fit of each spectrum with a single Gaussian, and we estimated the redshift of the line as $\nu_{\it obs}\ = \nu_{\it rest}/(1+z)$, where $\nu_{\it obs}$ is the mean of the Gaussian model and $\nu_{\it rest}$ is the line rest-frame frequency. Finally, all the estimates of $z$ obtained with this method were histogrammed and its distribution was fitted with a Gaussian model. We finally assumed the mean and the standard deviation of the best-fit model as the best value of redshift and its uncertainty, respectively. In addition, the fit of integrated spectra allowed us to determine the line FWHM and flux. In Appendix~\ref{sect:spectra_derived_quantities}, we use these quantities to derive the line luminosity, the [CII] mass ($M_{\rm[CII]}$), the total gas mass ($M_{\it gas}$) and the star formation rate (SFR) of the quasar host galaxies.

\subsection{Flux, velocity, and velocity-dispersion maps}
\label{ssect:maps}
The cube fitting procedure provides the best-fit values of the Gaussian parameters $(A, \nu_{\it obs}, \sigma)$ in each pixel. We used these values to obtain the line-integrated velocity and the velocity-dispersion maps along the line of sight (LOS).

In order to produce the maps, among all the spaxels in which we performed the line fit, we selected those satisfying the following conditions: %
(1) the peak of the best-fitting Gaussian is $\ge1.5\times rms$ in the corresponding channel; (2) the percentage relative error on the flux value is $\le50\%$.

Condition $1$ represents the signal-to-noise cut-off we used to reject pixels in which the line emission is not clearly detected. However,  in case of poor signal-to-noise ratio, the fit process possibly fails, resulting in a bad Gaussian model for which condition 1 could be still satisfied. Therefore, we also imposed condition $2$ in order to avoid this kind of effect and to consequently reject the corresponding pixels when producing the maps.

We also manually masked bad pixels far away from the galaxy centre, which are clearly associated with spikes of noise. Finally, we obtained error maps using the uncertainties on the best-fit Gaussian model parameters of each pixel. As an example, in Fig.~\ref{fig:3_maps}, we report the maps obtained from the continuum-subtracted cube of SDSS J0923+0247 (see also Appendix~\ref{sect:all_maps}).

\begin{figure}[!t]
        \centering
        \includegraphics[width=\hsize]{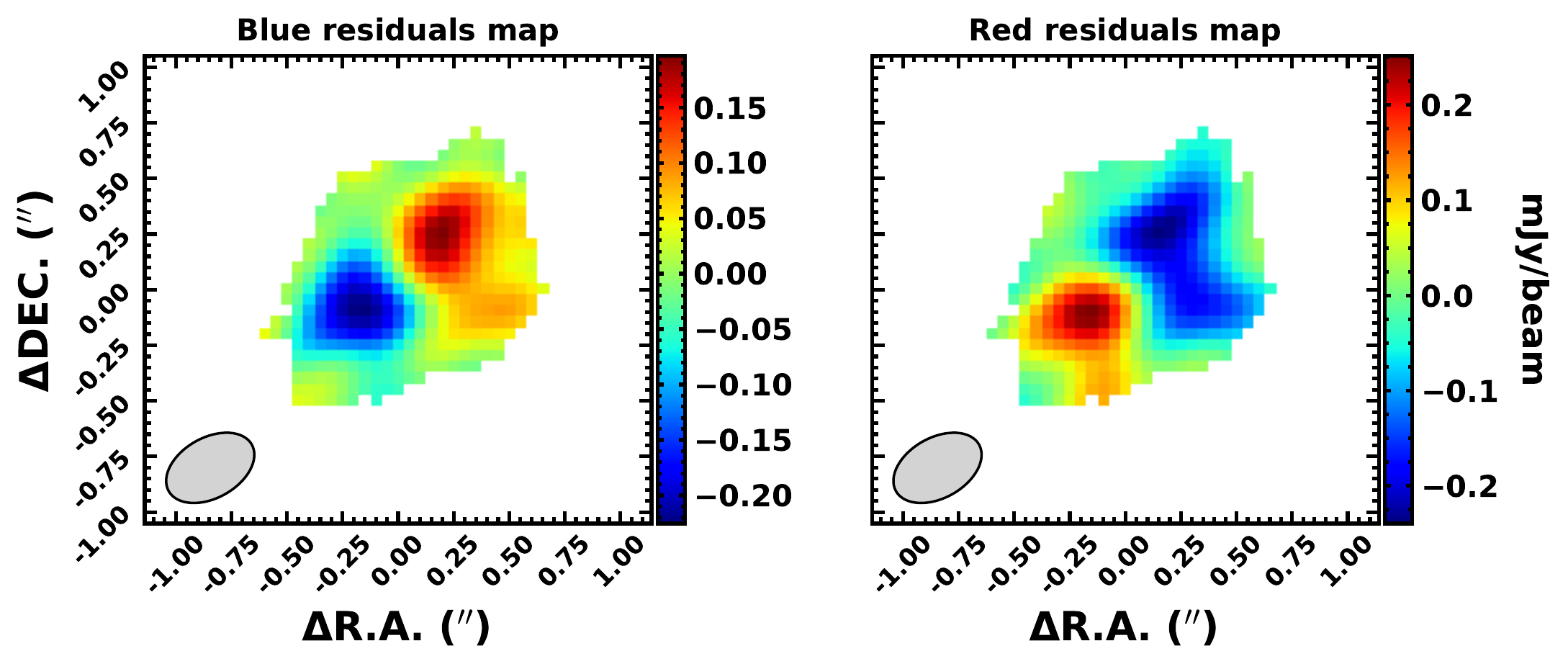}
        \caption{Red and blue residual maps of SDSS J0923+0247. This test reveals spatially resolved kinematics consistent with what we would expect from a rotating disc.}
        \label{fig:vel_grad}
\end{figure}

\subsection{Red and blue residuals maps}
\label{ssect:red-blue}
The angular resolution may not be high enough to spatially resolve the rotation of the emitting gas in host galaxies. In order to assess if the kinematics are spatially resolved or not, we performed the same analysis computed by \citet{Carniani+2013} for ALMA [CII] observations of a QSO at $z=4.7$ . We replicated the spaxel-by-spaxel fit of continuum-subtracted cubes with a single Gaussian component, using the amplitude ($A$) as the only free parameter, and by fixing the values of $\nu_{\it obs}$ and $\sigma$ to the best values obtained from the fit of the integrated spectrum (see Sect.~\ref{ssec:integrated_spectrum}). Then, we computed the residuals of fits in each channel, which is the $data - model$,  and we obtained %
two maps by collapsing all channels in the blue-shifted and red-shifted (with respect to the central frequency) side of the residual spectrum. If the kinematics are consistent with a spatially resolved rotating disc, we expect the blue and red residual maps to show two symmetric lobes: a positive and negative lobe at the opposite side with respect to the map centre  (e.g. as shown in Fig.~\ref{fig:vel_grad}). Otherwise, if rotation is not resolved then we expect a random distribution of negative and positive residuals on both the maps. After performing this test on the 32 objects listed in Table~\ref{tbl:sample}, we conclude that 14 of them $(\sim45\%)$ show no evidence of spatially resolved kinematics. We therefore excluded them from the final sample (see Sect.~\ref{sect:subselections} for a summary of sample sub-selections). In Sect.~\ref{ssect:potential_biases}, we investigate biases possibly occurring while excluding these objects.

\begin{figure}[!t]
        \centering
        \includegraphics[width=\hsize]{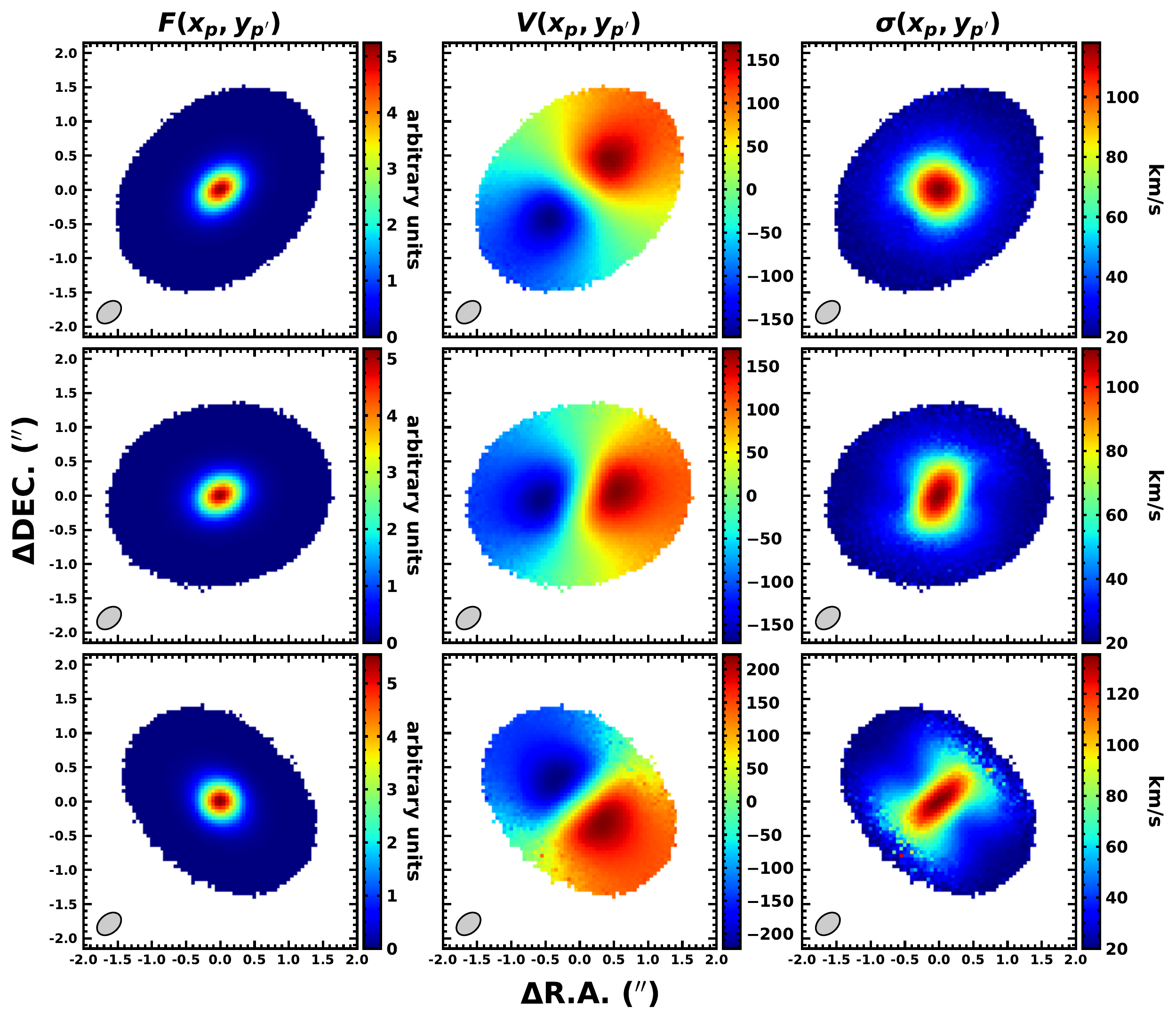}
         \caption{Simulated maps obtained with the kinematical model for a galaxy thin disc with an exponential brightness profile defined by a scale radius $R_D=0.125\si{\arcsecond}$ and dynamical mass  $M_{dyn}=5.0\times10^{10}M_{\astrosun}$. The flux, velocity, and velocity-dispersion maps along the LOS are indicated by $F(x_p,y_{p'})$, $V(x_p,y_{p'})$, and $\sigma(x_p,y_{p'}),$ respectively. From the top to the bottom panel %
         disc inclination and the position angle $(\beta,\gamma)$ are, respectively, equal to $(40,-45)$ deg, $(40,0)$ deg, and $(60,45)$ deg. The other parameters defining the model are: the FWHM of the synthesised beam ($0.4\si{\arcsecond}\times0.275\si{\arcsecond}$), the position angle of the beam ($B_{\rm{PA}}=-50$ deg), the FWHM of the LSF ($\sigma_{\rm{LSF}}=20\,\si{km\,s^{-1}}$), the bin size ($0.05\si{\arcsecond}\times0.05\si{\arcsecond}$), and the angular radius of the disc ($R=1.25\si{\arcsecond}$). The number of clouds used in this model is $N_{p}=7.5\times10^6$.}
         \label{fig:model_examples}
\end{figure}

\begin{figure}[!t]
        \centering
        \includegraphics[width=\hsize]{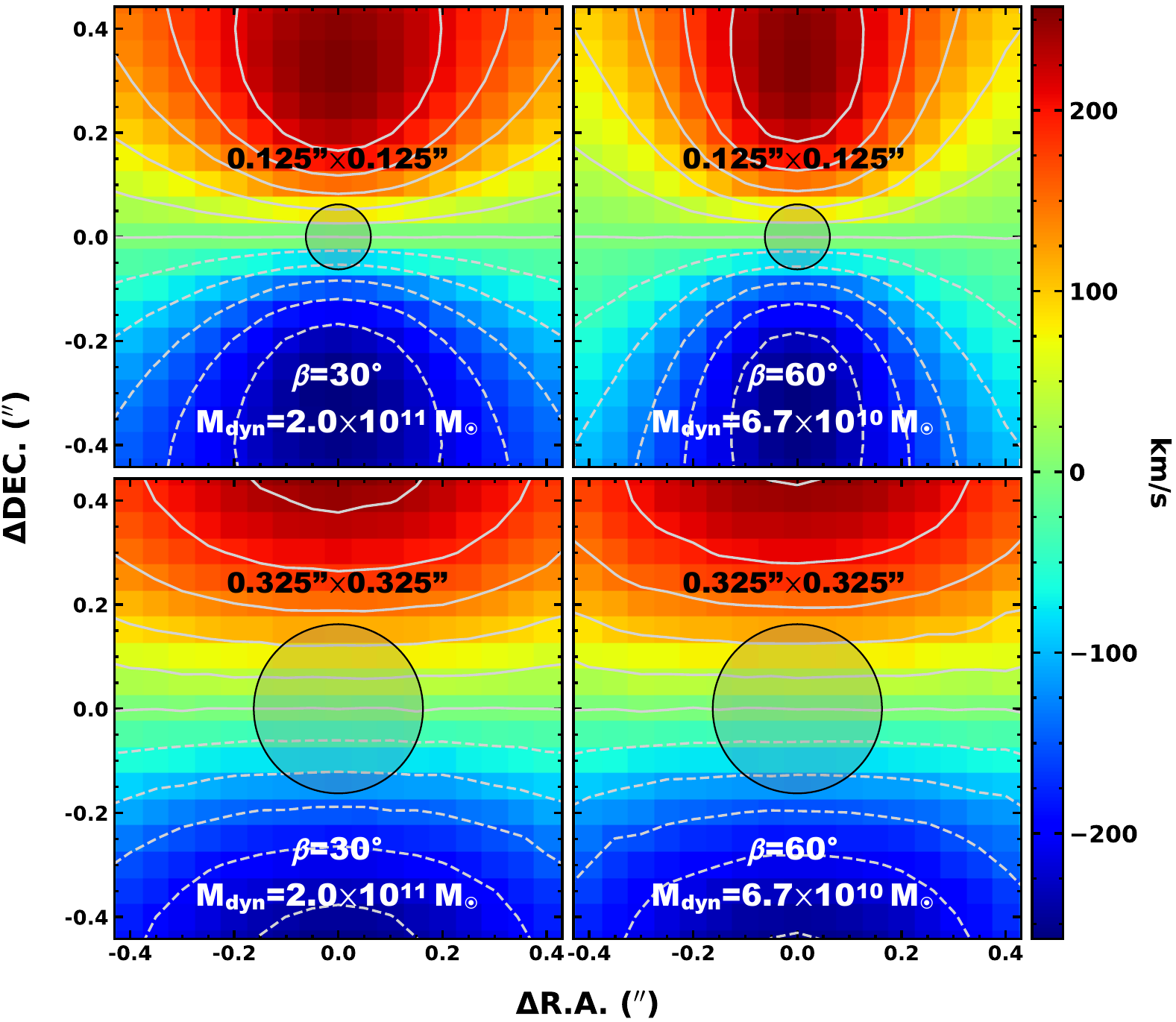}
         \caption{Simulated velocity field corresponding to four different couples of inclination angle $\beta$ and dynamical mass $M_{dyn}=5.0\times10^{10}/\sin^2\beta$. Increasing the dimension of the beam FWHM (shadowed area), the iso-velocity curves are increasingly smoothed, and different configurations of the disc appear almost indistinguishable. The simulated maps are obtained with the kinematical model for a galaxy thin disc with $R_D=0.125''$, $\gamma=-90$ deg. The other parameters defining the model are set as equal to the model in Fig.~\ref{fig:model_examples}.}
         \label{fig:beam_smearing}
\end{figure}

\section{Kinematical modelling}
\label{sect:kinematical_model}
In Sect.~\ref{sect:methods}, we obtained all the necessary information about galaxy morphology (line-integrated maps) and kinematics (velocity and velocity-dispersion maps). In order to measure the dynamical masses of the host galaxies, we designed a kinematical model to perform a 2D fit of the maps. Therefore, we assumed that: %

\begin{enumerate}
\item The observed line emission (i.e. [CII]$_{158\mu m}$ or CO transition) traces cold gas distributed in a rotating thin disc.
\item The gas mass surface density $\Sigma(r)$, is an exponential distribution that also tracks the distribution of surface brightness $I(r)$, that is:
\eq{\Sigma(r)\propto I(r)= I_0\exp\tonda{-r/R_D},\label{eq:exp_profile}}
where $I_0$ is a normalisation constant and $R_D$ is the scale radius.
\item The galaxy stellar mass is distributed as the gas mass component (Eq.~\ref{eq:exp_profile}).
\item The contribution of the dark matter is negligible.
\end{enumerate}
Under these assumptions, \citet{Freeman+1970}  showed that the corresponding circular velocity is given as:
\eq{V^2(r) = 4\pi G\Sigma_0R_Dy^2\quadra{\mathcal{I}_0(y)\mathcal{K}_0(y)-\mathcal{I}_1(y)\mathcal{K}_1(y)},\label{eq:freeman_law}}
where $\mathcal{I}$ and $\mathcal{K}$ are the modified Bessel functions evaluated at $y=r/2R_D,$ and $\Sigma_0$ is the normalisation constant of the mass distribution that accounts for both gas and stars contribution.

The total mass of the disc, which is the dynamical mass of the galaxy ($M_{dyn}$), is thus obtained by integrating the mass surface density over all the radii; $M_{dyn} = 2\pi R_D^2\Sigma_0$. By inserting this expression in Eq.~\ref{eq:freeman_law}, we can relate the total dynamical mass with the velocity curve: $V^2(r) = 2(M_{dyn}/R_D)G y^2\quadra{\mathcal{I}_0(y)\mathcal{K}_0(y)-\mathcal{I}_1(y)\mathcal{K}_1(y)}$. Therefore, by estimating $R_D$ from the flux map, we can infer the galaxy dynamical mass by performing a 2D fit of the velocity field.

\begin{figure*}[!htbp]
        \centering
        \includegraphics[width=\textwidth,trim={0 2cm 0 1cm},clip]{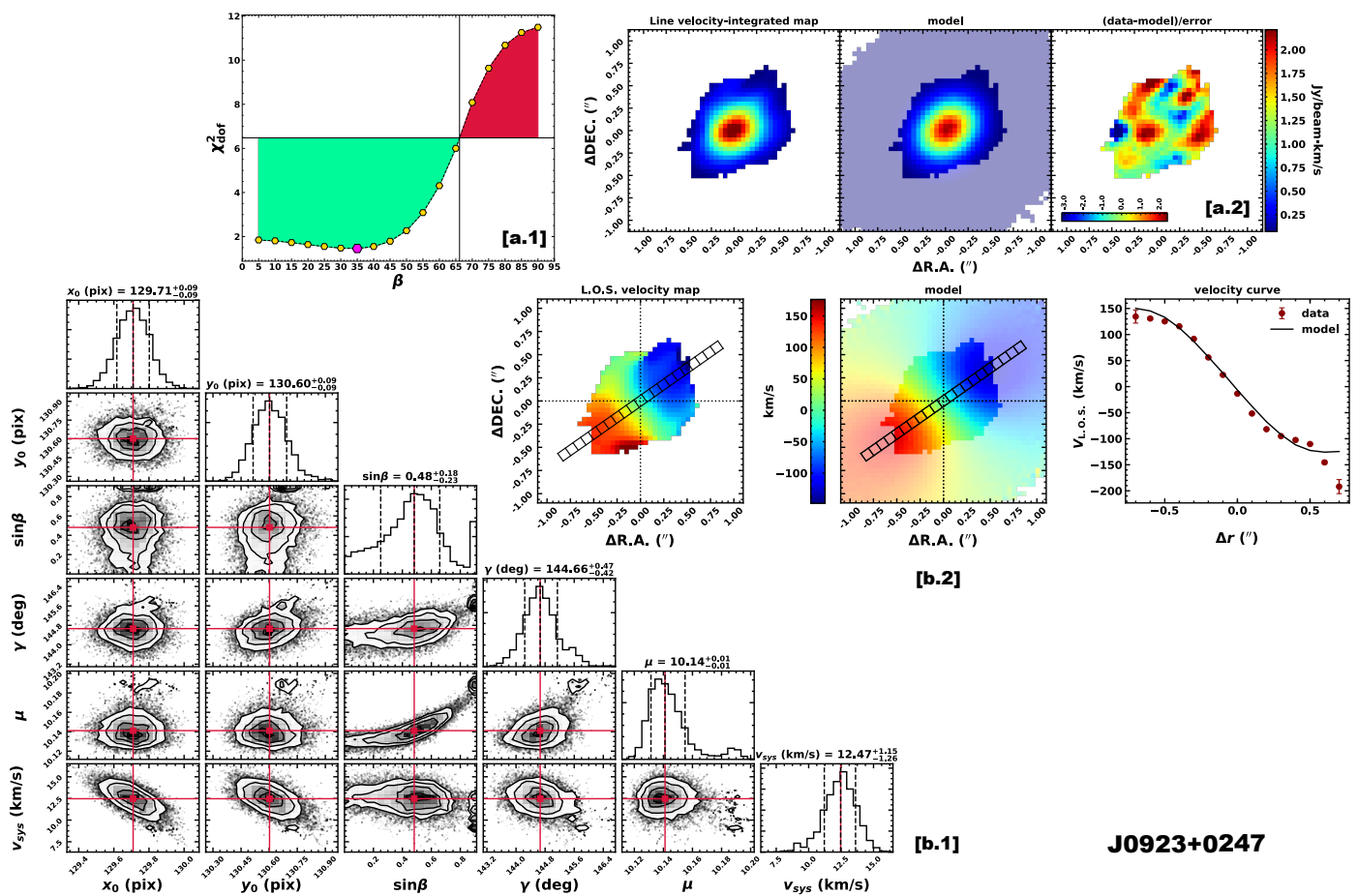}
         \caption{Figure shows the kinematical modelling performed on SDSS J0923+0247. The upper panels [a.1] \& [a.2] show the fit result of the flux map, while the bottom panels [b.1] \& [b.2] show the result of the 2D velocity field fit. {\it [a.1]:}  the curve of $\chi^2$ minima as a function of the disc inclination (see Sect.~\ref{ssec:fit_flux_maps} for details). The magenta hexagon indicates the absolute minimum ($=35\,{\rm deg}$) of $\chi_{m}(\beta)$. The green area (see Sect.~\ref{ssec:fit_velocity_maps} for the definition) indicates the allowed inclination values used as a prior in performing a 2D velocity field fit. {\it [a.2]:} 2D best-fit model of the flux map. From left to right, we report the observed map, the model, and the residuals (\textit{[data-model]/error}). {\it [b.1]:} the posterior probability distributions of the free parameters in 2D velocity field fits retrieved with the MCMC algorithm with the best values and their uncertainties. {\it [b.2]:} 2D best-fit model of the kinematical map. From left to right, we report the observed velocity map, the model, and the velocity curves extracted from a long-slit of two pixels in width aligned with the line of nodes. The slit is superimposed on maps; red circles and solid black lines in the right panel indicate the observed flux-weighted velocity values in each bin of the slit and the model, respectively.}
         \label{fig:fitted_maps}
\end{figure*}

\subsection{Details on the kinematical model and the strategy of the analysis}
\label{ssect:fitting_strategy}
The kinematical model is calculated using Monte Carlo methods. At first, the 3D space is randomly filled with $N\gg1$ point-like sources uniformly distributed in a thin disc. Each source represents a "cloud" that contributes with a unit of flux in the computation of the total observed flux. Then, the 3D disc model is projected on the sky plane and convolved with the appropriate instrumental point spread function (PSF) and line spread function (LSF) of the observation. Thus, the flux map, the flux-weighted velocity map and the velocity-dispersion map are obtained through 2D-weighted histograms by using the pixel size of the corresponding observed map as the bin width. By properly choosing the weights, we can set the flux contribution of each cloud forming the model in order to reproduce any brightness (density) and velocity profiles. We set the weights in order to create an exponential thin disc defined by Eq.~\ref{eq:exp_profile} and Eq.~\ref{eq:freeman_law}. As an example, in Fig.~\ref{fig:model_examples}, we show the simulated flux, velocity %
and velocity-dispersion maps obtained with our kinematical model in three different geometrical configurations.

In order to recover the galaxies dynamical masses, we basically adopted the same method used in \citet{Cresci+2009} and \citet{Carniani+2013}. %
We first performed a 2D fit of the observed flux map using a thin disc model with an exponential brightness profile, and we recovered the best value of scale radius $R_D$ (see Eq.~\ref{eq:exp_profile}); then, by using the resulting $R_D$ value, we computed the velocity field of our disc model accordingly with Eq.~\ref{eq:freeman_law}, and we performed a 2D fit of the LOS velocity map, thus recovering the best estimate of $M_{dyn}$.

\setcounter{magicrownumbers}{0}
\begin{table*}
\setcellgapes{1.2pt}\makegapedcells
\caption{Key parameters estimated from integrated spectra and kinematical modelling.}             
\label{tbl:results_data}      
\centering          
\begin{tabular}{l l c c c c}
\hline\hline       
No.$^{a}$       &Object ID              &               $z_{\rm{[CII]}}$$^{b}$&          $R_D$$^{c}$                     & $M_{dyn}$$^{d}$                               &$\sin\beta$\\
        &                               &                                               &               ($\si{kpc}$)                     &($10^{10}$ M$_{\astrosun}$)            &\\
\hline
2               &\object{VHS J2101-5943}$^{\dagger}$     &       $2.307262\pm0.000007$&  $2.95 \pm 0.11$&              $2.6^{+1.5}_{-0.9}$&                            $0.73^{+0.17}_{-0.15}$\\

6               &\object{SDSS J1328-0224}&       $4.64616\pm0.00002$&            $0.776 \pm 0.015$&             $0.26^{+0.03}_{-0.02}$&                 $0.84^{+0.02}_{-0.03}$\\

7               &\object{SDSS J0923+0247}&       $4.654876\pm0.000015$&  $0.850 \pm 0.008$&         $0.6^{+1.6}_{-0.3}\times10^{1}$&        $0.48^{+0.18}_{-0.23}$\\

9               &\object{SDSS J0129-0035}        &       $5.778883\pm0.000010$&  $0.626\pm0.006$&                $>0.78\times10^{1}$&                    $<0.16$\\

10              &\object{SDSS J1044-0125}        &       $5.78440\pm0.00006$&            $1.14\pm0.07$&          $3.7^{+3.2}_{-0.3}$&                            $0.90^{+0.03}_{-0.26}$\\

11              &\object{SDSS J1306+0356}        &       $6.03332\pm0.00003$&            $3.21\pm0.09$&          $1.8^{+0.9}_{-0.4}$&                            $0.70^{+0.08}_{-0.13}$\\

12              &\object{SDSS J2310+1855}        &       $6.002841\pm0.000011$&  $0.876\pm0.006$&                 $3.3^{+0.9}_{-0.9}$&                            $0.58^{+0.11}_{-0.06}$\\

14              &\object{SDSS J2054-0005}        &       $6.038828\pm0.000009$&  $0.595 \pm 0.007$&             $0.7^{+4.5}_{-0.3}$&                            $0.64^{+0.21}_{-0.42}$\\

19              &\object{ULAS J1319+0950}        &       $6.13334\pm0.00004$&            $1.50 \pm 0.03$&              $>4\times10^{2}$&                               $<0.12$\\

22              &\object{PSO J308-21}&           $6.23265\pm0.00004$&            $0.9\pm0.02$&                   \multicolumn{2}{c}{\textsc{bimodal distributions}}\\

25              &\object{PSO J183+05} &          $6.43835\pm0.00002$&            $1.30\pm0.01$&           $>2\times10^{1}$&                               $<0.24$\\

26              &\object{PSO J167-13}&           $6.514770\pm0.000015$&  $4.28 \pm 0.08$           &$1.69^{+0.14}_{-0.11}\times10^{1}$     &$0.83^{+0.03}_{-0.04}$\\

28              &\object{VIKING J0305-3150} &    $6.61434\pm0.00002$&            $1.231\pm0.019$&                \multicolumn{2}{c}{\textsc{bimodal distributions}}\\

\hline                  
\end{tabular}
\tablefoot{$^{a}$Source identification numbers in agreement with those listed in the first column of Table~\ref{tbl:sample}. %
$^{b}$Redshift estimates obtained from the line integrated spectra. $^{\dagger}$This source is observed in CO(3-2), therefore the redshift estimate refers to this line. $^{c}$Uncertainties on scale radius are statistical errors provided from $\chi^2$-minimisation algorithm (see Sect.~\ref{ssec:fit_flux_maps}). $^{d}$Uncertainties on dynamical masses are statistical errors computed using the posterior probability distributions of $M_{dyn,}$ ignoring any possible systematic biases: lower limits, nominal values and the upper limits correspond to 16th, 50th, and 84th percentiles, respectively.  Bimodal posterior probability distributions of the inclination angle (and dynamical mass) are explicitly indicated. In the case of bimodal distributions, no values are provided and the corresponding object is rejected from the final sample.}
\end{table*}

\subsection{Estimation of $R_D$: 2D fit of the flux maps}
\label{ssec:fit_flux_maps}
Following the method illustrated in the previous section, we first estimated the scale radius $R_D$ on the 18 flux maps of the sources with spatially-resolved kinematics. The typical angular extension of the observed maps is $\sim1''$, with a pixel size depending on the beam size of the observation (see Sect.~\ref{sect:data}) resulting in a typical map size of $\sim15-20$ pixels in linear diameter. We thus generated simulated maps using a 3D disc model with radius of $R=20$ pixels filled by $N_{p}=5\times 10^6$ clouds. These values turned out to be the best compromise to smooth the stochastic oscillations of the cloud's numerical density and to avoid spurious numerical effects at the boundary of the model, while simultaneously keeping the computational time relatively short.

The 3D disc model is then projected on the sky plane, and, according to Eq.~\ref{eq:exp_profile}, the observed image of the simulated flux map depends on the normalisation constant $I_0$, the scale radius $R_D,$ and on the geometrical parameters: the coordinates of the map centre $(x_0,y_0)$, the inclination with respect to the sky plane ($\beta$) and the position angle of the line of nodes ($\gamma$) measured clockwise with respect to the east. We note that, for the purposes of this work, we were not interested in the physical value of $I_0$.

The aforementioned parameters were variable during the fit procedure. Thus, to retrieve their best estimations, we carried out the 2D map fit using the \texttt{cap-mpfit} PYTHON procedure part of the \texttt{pPXF} package by \citet{Cappellari+2004} based on \texttt{minpack-1} \citep{More+1980}, performing a Levenberg-Marquardt least-squares minimisation between the data and the model.
For each map, the best model minimised the following function:
\eq{\chi^2 = \sum_{p,p'}\quadra{\frac{\tilde{F}(x_p,y_{p'})-F(x_p,y_{p'};[I_0,R_D,\beta,\gamma,x_0,y_0])}{\sigma_{F}(x_p,y_{p'})}}^2+\mathcal{P},\label{eq:chi_flux}}
where $\tilde{F}$, $F$ and $\sigma_F$ are, respectively, the observed flux map, the flux model map, and the flux error map. 
We note that, in addition to the standard $\chi^2$ function, we inserted a penalty term $\mathcal{P}=5\times N_{out}$ in Eq.~\ref{eq:chi_flux}, where $N_{out}$ is the number of pixels defined in the data but not in the model. Indeed, the sum in Eq.~\ref{eq:chi_flux} is computed only taking into account the pixels $(x_p,y_{p'})$ on the observed map in which the model is defined.
Therefore, unless the model is defined in all the pixels in which the observed data are present, during the minimisation process, the penalty term $\mathcal{P}$ ensures the adequate penalisation of configurations for which the model cannot reproduce the data in all the points (e.g. since the disc model is thin, completely edge-on disc configurations are highly unlikely unless the PSF is large enough).

In order to obtain a robust $\chi^2$ minimisation avoiding the convergence to a possible local minimum, the 2D flux map fit is performed multiple times by fixing the disc inclination angle ($\beta$) to $5\,{\rm deg}$ and up to $90\,{\rm deg}$ with a step size $\Delta\beta = 5\,{\rm deg}$.
For each value of $\beta$, the $\chi^2$ function in Eq.~\ref{eq:chi_flux} is minimised with respect to the free parameters $[I_0,R_D,\gamma,x_0,y_0]$. At the end of each step, we retrieve the minimum of $\chi^2(\beta)$ function and the corresponding set of best values of free parameters. We then used them as starting points for the next step. Once the minimisation is performed for all the inclination values in the range $[5,90]\,{\rm deg}$, we sampled the curve of the minima as a function of the disc inclination angle, which is $\chi^2_m(\beta)$. Finally, by finding the absolute minimum of $\chi^2_m(\beta)$, we retrieved the best set of $[I_0,R_D,\beta,\gamma,x_0,y_0]$. Following the method illustrated above, we estimated the best value of $R_D$ measured in arcseconds for all 18 objects indicated in Table~\ref{tbl:sample} with flag [rot]="r". Finally, we computed $R_D$ values in physical length, using the redshift estimates obtained in Sect.~\ref{ssec:integrated_spectrum}. 

In the next section, we use the $R_D$ estimates to compute the kinematical model in order to perform the 2D fit of the velocity fields. For this purpose, we use the $\chi^2_m(\beta)$ curve as a prior knowledge on disc inclination angle. As an example, in Fig.~\ref{fig:fitted_maps}, we show flux map modelling results and the correspondent $\chi_{m}^2(\beta)$ curve for SDSS J0923+0247. The $R_D$ values are listed in Table~\ref{tbl:results_data} for those objects with constraints on dynamical mass (see also Appendix~\ref{sect:all_maps}).

\subsection{Estimation of $M_{dyn}$: 2D fit of the velocity maps}
\label{ssec:fit_velocity_maps}
In order to estimate the dynamical masses of the quasars sample, we performed the 2D fit of LOS velocity maps. At high-$z$, uncertainties on the dynamical mass estimates are mainly driven by the poor angular resolution of observations. As the integrated flux map of line emission, even the velocity maps are affected by beam smearing, thus introducing additional uncertainties in the fitting parameters \citep[as also pointed out by other authors, see e.g.][]{Bosma1978,Begeman1987,deBlok+1997,OBrien+2010,Swaters+2000,Epinat+2009, Epinat+2010,Swaters+2009,Carniani+2013,Kamphuis+2015}. This effect leads to the disc inclination angle and the dynamical mass becoming almost degenerate parameters, meaning the observed velocity field can be similarly reproduced by different couples ($\beta$, $M_{dyn}$) with similar $M_{dyn}\sin^2\beta$, thus providing very near values of $\chi^2$ function \citep[see also e.g.][]{Epinat+2010}. In Fig.~\ref{fig:beam_smearing}, we show the effect of the beam smearing on the iso-velocity curves of simulated velocity fields.

Consistently with the method illustrated in Sect.~\ref{ssect:fitting_strategy}, we performed fits of velocity maps using thin rotating disc models defined by exponential mass distributions with the $R_D$ values estimated in Sect.~\ref{ssec:fit_flux_maps}. In order to retrieve the best-fitting model and to estimate the parameter uncertainties, we used the PYTHON affine invariant Markov chain Monte Carlo (MCMC) ensemble sampler \texttt{emcee} \citep{Foreman+2013}. For this purpose, we defined the likelihood function of a model, given the data, as:
\begin{align}\nonumber\ln &\,p =-\frac{1}{2}\tonda{\frac{\sopra{x_0}-x_0}{\Delta x_0}}^2-\frac{1}{2}\tonda{\frac{\sopra{y_0}-y_0}{\Delta y_0}}^2+\\
&\nonumber-\frac{1}{2}\sum_{p,p'} \Biggl\{\frac{\quadra{\tilde{V}(x_p,y_{p'})-V(x_p,y_{p'};[\mu,\gamma,\sin\beta,x_0,y_0])-V_{sys}}^2}{\sigma_{V}^2(x_p,y_{p'})} +\\
&+\ln\biggl [2\pi\tonda{\sigma_V^2(x_p,y_{p'})}\biggr]\Biggr\}-\frac{1}{2}\mathcal{P},\label{eq:lnlike}\end{align}
where $\tilde{V}$, $V,$ and $\sigma_{V}$ are the observed velocity map, simulated velocity map, and velocity error map, respectively. 

The simulated field depends on geometrical parameters, the scale radius, and the dynamical mass (see Eq.~\ref{eq:freeman_law}). However, instead of $M_{dyn,}$ we used $\mu=\log(M_{dyn}\sin^2\beta)$ as free parameter, since it is decoupled from disc inclination, thus it uniquely determines the intrinsic velocity field. Here, we also took into account the systemic velocity of the galaxy, which is $V_{sys}$, as an additional free parameter. Furthermore, defining the likelihood function in Eq.~\ref{eq:lnlike}, we assumed Gaussian priors for the coordinates of the galaxy centre with maximum probability corresponding to $(x_0,y_0)$ obtained from the flux map modelling, as illustrated in Sect.~\ref{ssec:fit_flux_maps}. Here, $\Delta x_0$ and $\Delta y_0$ are the standard deviations assumed equal to $0.1$ pixel. In addition, we assumed a box-like prior on the position angle of the disc ($-180 \le\gamma(\text{deg})\le180$) and on the dynamical mass ($M_{\rm{dyn}}>0$). Finally, we accounted for prior knowledge on disc inclination from morphology by imposing that, during the fitting process, $\Delta \chi_m^2(\beta) < \max\{\Delta \chi_m^2(\beta)\}/2 $; where $\Delta \chi_m^2(\beta)=\chi^2_m(\beta)-\chi^2_M$ with $\chi^2_M$ the absolute minimum of $\chi^2_m(\beta)$ curve resulting from the fit of the flux map (see Sect.~\ref{ssec:fit_flux_maps}).

We thus maximised Eq.~\ref{eq:lnlike} using $[V_{sys},\mu,\gamma,\sin\beta, x_0, y_0]$ as free parameters, and we recovered %
their posterior probability distributions\footnote{We set up the MCMC procedure with 50 walkers performing 1000 steps each for a total of $5\times10^{4}$ evaluations of the log-likelihood function.}. Finally, the best values of parameters and related uncertainties were estimated by computing the 16\textit{th}, 50\textit{th,} and 84\textit{th} percentile of the distributions. As an example, Fig.~\ref{fig:fitted_maps} shows the kinematical modelling of SDSS J0923+0247 (see also Appendix~\ref{sect:all_maps}).

Due to the poor spatial resolution of some observations, we successfully constrain the disc inclination and the dynamical mass for only 13 objects of the sample (see Table~\ref{tbl:results_data}). We note that for two of them we find bimodal distributions for $\sin\beta$ and $\mu$, thus not permitting us to define unique values.

\section{Determination of the black hole masses}
\label{sect:BH_masses}
Currently, the only possible technique to carry out black hole mass estimates at high-$z$ is the use of single epoch (SE) virial relation, which combines the FWHM or the line emission that originated in the broad line region (BLR) of quasar,  with the continuum luminosity emitted from the BH accretion disc (e.g. \citealt{McLure+2004,Vestergaard+2006,Vestergaard+2009}; but see also e.g.
\citealt{Trevese+2014,Grier+2019}). This approach assumes that the BLR is virialised and that there is a tight relation between the BLR radius ($R_{\rm{BLR}}$) and the continuum luminosity of the AGN ($L_{\rm{AGN}}$) \citep[e.g.][]{Kaspi+2005, Bentz+2006,Bentz+2009}. Under these assumptions, $L_{\rm{AGN}}$ and the FWHM of the broad emission lines are used as proxies for $R_{\rm{BLR}}$ and virial velocity, respectively. 

To date, thanks to the effort of various groups \citep[e.g.][]{McLure+2002,McLure+2004,McGill+2008,WangJ+2009,Shen+2011b,Denney2012,Park+2013,Coatman+2017}, many relations have been calibrated by employing different broad lines in order to infer the $\Mbh$ and by assuming that such BH-mass estimates are in agreement with reverberation mapping BH masses \citep{Vestergaard+2006}, which, in turn, are in agreement with local BH-galaxy scaling relations for normal galaxies \citep{Onken+2004}. This is motived by the unknown geometry and kinematics of the BLR \citep[e.g.][]{McLure+2001,Onken+2004}. In high-redshift quasars, the atomic transitions of MgII and CIV are the most common and brightest BLR lines that are observed in optical (rest-frame UV) spectra, and thus they are extensively used as virial mass estimators with the corresponding continuum luminosity measured by convention at $3000\AA$ and $1550\AA$ for MgII and CIV, respectively. However, the reliability of CIV line is still strongly debated. Firstly, the CIV scaling relation is based on very few measurements \citep{Kaspi+2007,Saturni+2016, Park+2017}; secondly the CIV line is often associated with broad and blueshifted wings likely resulting from outflows \citep[e.g.][]{Richards+2011, Denney2012} that may affect the measurement of the line width biasing the BH-mass estimates. We note that the aforementioned calibrations have intrinsic uncertainties of $\sim0.3-0.4$ dex \citep[see, e.g.][]{Vestergaard+2006,Denney2012, Park+2017} that are usually larger than the errors associated with line width and flux measurements.

\setcounter{magicrownumbers}{0}
\begin{table*}
\caption{Black hole masses retrieved from literature and spectroscopic data used to estimate $\Mbh$ through MgII-based virial relation.}             
\label{tbl:spec_data}      
\centering          
\begin{tabular}{l l c c c r}
\hline\hline       
No.$^{a}$       &Object ID & FWHM(MgII) & $\lambda L_{\lambda}(3000\AA)$                 &$\Mbh$ $^{b}$                                  &References $^{c}$\\ 
&               &(km s$^{-1}$)& ($10^{46}$ erg s$^{-1}$)                                                       &(M$_{\astrosun}$)                                &\\
\hline
2               &\object{VHS J2101-5943}                         &-                                      &-                            &$3.2\pm0.7\times{{10}^{10}}$              &Ban2015\\
6               &\object{SDSS J1328-0224}                                &$3815\pm954$           &$1.9\pm0.4$         &$8\pm4\times{{10}^8}$                  &Tra2011\\
7               &\object{SDSS J0923+0247}                                &$2636\pm264$           &$1.4\pm0.3$          &  $3.3\pm1.0\times{{10}^{8}}$             &Tra2011\\
9               &\object{SDSS J0129-0035}                                &-                                      &-                            &$1.7_{-1.1}^{+3.1}\times {{10}^{8}}$      &Wan2013\\
10              &\object{SDSS J1044-0125}                                &-                                      &-                            &$5.6\pm0.5\times {{10}^{9}}$              &She2019\\
11              &\object{SDSS J1306+0356}                                &$3158\pm145$           &$2.45\pm0.06$     &$6.3\pm1.5\times{{10}^{8}}$                &DeR2011\\
12              &\object{SDSS J2310+1855}                                &$4497\pm352$           &$6.027\pm0.018$ &$2.0\pm0.6\times{{10}^{9}}$            &She2019\\
14              &\object{SDSS J2054-0005}                                &-                                      &-                            &$0.9_{-0.6}^{+1.6}\times {{10}^{9}}$      &Wan2013\\
19              &\object{ULAS J1319+0950}                                &$3675\pm17$                    &$3.8\pm1.0$         &$1.1\pm0.2\times{{10}^{9}}$            &Sha2017\\
26              &\object{PSO J167-13}                                    &$2350\pm470$           &$1.5\pm0.7$            &$2.7\pm1.4\times {{10}^{8}}$             &Ven2015\\   
28              &\object{VIKING J0305-3150}                      &$3189\pm85$                    &$1.66\pm0.02$     &$5.2\pm1.2\times{{10}^8}$          &DeR2014\\
\hline                  
\end{tabular}
\tablefoot{$^{a}$Source identification numbers in agreement with those listed in the first column of Table~\ref{tbl:sample}. $^{b}$Black hole mass estimated using SE virial relation in Eq.~\ref{eq:se_virial_BH} when possible, otherwise, the value provided here is the one available in literature (see references for full details). The uncertainties we report do not include the systematic uncertainties intrinsic to the $\Mbh$ estimators. $^{c}$References: Ban2015 - \citet{Banerji+2015}; DeR2011, DeR2014 - \citet{DeRosa+2011,DeRosa+2014}; Sha2017 - \citet{Shao+2017}; She2019 - \citet{Shen+2019}; Tra2011 - \citet{Trakhtenbrot+2011}; Ven2015 - \citet{Venemans+2015}; Wan2013 - \citet{Wang+2013}}
\end{table*}

\subsection{Black hole masses from the literature}
In this work, we adopted a unique SE virial relation to estimate the BH masses of our sample homogeneously. In detail, we used the relation by \citet{Bongiorno+2014}, which was calibrated by assuming the BH-galaxy scaling relations by \citet{Sani+2011}. The latter is consistent with the relation used as a $z=0$ reference for studying the redshift evolution \citep[e.g.][]{Kormendy+2013,DML2019}:
\eq{\begin{aligned}\log\tonda{\frac{\Mbh}{M_{\asun}}}=6.6&+2\log\tonda{\frac{\text{FWHM(MgII)}}{10^3\si{km  \,s^{-1}}}}+\\&+0.5\log\tonda{\frac{\lambda L_\lambda(3000\AA)}{10^{44}\si{erg\,s^{-1}}}}.\end{aligned}\label{eq:se_virial_BH}}
Thus, where available, we retrieved MgII FWHM and the continuum luminosity estimates from the literature and, if they were unavailable, we assumed $\Mbh$ estimates as provided in the literature.%

In summary, we recover BH masses from the literature for everything except PSO J308-21 and PSO J183+05, using H$\beta$, MgII, and CIV BLR lines. For SDSS J0129-0035 and SDSS J2054-0005, the black hole masses were estimated from the bolometric luminosity ($L_{bol}$) assuming Eddington accretion ($L_{bol}/L_{Edd}=1$). We note that this assumption is supported by some evidences showing that black holes at $z\apprge6$ accrete matter at a rate comparable to the Eddington limit \citep{DeRosa+2014, Mazzucchelli+2017}. The data are listed in Table~\ref{tbl:spec_data}. The BH masses computed in this work using \citet{Bongiorno+2014} calibrations are a factor $\sim2$ smaller than those reported in literature using different calibrations (see references in Table~\ref{tbl:spec_data} for full details). However, both estimates are consistent within the typical uncertainties ($\sim0.4$ dex).

\subsection{Black hole mass from LBT data}
The study of the redshift evolution of $\Mbh-M_{dyn}$ relation can be severely affected by reliability of $\Mbh$ estimates; in particular, the available spectroscopic information for our sample did not allow us to derive $\Mbh$ measurements with a unique method. We therefore started an observational campaign with the LBT (Large Binocular Telescope), of those sources with estimated $\Mbh$ assuming $L_{bol}/L_{Edd} =1$ (SDSS J0129-0035, SDSS J2054-0005, and SDSS J2310+1855), and of two additional targets without previous $\Mbh$ estimates from the literature (PSO J308-21 and PSO J138+05; even though dynamical masses of these two sources are tentative). We thus obtained NIR spectra of the quasars with LUCI (LBT Utility Camera in the Infrared) targeting the CIV line and the adjacent continuum, which are redshifted into the $zJ$ filter (central $\lambda=1.17\,\si{\mu m}$), for everything except PSO J183+05. In fact, for all these sources, the MgII line falls in a spectral region with a very low atmospheric transmission. For PSO J183+05, instead, we targeted the BLR MgII line, which is redshifted at $\approx2.0817\,\si{\mu m}$ and can be observed with the $HK$ filter (central $\lambda=1.93\,\si{\mu m}$).

Unfortunately, due to poor weather conditions, we did not achieve the requested sensitivities. No BLR emission lines have been detected in any of the quasars except J2310+1855, from which we derived a CIV-based $\Mbh$ estimate of $6\times10^{9}\,M_{\astrosun}$ (see Appendix~\ref{sect:LBT_observations} for full details of the observations). Our estimation is consistent, within the error, with that reported by \citet{Feruglio+2018} and \citet{Shen+2019}. We note that \citet{Shen+2019}, who published NIR spectra of a large sample of $z\sim5.7$ QSOs, also provide a $\Mbh$ measurement for J2310+1855 through virial relation based on MgII as well. In the following, we refer to the MgII estimate from \citet{Shen+2019}, because of the aforementioned issues related to CIV-based measurements.

\section{Summary of sample sub-selections}
\label{sect:subselections}
The data analysis described in Sects.~\ref{sect:methods} and~\ref{sect:kinematical_model} was performed on the 32 continuum-subtracted cubes of the sources listed in Table~\ref{tbl:sample}. Each step of the analysis has led to the rejection of a number of objects that turned out not to be suitable for the method adopted in this work. Here, we briefly summarise the different sub-selections used throughout this work:

\begin{enumerate}
\item By inspecting the velocity maps and red/blue residuals maps (see Sect.~\ref{ssect:red-blue} for details), we found that 14 out of 32 objects ($\sim45\%$, flagged with [rot]="u" in Table~\ref{tbl:sample}) do not show spatially resolved kinematics or rotating disc kinematics. This is possibly due to the presence of outflows or merging events, or of a companion located in proximity (projected on the sky plane) of the quasar. For this purpose, velocity-dispersion maps provide additional information on the kinematics. However, a comprehensive interpretation of the complex velocity fields observed in these sources is beyond the scope of this work. As a result of this analysis, the sample has been reduced from 32 to 18 objects. On the other hand, excluding these objects from the final sample may introduce bias in the final results (see Sect.~\ref{ssect:potential_biases}). %
\item We then performed the fits of the flux and velocity maps (see Sects.~\ref{ssec:fit_flux_maps} and~\ref{ssec:fit_velocity_maps} for details) on the remaining 18 objects selected in the previous step. As a result, for five objects ($\sim 30\%$), the kinematical modelling has not enabled us to constrain the disc inclination, and consequently the dynamical mass. It is possible that incorrect assumptions on the mass distribution (see Eq.~\ref{eq:exp_profile}) and/or the poor angular resolution of the observations making inclination and dynamical mass almost degenerate parameters (see Sect.~\ref{ssec:fit_velocity_maps}) have prevented the determination of the mass in these host galaxies. In addition, the iso-velocity curves of the kinematical fields are typically distorted, suggesting the presence of non-circular motion. In particular, for two objects (PSO J308-21, VIKING J0305-3150), the posterior probability distributions of the inclination angle $\beta$ (and $\mu$) are bimodal, preventing us from constraining these parameters. For three objects (PSO J183+05, SDSS J0129-0035, ULAS J1319+0950), we derived an upper limit on the disc inclination, which is a lower limit on mass $M_{dyn}$. In summary, we obtained $8$ constrained, $2$ bimodal, and $3$ lower limit measurements of the dynamical mass (see Table~\ref{tbl:results_data}).
\item The final step is to relate our dynamical mass estimates with $\Mbh$ retrieved from the literature. We illustrate this step in Sects.~\ref{sect:BH_masses} and ~\ref{sect:BH-galaxy_relation}. Despite several studies performed in this field aiming to estimate $\Mbh$ even for high-redshift quasars, for two objects ($\sim15\%$ of the remaining 13 resulted from the previous step), black hole mass estimates were not available at the time this paper was written. Therefore, we rejected these objects from the final sample (see Table~\ref{tbl:spec_data}). These quasars are flagged with [$\Mbh$] = "u" in Table~\ref{tbl:sample}.
\end{enumerate}

\begin{figure}[!t]
        \centering
        \includegraphics[width=\hsize]{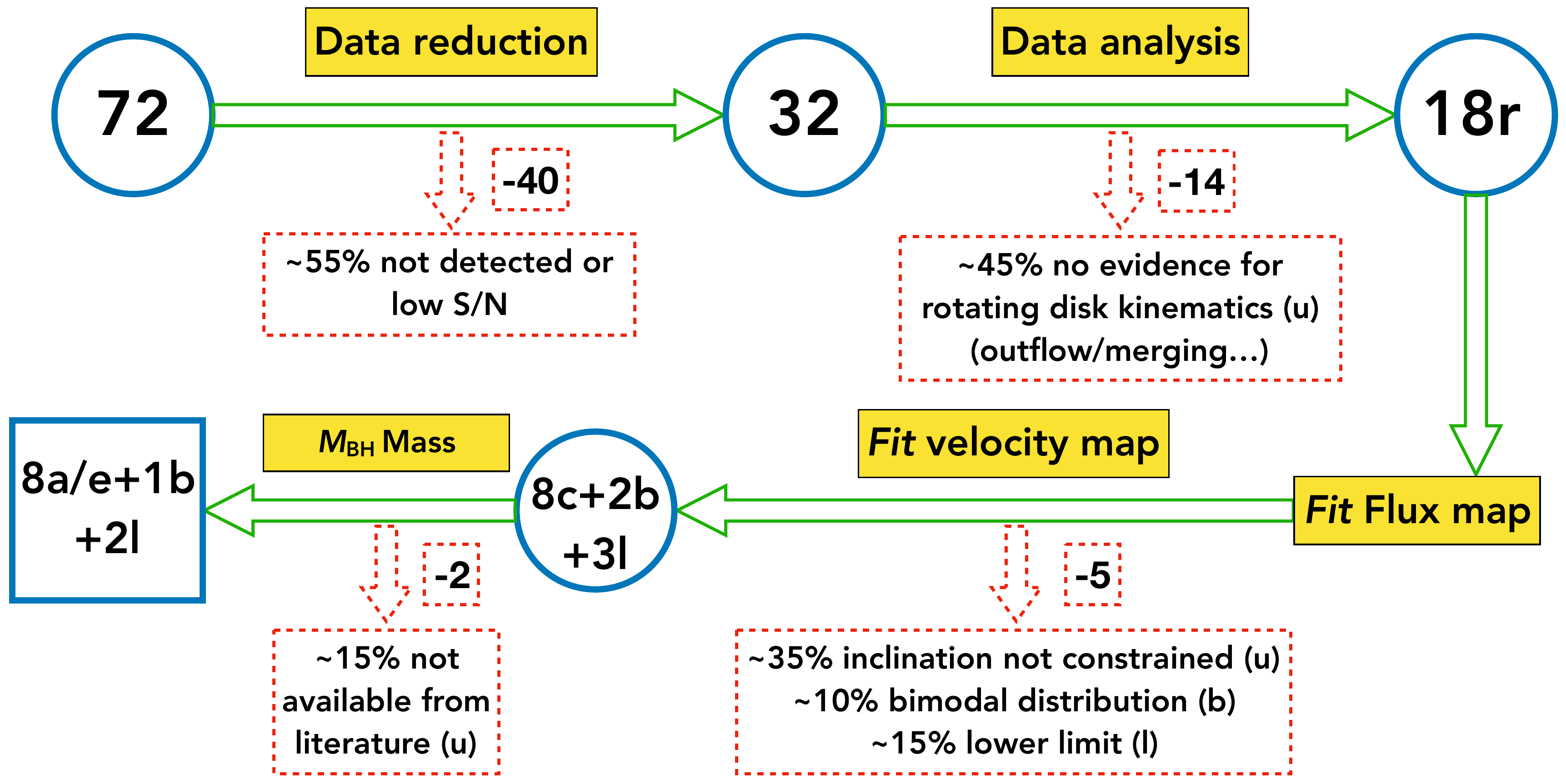}
         \caption{Scheme summarises the different sub-selections used %
         throughout this work. Starting from an initial sample of 72 quasar host galaxy observations extracted from the ALMA data archive, the final sample is composed of ten high-$z$ objects for which we study the BH-galaxy relation. The characters indicating the selection type are in accordance with Table~\ref{tbl:sample}. The number of objects rejected from the sample is also indicated at each step.}
         \label{fig:selection_scheme}
\end{figure}

These selection steps are shown in the scheme drawn in Fig.~\ref{fig:selection_scheme}. %
Overall, we were able to obtain a measurement of galaxy dynamical mass and retrieve black hole mass only for eight sources, that is $\sim10\%$ of the initial sample of 72 QSOs.

\section{Comparison of results and uncertainties on dynamical mass estimates}
\label{sect:comparison}
In Sect.~\ref{ssect:other_modelling}, we compare our results with those obtained by other authors who attempted to perform a full kinematical modelling of individual sources that belong to our sample.
Other works highlighted the presence of companion sources in the close environment of a few QSOs analysed in this work. Such satellite galaxies can disturb the gas kinematics of the host through tidal interaction, thus introducing additional uncertainties in measuring the galaxy dynamical mass. We discuss this point in Sect.~\ref{ssect:companion_presence}.
In Sect.~\ref{ssect:assumptions}, we discuss the limit of validity of the assumption of a rotating disc model and the possible mass contribution arising from random motions throughout the galaxy (see also Appendix~\ref{sect:mass_sigma}).
Finally, in Sect.~\ref{ssect:potential_biases}, we investigate observational biases possibly arising from the sub-selection of the sample.

\subsection{Comparison of results from other kinematical modelling in the literature\\ }
\label{ssect:other_modelling}
In the following sections, we discuss the results obtained from the kinematical modelling of J1319+0950, J0305-3150, J1044-0125, and J0129-0035, making a comparison between our results and those obtained in previous works.

\subsubsection{ULAS J1319+0950}
\citet{Jones+2017} and \citet{Shao+2017} performed a kinematical characterisation of [CII] emission of J1319+0950 by using a tilted rings warpless model and assuming purely circular rotation. They inferred, respectively, a dynamical mass of $15.8\times10^{10}\,M_{\astrosun}$ and $13.4\times10^{10}\,M_{\astrosun}$ and an inclination angle of $29\,\si{deg}$ and $33\,\si{deg}$ (which are roughly consistent with the value estimated by \citet{Wang+2013}) by using the axial ratio of the [CII] flux map to estimate the disc inclination and the [CII] FWHM as estimate of maximum circular velocity. Furthermore, by fitting the dust continuum emission in UV plane, \citet{Carniani+2019} inferred an inclination angle of $\sim15\,\si{deg}$. In contrast, we are not able to constrain the disc inclination angle from our kinematical modelling, resulting in a lower limit on dynamical mass ($>4\times10^{12}\,M_{\astrosun}$). The disagreement between our result and the previous ones may arise from the beam smearing effect that is not taken into account in the model used  by \citet{Jones+2017} and \citet{Shao+2017}. As discussed in Sect.~\ref{ssec:fit_velocity_maps}, beam smearing strongly affects the observed velocity gradients and introduces additional uncertainties in the fitting parameters. In addition, \citet{Shao+2017} mentioned that  current data cannot fully rule out the presence of a bidirectional outflow, which introduces additional uncertainty regarding the dynamical mass. In such a case, the strong deviation of the ratio $\Mbh/M_{dyn}$ could be associated with an incorrect kinematical modelling of the observed data, for which we also assume rotating disc kinematics, like in \citet{Jones+2017} and \citet{Shao+2017}.

\subsubsection{VIKING J0305-3150}

High angular resolution ($0.076\si{\arcsecond}\times0.071\si{\arcsecond}$) ALMA [CII] observations of J0305-3150 were recently presented and analysed by \citet{Venemans+2019}. The resulting analysis highlights that the distribution and kinematics, as traced by the [CII] emission, are highly complex and include the presence of cavities and blobs.

\citet{Venemans+2019} attempted to model the kinematics adopting different 3D models (thin rotating disc with constant velocity, Keplerian disc, truncated disc, and a simple AGN model embedded in a uniform rotating spherical gas) taking into account beam smearing effects and pixel correlation within the beam with a Bayesian approach. The results show that the gas kinematics in J0305-3150 appear to be dispersion-dominated, with some overall rotation in the central kiloparsecs, and cannot easily reproduced by a simple rotating disc model with the implication that most of the gas has not settled in a disc yet. In addition, authors point out that energy injection into the ISM produced by AGN feedback processes, and the presence of a companion in the close environment, may play an important role in producing the observed [CII] cavities and in perturbing the gas kinematics. In conclusion, a simple model of a rotating disc is not sufficient to match the [CII] observations of J0305-3150 also derived by our analysis, where $M_{dyn}$ is unconstrained by the simple model assumed.

\subsubsection{SDSS J1044-0125 \& SDSS J0129-0035}
In the work by \citet{Wang+2019}, authors carried out observations of J1044-0125 and J0129-0035 through the ALMA program 2012.1.00240.S (the same dataset used in this work for the latter source) with angular resolution of $\sim0.2\si{\arcsecond}$. The authors show that gas in J1044-0125, as traced by [CII] emission, does not show a clear sign of rotation, suggesting a very turbulent gas velocity field. Furthermore, the [CII] spectrum reveals offset components that could be associated with a node of outflowing gas or the dense core of a satellite galaxy, which contribute to increasing the velocity-dispersion %
component of the gas in the host galaxy. On the other hand, the lower angular resolution data used in our work ($\sim0.6\si{\arcsecond}\times0.5\si{\arcsecond}$, ALMA programme 2011.0.00206.S, see \citealt{Wang+2013}) reveal the presence of a velocity gradient. This could be the result of beam smearing effects producing a smoothing of the rapidly changing velocity gradients. In the case of J1044-0125, we find that the observed velocity field is roughly consistent with a nearly edge-on rotating disc model. Therefore, we conclude that our dynamical mass estimate is tentative. We also note that \citet{Wang+2019} show that [CII] and dust emissions in the nuclear region of J1044-0125 and J0129-0035 follow an exponential light profile, in accordance with the hypothesis at the base of our model.

In the case of J0129-0035, the observations analysed in \citet{Wang+2019} reveal that [CII]-emitting gas shows clear velocity gradients likely associated with a rotating disc with additional gas clumps, thus suggesting complex kinematics in the nuclear region. They attempted to constrain the host galaxy dynamics adopting the same method as in the works of  \citet{Jones+2017} and \citet{Shao+2017}. The results show that the kinematics are consistent with a nearly face-on rotating disc with an inclination angle of $\beta = (16 \pm 20)\,\si{deg}$ and a lower limit on the dynamical mass equal to $M_{dyn} = 2.6\times10^{10}\,M_{\astrosun}$. The results are consistent with what we found in this work. The BH mass of J0129-0035 is estimated as in \citet{Wang+2013}, assuming Eddington accretion, and is the same one that we used in this work. Hence, \citet{Wang+2019} estimated an SMBH to host a galaxy dynamical mass ratio of $\Mbh/M_{dyn}=0.0066$ to be compared with $\Mbh/M_{dyn}=0.0022,$ which is roughly consistent with the local ratio predicted in \citet{Decarli+2010}, unlike the most luminous quasars with massive BHs ($10^{9}-10^{10}\,M_{\astrosun}$) at this redshift that show ratios from a few to $\apprge10$ times higher \citep{Venemans+2016, Decarli+2018}. Therefore, as pointed out by \citet{Wang+2019}, this result may suggest that the BH-galaxy coevolution of a less massive system ($\Mbh\sim10^{7}-10^{8}\,M_{\astrosun}$) in the early Universe %
is closer to the trend of local galaxies (see also, \citealt{Willott+2010,Willott+2015mbh,Willott+2017,Izumi+2018,Izumi+2019}).

\subsection{Possible contamination due to the presence of companion sources in the quasar's local environment}
\label{ssect:companion_presence}
\citet{Decarli+2017} serendipitously discovered companion galaxies in the ALMA field of SDSS J0842+1218, CFHQS J2100-1715, PSO J231-20, and PSO J308-21. Such companions appear similar to the host galaxies of quasars in terms of [CII] brightness and implied dynamical mass, but do not show evidence of AGN activity. In our work, we analysed the same dataset as \citet{Decarli+2017} (ALMA programme 2015.1.01115.S), %
concluding that the kinematics are unresolved (flag [rot]="u"; see Table~\ref{tbl:sample}) in the case of J0842+1218 and J2100-1715 (beam size of $\sim 1.0\si{\arcsecond}\times0.9\si{\arcsecond}$ and $\sim 0.7\si{\arcsecond}\times0.6\si{\arcsecond}$, %
respectively); and marginally resolved (disc inclination angle is unconstrained; flag [$\beta$]="u"; see Table~\ref{tbl:sample}) in the case of J231-20 (beam size of $\sim 1.0\si{\arcsecond}\times0.8\si{\arcsecond}$). This last source together with J308-21 has a [CII]-bright companion at small projected separation of $\sim10\,\si{kpc,}$ suggesting a strong gravitational interaction between quasar and companion able to alter the disc kinematics increasing the velocity-dispersion component of the gas. In particular, \citet{Decarli+2017} show that the [CII] emission of J308-21 stretches over about $4\si{\arcsecond}$ ($\approx25\,\si{kpc}$) and more than $1000\,\si{km\,s^{-1}}$ connecting the companion source suggesting that is undergoing a tidal disruption due to the interaction or merger with the quasar host. This scenario is successively supported by higher angular resolution ($\sim0.3\si{\arcsecond}$; ALMA programme 2016.A.00018.S) follow-up observation of J308-21 presented in \citet{Decarli+2019}; the same dataset analysed in this work. However, the bulk of [CII] emission of the quasar host galaxy shows a spatially resolved velocity gradient, which, in our work we attempt to model with a rotating disc by excluding pixels that are clearly not associated with the quasar host. Nevertheless, our analysis leads to a bimodal posterior probability distribution of disc inclination angle and dynamical mass parameters of J308-21. We can thus conclude that the complex kinematics of this system highlighted in the previous analysis presented in \citet{Decarli+2017, Decarli+2019}, cannot be easily interpreted with a simple rotating disc, likely due to the perturbed kinematics caused by the strong interaction with the satellite galaxy.

\citet{Willott+2017} analysed the source PSO J167-13 observed in ALMA Cycle 3 project 2015.1.00606.S; the same dataset analysed in this work. This source shows an asymmetric continuum emission that is more extended to the south--east than north--west of the peak. This excess is located at $\approx 0.9\si{\arcsecond}$ (projected distance $\approx5.0\,\si{kpc}$), and it is associated with a companion galaxy whose [CII] blueshifted ($~270\,\si{km\,s^{-1}}$) emission corresponds to about 20\% of the QSO [CII] luminosity. The P-V diagram of the source shows a positive velocity gradient, suggesting a rotating disc geometry. With this assumption, \citet{Willott+2017} infer the dynamical mass of $M_{dyn}=2.3\times10^{11}\,M_{\astrosun}$ using the axial ratio of the quasar (excluding the companion source) [CII] flux map as an estimate of the disc inclination angle. The black hole mass of J167-13, $\Mbh=(4.0\pm2.0)\times10^8\,M_{\astrosun}$, is estimated in \citet{Venemans+2015} using calibration based on MgII broad emission line \citep{Vestergaard+2009}. By comparison, we measure a dynamical mass of $1.67^{+0.14}_{-0.11}\times10^{11}\,M_{\astrosun}$ and a black hole mass of $\Mbh=(2.7\pm1.4)\times10^8\,M_{\astrosun}$ (see, Sect.~\ref{sect:BH_masses} for details) resulting in a ratio of $\Mbh/M_{dyn}=0.0016,$ which is completely consistent with the value found by \citet{Willott+2017} ($\Mbh/M_{dyn}=0.0017$) and with the prediction of the local relation \citep{Decarli+2010,Kormendy+2013}.

\citet{Neeleman+2019} further investigated the aforementioned four quasar host-companion galaxy pairs of J0842+1218, J2100-1715, J231-20, J167-13 by analysing high angular resolution ($\sim0.4\si{\arcsecond}\times0.3\si{\arcsecond}$) ALMA observations of [CII] emission. They observe tidal interactions disturbing the gas in these high-$z$ galaxies making the ISM turbulent and thus confirming previous results of \citet{Decarli+2018,Decarli+2019} and \citet{Willott+2017}. Furthermore, these high angular resolution observations reveal that [CII] emission of SDSS J1306+0356 arises from two spatially and spectrally distinct sources with a physical separation of $5.4\,\si{kpc}$ that are interacting tidally. \citet{Neeleman+2019} modelled the [CII] kinematics of the galaxy pairs with a rotating thin disc model, taking into account the beam smearing and the correlation between pixels. They obtained upper limits on dynamical masses for all the sources except J167-13 and J2100-1715. In particular, they measured a dynamical mass of $(3.5\pm0.4)\times10^{10}\,M_{\astrosun}$ for the J167-13 quasar. This value is about one order of magnitude lower than the result of \citet{Willott+2017} and our work. In our estimate, we also take into account the emission from the companion galaxy, thus possibly overestimating the quantities derived from the total integrated spectrum (FWHM$_{\rm[CII]}$, [CII] flux, luminosity, etc. see Table~\ref{tbl:line_fit_results} in Appendix~\ref{sect:spectra_derived_quantities}), the scale radius of the mass profile and the total mass content of the quasar host galaxy. This could explain the inconsistency in our dynamical mass measurements of J167-13 compared to the works of \citet{Willott+2017} and \citet{Neeleman+2019}.

\subsection{Limit on the assumption of thin rotating disc}
\label{ssect:assumptions}
The comparison of our results discussed in Sects.~\ref{ssect:other_modelling} and ~\ref{ssect:companion_presence} highlights that, at least in some cases, the assumption of a thin rotating disc is too simplified to properly describe the observed complex kinematic field. Furthermore, from the analysis of the velocity maps, we find extreme cases in which the disc inclination with respect to the sky plane is very low (e.g. ULAS J1319+0950), compatible with a face-on configuration. However, for these objects, the observed velocity dispersions are still high ($\sim 100-200\,{\rm km\,s^{-1}}$, as is clear from the figures in Appendix~\ref{sect:all_maps}), which is not expected for thin face-on discs. 
 
The observed velocity dispersion can be produced both by instrumental effect and random motions throughout the galaxy \citep[see e.g.][]{Flores+2006, Weiner+2006, Epinat+2010}. Different authors \citep[e.g.][]{Cresci+2009, Epinat+2009, Taylor+2010} pointed out that random motions can support part of the mass. In this case, modelling the kinematics with a rotating disc could result in underestimating the galaxy's dynamical mass. In Appendix~\ref{sect:mass_sigma}, we investigate the contribution of random motions to the dynamical mass and find that, in our sample, the mass supported by non-rotational motions is negligible, meaning it is included within the dynamical mass uncertainties. Therefore, we conclude that the rotating gas disc model provides an overall good description for the gas kinematics of our QSO host galaxies.

\subsection{Potential observational biases in excluding the unresolved objects}
\label{ssect:potential_biases}
In Sect.~\ref{ssect:red-blue}, we investigated whether the observed kinematics are spatially resolved. Out to 32 objects with a line detection, 14 ($\apprge 40\%$) were rejected from the final sample (see also Sect.~\ref{sect:subselections}). Excluding those objects that are spatially unresolved could result in an observational bias. In fact, if these sources were significantly less massive than the others, the final results might therefore be biased towards more massive host galaxies.

The observed size of the line emitting region may depend on both the achieved sensitivity and the angular resolution. Therefore, in the case of spatially unresolved emission, it is difficult to asses if this is due to the intrinsic compact size of the galaxy or to the low sensitivity level of the observations. For this purpose, deeper observations of these objects with similar observational setups could help us to make a fair comparison of the observed size. However, we do not observe a strong correlation between the spatial size of the FIR line emission and the dynamical mass of the galaxy (see Table~\ref{tbl:results_data}). Therefore, we conclude that we cannot safely argue that a possible bias is introduced in rejecting the spatially unresolved objects.

\begin{figure}[!t]
        \centering
        \includegraphics[width=\hsize]{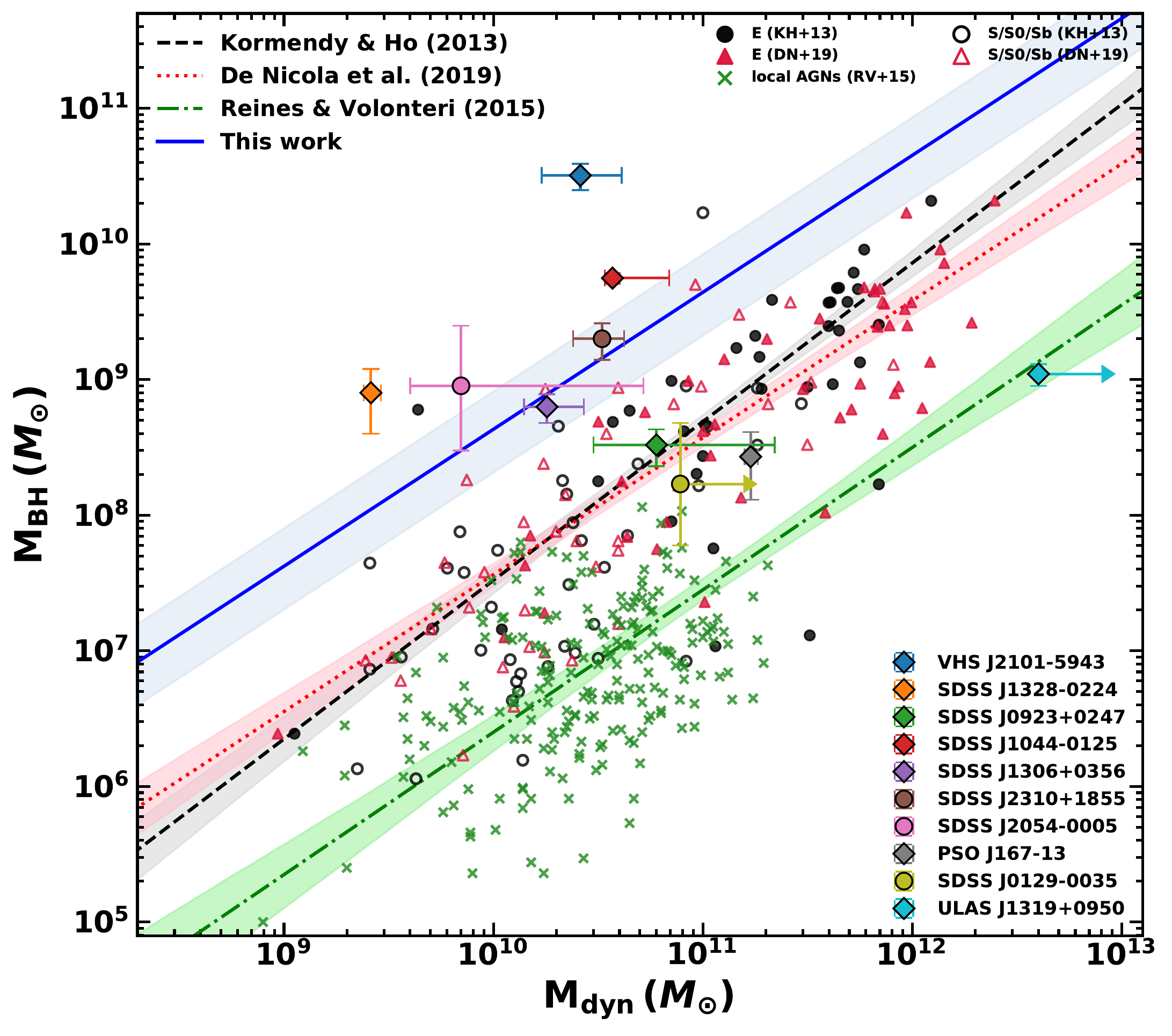}
         \caption{High-redshift relation between the black hole mass ($\Mbh$) and the dynamical mass of the host galaxy ($M_{dyn}$). The dashed black line and the dotted red line represent the reference local relation inferred using samples of local galaxies (E=ellipticals, S/S0/Sb=spirals) shown as black dots \citep[][also indicated as KH+13]{Kormendy+2013} and red triangles \citep[][or DN+19]{DML2019}. The green line is the relation found by \citet[][or RV+15]{Reines+2015} by measuring the total stellar mass in a sample of the local AGN (green crosses). The solid blue line is the best fit to our data. The shadowed areas show the $1\sigma$ uncertainty. In the case of SDSS J0129-0035 and ULAS J1319+0950, we inferred a lower limit on the dynamical mass. We do not take these data into account in the fit. The circles of our data points indicate the sources for which the BH masses are estimated from bolometric luminosity assuming Eddington accretion.}
         \label{fig:mbh-mdyn-relation}
\end{figure}

\begin{figure*}[!t]
        \centering
        \includegraphics[width=\textwidth]{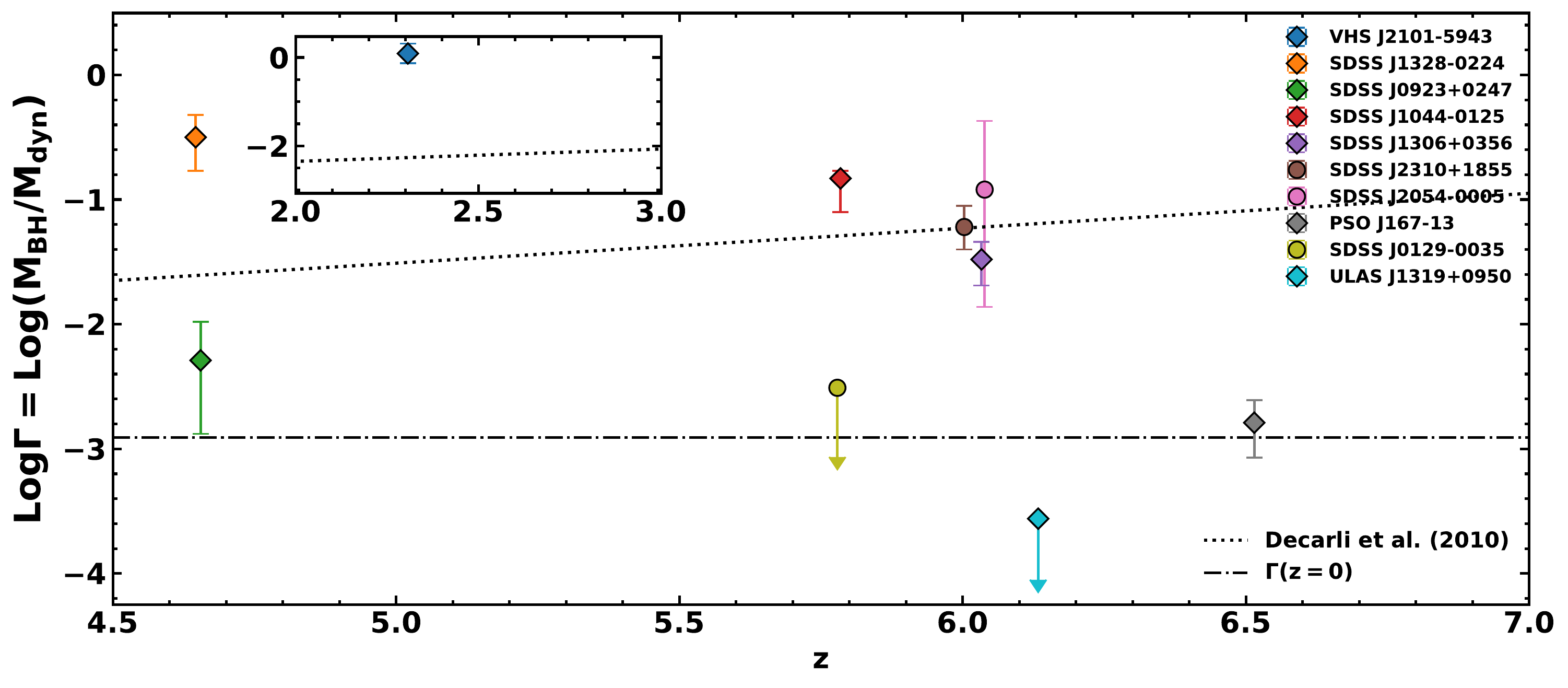}
         \caption{Evolution of $\Gamma=\Mbh/M_{dyn}$ as a function of redshift $z$. The black dotted and dash-dotted lines represent, respectively, the relation found by \citet{Decarli+2010} at $z\apprle 3,$ and the corresponding ratio at $z=0$. The inset panel shows the same plot at $2<z<3$. In the case of SDSS J0129-0035 and ULAS J1319+0950 we inferred lower limits on dynamical masses (i.e. an upper limit on the ratio $\Gamma$). The circles indicate those sources which BH masses are estimated from bolometric luminosity assuming Eddington accretion.} 
         \label{fig:gamma-z-relation}
\end{figure*}

\section{The $\rm\Mbh-M_{dyn}$ relation at high redshift}
\label{sect:BH-galaxy_relation}
In order to trace the relation between black hole mass and dynamical mass for the final high-$z$ QSOs sample, we compared the $M_{dyn}$ measurements obtained through kinematical modelling illustrated in Sect.~\ref{sect:kinematical_model} (see Table~\ref{tbl:results_data}), with the black hole masses obtained from literature as we explained in Sect.~\ref{sect:BH_masses} (see Table~\ref{tbl:spec_data}). The relation is shown in the plot of Fig.~\ref{fig:mbh-mdyn-relation}. We also report two reference relations obtained with samples of local quiescent galaxies \citep{Kormendy+2013,DML2019} and AGN \citep{Reines+2015}. In order to infer the average redshift evolution of the $\Mbh-M_{dyn,}$ we adopted the relation $\log\Mbh = \alpha+\beta(\log M_{dyn}-10.8),$ and we performed the fit assuming fixed slope $\beta = 1.01\pm 0.07$ as found by \citet{DML2019}, and the normalisation $\alpha$ as the only free parameter. Furthermore, to reduce the impact of any possible outliers, we executed the fit adopting the bootstrap method on the standard $\chi^2$ minimisation.

Using $10^{4}$ bootstrap iterations, we obtained the best value of $\alpha$ and its uncertainties by computing the $16th$, $50th,$ and $84th$ percentiles, respectively:
\eq{\alpha = 9.4 \pm 0.3.}
Our result is in agreement with those reported by other high-$z$ works \citep[e.g.][]{Decarli+2010,Decarli+2018,Trakhtenbrot+2015,Trakhtenbrot+2017,Venemans+2016,Venemans+2017} suggesting that the $\Mbh-M_{dyn}$ relation evolves with redshift. It should be noted that the local reference relation \citep[e.g.][]{Kormendy+2013, DML2019} is obtained using bulge stellar mass in spiral and elliptical galaxies (where, in the latter case, bulge stellar mass corresponds to the total stellar mass). As a result, the galaxy dynamical masses estimated in this work should be treated as an upper limit of the total stellar mass. By comparing our results with the relation by \citet{Reines+2015}, who adopted the total stellar mass of the AGN host galaxy (green line in Fig.~\ref{fig:mbh-mdyn-relation}), we find an even stronger evolution with redshift.

\subsection{The evolution of $\Mbh/M_{dyn}$ across the cosmic time}
The evolution of the ratio $\Gamma = \Mbh/M_{dyn}$ as a function of redshift provides key information about the relative time scale between black hole growth and galaxy mass assembly. For this purpose, we show the $\Mbh/M_{dyn}$ ratios as a function of $z$ estimates obtained from the integrated spectra of the lines provided in Table~\ref{tbl:results_data}. The final result is shown in Fig.~\ref{fig:gamma-z-relation}, where we also overplot the relation found by \citet{Decarli+2010} using galaxy stellar masses of a sample of quasars at $z\apprle3$, extrapolated up to $z=7$:
\eq{\log\Gamma(z) = \tonda{0.28\pm0.06}z-\tonda{2.91\pm0.06}.
\label{eq:gamma-z}}
We conclude that the trend of $\Gamma$ that we inferred at high redshift is roughly consistent with that of Eq.~\ref{eq:gamma-z}, and therefore this result confirms the evidence that $\Mbh-M_{dyn}$ appears to evolve with the redshift, as has been highlighted in previous works \citep{Walter+2004, Decarli+2010,Decarli+2018, Merloni+2010, Venemans+2012,Venemans+2017}. The ratio $\Gamma$ appears to be $\sim10\times$ the local value $\log\Gamma(z=0) = 0.28$ at $z\sim4-6$. However, from Fig.~\ref{fig:gamma-z-relation}, we can infer that for SDSS J0923+0247, SDSS J0129-0035, ULAS J1319+0950, and PSO J167-13 at redshift $4.6\apprle z \apprle 6.6,$ the $\Mbh/M_{dyn}$ is consistent with the value observed in galaxies in the local Universe. Although a very preliminary result, this possibly suggests a decreasing of $\Gamma$ at $z\apprge6$.

The discussions reported in Sects.~\ref{ssect:other_modelling} and ~\ref{ssect:companion_presence} point out that at least some of the galaxy masses estimated in this work could suffer from large uncertainties associated with the simple assumptions that are the basis of the fitting method. Therefore, although $\Mbh$ estimates are strongly affected by the large (systematic) uncertainties associated with $\Mbh$ measurements at high-$z$ (up to $\sim 0.4$ dex, see Sect.~\ref{sect:BH_masses}), we conclude that the observed $\Mbh/M_{dyn}$ values are also likely affected by uncertainties in $M_{dyn}$.

Our method did, however, allow us to obtain accurate galaxy dynamical mass estimates at such high redshift. We find that the spread in $\Mbh/M_{dyn}$ values at $z\sim4-7$ is much greater compared with that of local galaxies. This suggests that the observed spread could not arise from the large uncertainties associated with rough galaxy (virial) mass estimates at high-$z$ usually adopted, but it could have a physical reason.

\section{Discussion}
\label{sect:discussion}
In order to extend the context of our work, we compare our results with both observational and theoretical predictions of the BH-galaxy relation obtained in other works. %
Then, we investigate observational biases possibly affecting the results, and we test the reliability of galaxy virial mass estimates.
 
In Sect.~\ref{ssect:biases}, we compare the $\Mbh-M_{dyn}$ and $\Gamma-z$ relation presented in this work with other results on the high-$z$ BH-galaxy relation and we discuss the effect of observational biases.
In Sect.~\ref{ssect:theoretical_simulations}, we compare the BH-galaxy relation prediction from recent simulations of galaxy evolution, and we discuss potential issues on dynamical mass estimates at high redshift outlined from theoretical models.
Finally, in Sect.~\ref{sect:virial_mass}, we test the reliability of virial mass estimates by comparing them with the dynamical mass measurements.

\subsection{Observational biases: comparison with other results on the early BH-galaxy relation}
\label{ssect:biases}
Previous works that aimed to study the BH-host galaxy co-evolution in the early epochs show that $z\apprge6$ luminous quasars have BH-to-host-galaxy-mass ratios $\sim10$ times larger than the typical value observed in the local Universe, implying that these SMBHs formed significantly earlier than their hosts \citep[e.g.][]{Walter+2004, Maiolino+2009, Merloni+2010, Decarli+2010,Decarli+2018,Venemans+2012,Venemans+2016,Wang+2013,Wang+2016, Trakhtenbrot+2015}. However, these results may be affected by observational biases.

\begin{figure}[!t]
        \centering
        \includegraphics[width=\hsize]{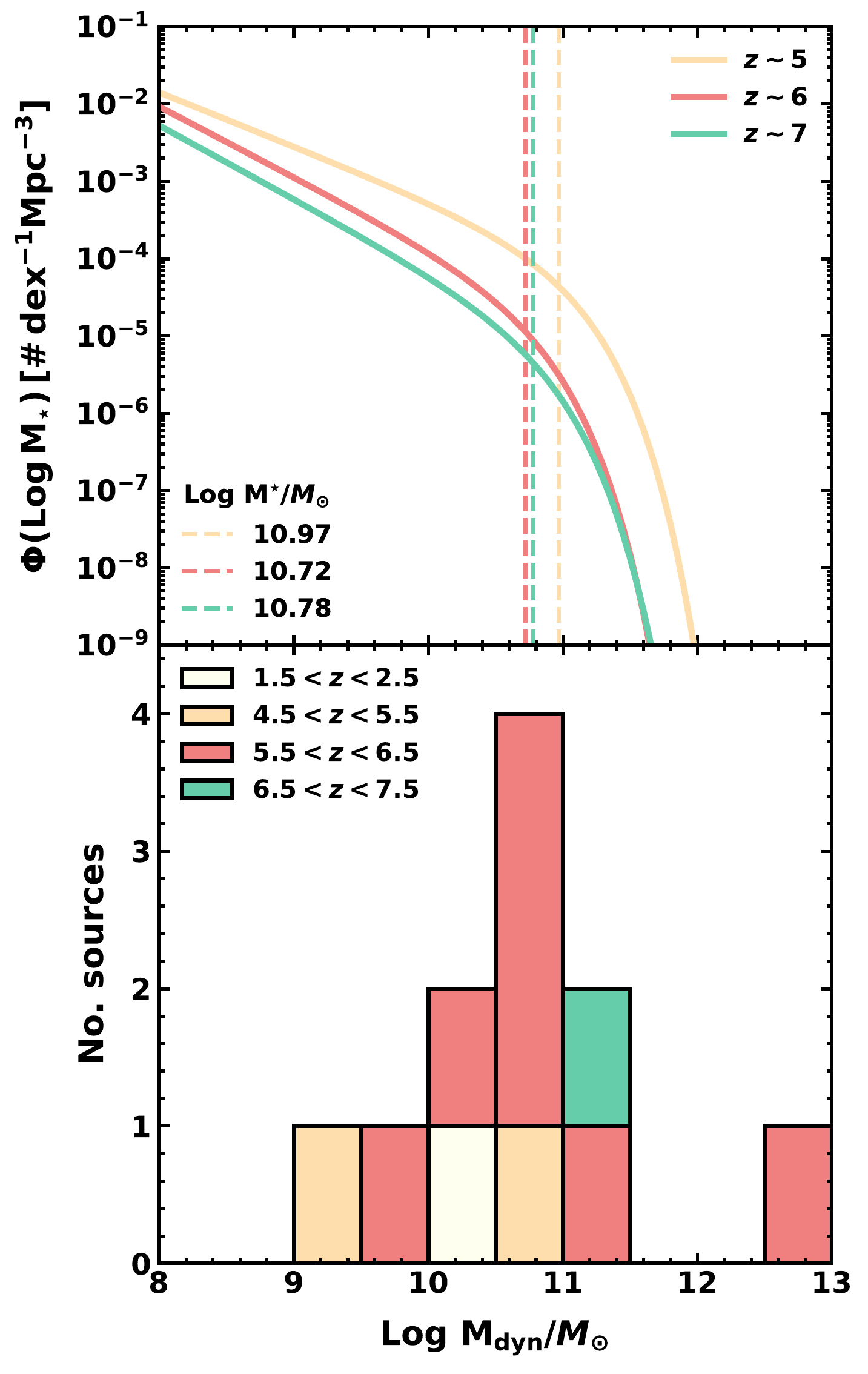}
         \caption{Comparison between our dynamical mass measurement distribution and galaxy stellar mass function. \textit{Upper panel:} Galaxy stellar mass function at different redshifts \citep{Song+2016}. The dashed vertical lines show the position of the $M^{\star}$ values that represent the 'knee' of the mass function. \textit{Lower panel:} Stacked histogram of our dynamical mass estimates.}
         \label{fig:m-function}
\end{figure}

Because most luminous quasars are powered by the most massive BHs at high redshift, if there is a scatter in BH-host galaxy mass relation, for a given $\Mbh$, the selection of objects with low galaxy mass is favoured due to the steepness of the galaxy mass function at its high-mass end \citep[see e.g.][]{Grazian+2015,Song+2016}, thus producing an artificially high average $\Mbh/M_{dyn}$ \citep{Lauer+2007, Schulze+2014}. In order to investigate this selection bias effect, in Figure~\ref{fig:m-function}, we compare the distribution of our dynamical mass estimates, with the galaxy stellar mass function at different redshift. For this purpose, our dynamical mass measurements represent upper limits on galaxy stellar masses ($M_{\star}$). Most of the quasars are at the knee of the quasar luminosity function \citep{Song+2016}, indicating that they represent the bulk of the quasar population at such high redshifts. On the other hand, J183+05, J167-13, and the most extreme J1319+0950 are at the massive end of the $M_{\star}$-function, and they may be more affected by the 'Lauer' bias. Notwithstanding all considerations of the reliability of dynamical mass estimates of the aforementioned objects, we can conclude that these three quasars may be considered the most evolved system known at $z\sim6$ in terms of galaxy mass.

Interestingly, J167-13 together with J0923+0247 and J0129-0035 in our sample have a black hole mass  $\Mbh<5.0\times10^{8}\,M_{\astrosun}$ (see Table~\ref{tbl:spec_data}), and all of them are consistent with the local BH-galaxy relation at $z\sim0$. As discussed in \citet{Venemans+2016}, \citet{Wang+2016}, and \citet{Willott+2015mbh, Willott+2017}, this may suggest that, while the $z\sim6$ quasars with BH masses to the order of $10^{8}\,M_{\astrosun}$ are close to the relation valid for their local counterpart, the most massive BHs ($\Mbh>10^9\,M_{\astrosun}$) at the earliest epochs grow faster than the quasar host galaxy and tend to be above the trend of local galaxies. This may imply that actually, there is no strong correlation between the two properties in high-redshift quasars, but that the scatter was %
much larger in the early Universe than today. However, to confirm this conclusion, a wide range of BH masses and larger sample are required to overcome the observational bias due to the intrinsic scatter of the BH-galaxy relation. In this context, \citet{Izumi+2018,Izumi+2019} studied the $\Mbh-M_{dyn}$ using a sample of seven $z\apprge6$ low-luminosity quasars (absolute magnitude at $1450\AA$, $M_{1450} > -25$) targeted in [CII]$_{158\mu m}$ by ALMA. They derived the quasar host galaxy dynamical (virial) masses assuming rotating disc geometry, and the axial ratio of [CII] flux map as a proxy of the disc inclination angle. Furthermore, they estimated $\Mbh$ through a SE virial relation or assuming Eddington-limited accretion. \citet{Izumi+2018,Izumi+2019}'s results show that while the luminous quasars ($M_{1450} <-25$) typically lie above the local relation \citep{Kormendy+2013} with BHs overmassive compared to local AGNs, the discrepancy becomes less evident at $M_{dyn}\apprge 10^{11}\,M_{\astrosun}$ \citep[see also e.g.][]{Trakhtenbrot+2017}. On the other hand, most of the low-luminosity quasars show comparable or even lower ratios than the local one, particularly at a range of $M_{dyn} \apprge4\times10^{10}\,M_{\astrosun}$. Therefore, \citet{Izumi+2018,Izumi+2019} show that, at least in this high $M_{dyn}$ range, previous works based on sample of luminous quasars might have been biased toward more massive black holes. If these results were confirmed with future follow-up observations, the observed evolution of $\Mbh/M_{dyn}$ ratio out to $z\sim2-3$ could be explained as the result of sample selection bias only. However, \citet{Izumi+2018, Izumi+2019} could be biased, due to the use of virial masses. As we illustrate in Sect.~\ref{sect:virial_mass}, the galaxy masses estimated in this work through a full kinematical model, are not correlated with those estimated through virial theorem. This fact suggests that the use of galaxy virial mass in studying the BH-galaxy relation, could be reflected in an increasing scatter in the observed relation.

\subsection{Comparison with recent theoretical models and simulations}
\label{ssect:theoretical_simulations}
The benchmark correlation in local galaxies is based on bulge stellar masses \citep{Kormendy+2013}. In high-redshift quasars, the host galaxies are completely outshone by the central emission, and the limited angular resolution of the current UV-based observations does not allow us to easily decouple the quasar from its host. Therefore, estimating the stellar mass content in a galaxy out to $z\sim2-3$ is very challenging (see Sect.~\ref{sect:introduction}). However, at such high redshifts, the galaxies' bulges may not have formed yet, or they cannot be detected. Therefore, in the high redshift studies, the dynamical mass of the host galaxy is estimated from the gas properties in the sub-mm observations, and it is usually used as a proxy of bulge stellar mass.

Beyond the observational biases affecting these studies (see Sect.~\ref{ssect:biases}), \citet{Lupi+2019} recently pointed out that the high-$z$ deviation from the local $\Mbh-M_{dyn}$ relation might be due to the different tracers used to estimate the mass of galaxies at high redshift, namely gas-based dynamical mass and the stellar mass. \citet{Lupi+2019} performed a high-resolution cosmological zoom-in simulation in order to investigate the evolution of quasar hosts by properly resolving both the distribution of the cold gas phase ($30\,\si{K}<T<3000\,\si{K}$) traced by [CII]$_{158\mu m}$ line emission, and stars traced by far-UV flux. Their results show that the gas settles in a well-defined dense thin disc extending out to $\sim1\,\si{kpc}$ already at $z\sim7$. Adopting the techniques used in observational studies, they derived dynamical mass estimates through the virial theorem under the assumption of rotationally supported systems. Comparing these dynamical (virial) mass estimates with the masses of the central BHs, they obtained an average BH-dynamical mass ratio of $\Mbh/M_{dyn}\sim0.017$, which is $\sim15$ times greater than local values \citep{Decarli+2010, Kormendy+2013, Reines+2015}, in agreement with previous high-$z$ observations and simulations \citep[see e.g.][]{Walter+2004,Venemans+2017b,Barai+2018,Decarli+2018}, and roughly consistent with our results. Additionally, they also compared galaxy stellar masses with BH masses, showing no sign of clear deviation with the local relation \citep{Reines+2015}. This result implies that dynamical mass estimated using the virial theorem underestimates the dynamical mass of the system. A similar discussion was recently reported by \citet{Kohandel+2019}, who analysed the kinematical properties of a simulated star forming galaxy at $z\sim7$. They show that using the virial theorem in a rotationally supported system, the dynamical mass estimates suffer from large uncertainties depending on the disc inclination.

The approach proposed in our work, in which we perform a kinematical modelling of observed velocity fields, allows us to infer both disc inclination and dynamical mass, thus reducing the uncertainties and biases on our estimates. However, larger samples with higher angular resolution observations are required to finally assess whether a deviation from the local relation exists or not.

\subsection{Comparison between virial masses and dynamical mass estimates}
\label{sect:virial_mass}
We tested the reliability of virial mass estimates by comparing them with our dynamical mass measurements. For this purpose, we made rough dynamical (virial) mass ($M_{vir}$) measurements of our galaxy sample following, e.g. \citet{Wang+2013, Willott+2015mbh, Decarli+2018}:
\eq{M_{vir} = G^{-1} R_{em}\,\tonda{0.75\,{\rm{FWHM}}_{line}/\sin\beta}^2,\label{eq:virial_mass}}
where $R_{em}$ is the radius of the emitting region, and ${\rm{FWHM}}_{line}$ is the full width at half maximum of the line emission.

\begin{figure}[!t]
        \centering
        \includegraphics[width=\hsize]{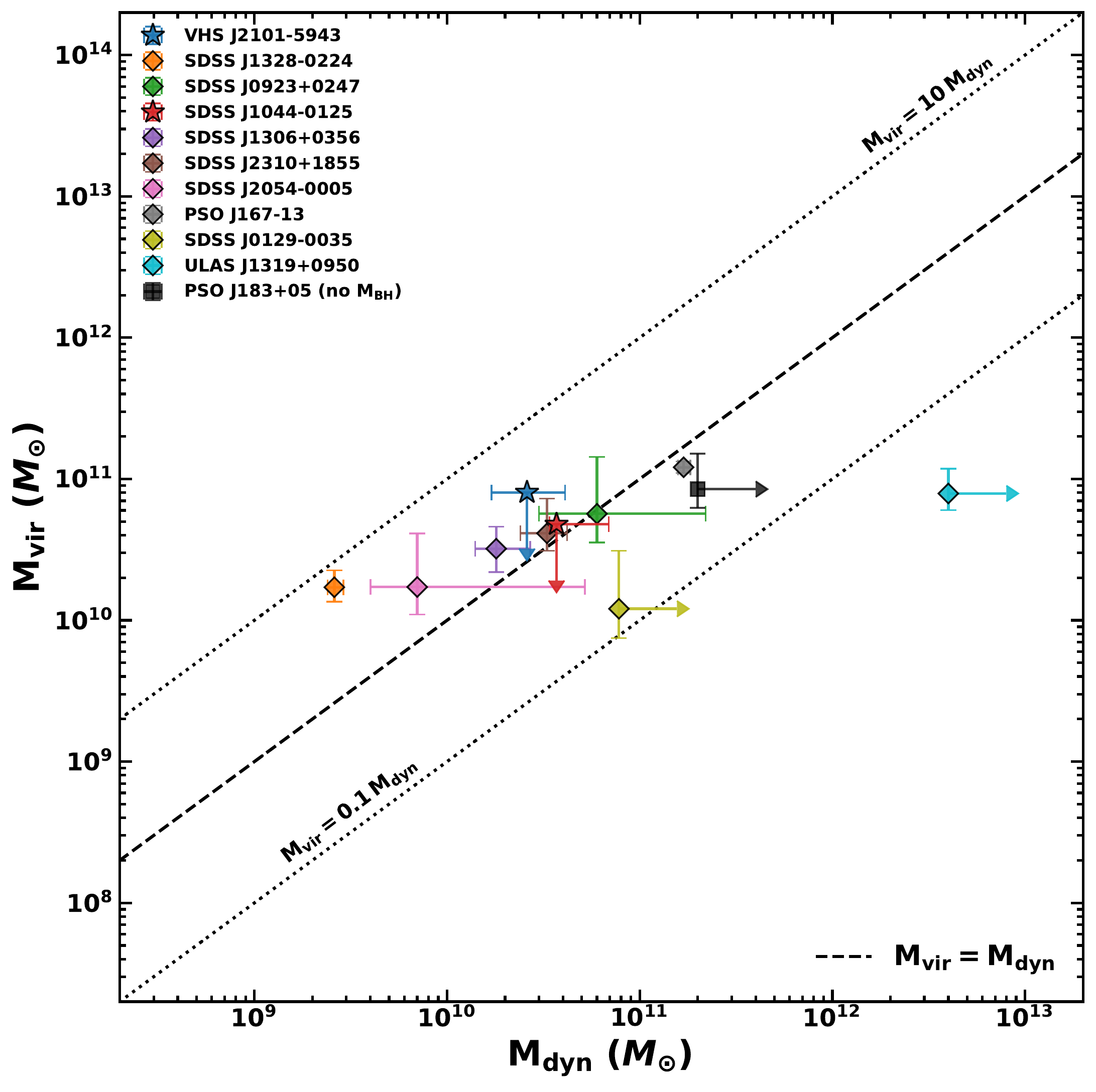}
        \caption{Comparison between dynamical virial mass ($M_{vir}$) computed using flux map properties and dynamical mass ($M_{dyn}$) obtained with the full kinematical model of the gas velocity field. The dashed line indicates the 1:1 relation, while dotted lines are $1\,{\rm dex}$ shifts. The objects represented with star symbols are upper limits on virial mass, since their emission appears consistent with a point-like source (as a result of 2D Gaussian fits within \textsc{CASA}). In these cases, following \citet{Willott+2015mbh}, we assume $\beta=55\,{\rm deg}$ as disc inclination angle. The black square indicates PSO J183-05, for which we retrieved a lower limit on $M_{dyn,}$ but for which $\Mbh$ is not available.}
        \label{fig:mvir_mdyn}
\end{figure}

We performed 2D Gaussian fits, within \textsc{CASA,} of the flux maps obtained in Sect.~\ref{sect:methods}, and we estimated the deconvolved major ($a_{maj}$) and minor axes ($b_{min}$) of the best model. Then, we computed $R_{em}$ as $a_{maj}/2$ in physical length using the redshift estimates obtained in Sect.~\ref{ssec:integrated_spectrum} and the disc inclination angle as $\sin^2\beta = 1-(b_{min}/a_{maj})^2$. Finally, we retrieved the ${\rm{FWHM}}_{line}$ from the Gaussian fit of the line spectra (see Table~\ref{tbl:line_fit_results}). The results of the 2D Gaussian fits, the $R_{em}$ values, and virial masses are listed in Table~\ref{tbl:results_data_spec}.
In Figure~\ref{fig:mvir_mdyn}, we compare dynamical virial masses estimated through Eq.\ref{eq:virial_mass} with the dynamical mass measurements obtained in this work through a full kinematical modelling (see Table~\ref{tbl:results_data}).

We conclude that $M_{vir}$ and $M_{dyn}$ are roughly in good agreement, but they appear not to be correlated, confirming that virial mass is not a reliable dynamical mass estimate of the host galaxy. We note that the errors on virial mass measurements are statistical errors, ignoring any intrinsic uncertainties and systematic biases associated to the virial assumption. 

In order improve the galaxy virial mass estimate, we can use the spectroastrometry method by \citet{Gnerucci+2011}. With this method, is possible to probe spatial scales smaller than the angular resolution, thus allowing a more accurate measurement of the dimension of the line-emitting region. On the other hand, the mass estimates are affected by uncertainties associated with the measurement of the galaxy disc inclination angle, for which it is possible to use the axial ratio from galaxy morphology. In Appendix~\ref{sect:spec-comparison}, we compare the mass factor $M_{dyn}\sin^2\beta$ obtained through the full kinematical model, the virial formula, and the spectroastrometry method. The results show that spectroastrometry is a robust proxy for the galaxy dynamical mass in contrast with the 'classical' virial estimates usually adopted.

\section{Conclusions}
\label{sect:conclusions}
In this work, we investigated the relation between supermassive black hole mass ($\Mbh$) and the dynamical mass of their host galaxies ($M_{dyn}$) of a sample of 10 quasars at $2.3\apprle z \apprle6.5$ targeted in either [CII]$_{158\mu m}$ or CO rotational transitions by ALMA. We then studied the evolution of $\Gamma=\Mbh/M_{gal}$ across cosmic time.

Previous works exploiting ALMA observations attempted to trace the $\Mbh-M_{dyn}$ relation at high-$z$ by estimating the galaxy mass through virial theorem and thus possibly introducing significant uncertainties and biases. To avoid such large uncertainties, we performed a kinematical modelling of the cold gas in the hosts taking into account the beam smearing effect.

In summary, we conclude that:
\begin{itemize}
\item The galaxy mass estimated using the virial theorem combining the axial ratio of flux map to estimate the disc inclination angle, and line FWHM as a proxy of circular velocity, suffers from large uncertainties and could underestimate the dynamical mass of the system \citep{Lupi+2019,Kohandel+2019}.
\item The beam smearing effect strongly affects the observed velocity field in the host galaxy, making the disc inclination angle and galaxy dynamical mass almost degenerate parameters. The more the angular resolution decreases, the more significant this effect becomes, and it should be taken into account in kinematical modelling.
\item The dynamical masses estimated from the kinematical modelling highlight evidence of the evolution of the $\Mbh-M_{dyn}$ relation consistently with previous works \citep[e.g.][]{Walter+2004,McLure+2006,Maiolino+2009,Bennert+2010, Bennert+2011, Decarli+2010, Decarli+2018, Merloni+2010, Wang+2010,Wang+2013,Wang+2016, Canalizo+2012, Targett+2012, Venemans+2012,Venemans+2016,Venemans+2017, Bongiorno+2014, Trakhtenbrot+2015}. In particular, we conclude that, on average, our sample is placed above the reference relation found for galaxies in the local Universe. The normalisation $\alpha$ of the BH-galaxy relation is such that, on average, for a given value of $M_{dyn}$, $\Mbh$ is $\sim10\times$ higher compared to that found by \citet{DML2019}, and $\sim150\times$ higher than what \citet{Reines+2015} found using the total stellar mass of local AGNs.
\item The ratio $\Gamma = \Mbh/M_{dyn}$ appears to be $\sim10\times$ the local value at $z\sim 4-6$, consistent with the result found by \citet{Decarli+2010} extrapolated up to $z=7,$ except for four objects at $z\sim4-6$ that show $\Gamma$ ratios consistent with the local one ($\Gamma(z=0)$). Despite the low statistics, this is the first evidence of a $\Gamma$ value decreasing at $z\sim6$. We are possibly witnessing the phase in which a black hole rapidly grows with respect to the galaxy mass.
\item The observed spread in $\Mbh/M_{dyn}$ values at $z\sim4-6$ is much greater compared to galaxies in the local Universe. Given the accurate galaxy dynamical mass estimates obtained in this work, the observed spread could be due to physical factors, and not associated with the large uncertainties affecting the galaxy virial mass estimates usually adopted in high-$z$ studies.
\item The sources in our sample with $\Mbh$ to the order of $10^8\,M_{\astrosun}$ are close to the relation found for galaxies in the local Universe \citep{Kormendy+2013, DML2019}, while the most massive BHs ($\Mbh > 10^9\,M_{\astrosun}$) lie above them, thus suggesting a faster evolution with respect to their host at $z\sim6$.
\item Most of our sample represents the bulk of the quasar population at $z>4$; thus, overall, the selection of our galaxy sample is not strongly affected by the 'Lauer' bias \citep{Lauer+2007, Schulze+2014}. However, a wide range of BH masses and a larger sample is required %
in order to avoid the observational bias resulting from the intrinsic scatter in the $\Mbh-M_{dyn}$ relation.
\end{itemize}

Based on our blind search, we conclude that one third of high-$z$ quasar hosts have gas kinematics consistent with rotating discs, but it is still very challenging to infer the dynamical mass due to the poor angular resolution and sensitivity of current observations. The typical angular resolution of the observations ($\sim0.5\si{\arcsecond}$) is frequently not good enough to constrain the dynamical parameters of the discs at $z\apprge5,$ and the fitting procedures cannot take into account possible distortions of the velocity field introduced by instrumental effects. As a result, we inferred the dynamical masses only for ten out of 72 quasars observed with ALMA so far.

On the other hand, for those quasars with deep ALMA observations and high angular resolution, this work shows that dynamical mass estimations are also feasible at $z\sim6$. Further ALMA high angular resolution observations of high-$z$ quasars are crucial to studying the evolution of the $\Mbh/M_{dyn}$ ratio and verifying whether $\Gamma(z)$ decreases at $z\apprge6$ as suggested by our preliminary results.

\begin{acknowledgements}
 We thank the anonymous referee for her/his careful reading of the manuscript and her/his comments which really helped us to improve the paper. This paper makes use of the following ALMA data: ADS/JAO.ALMA\#2011.0.00206.S, ADS/JAO.ALMA\#2012.1.00240.S, ADS/JAO.ALMA\#2012.1.00882.S, ADS/JAO.ALMA\#2013.1.00815.S, ADS/JAO.ALMA\#2013.1.01153.S, ADS/JAO.ALMA\#2015.1.00228.S, ADS/JAO.ALMA\#2015.1.00399.S, ADS/JAO.ALMA\#2015.1.00606.S, ADS/JAO.ALMA\#2015.1.01115.S, ADS/JAO.ALMA\#2015.1.01247.S, ADS/JAO.ALMA\#2016.1.00544.S, ADS/JAO.ALMA\#2016.A.00018.S. ALMA is a partnership of ESO (representing its member states), NSF (USA) and NINS (Japan), together with NRC (Canada), MOST and ASIAA (Taiwan), and KASI (Republic of Korea), in cooperation with the Republic of Chile. The Joint ALMA Observatory is operated by ESO, AUI/NRAO and NAOJ. We acknowledge the support from the LBT-Italian Coordination Facility for the
execution of observations, data distribution and reduction. The LBT is an international collaboration among institutions in the United States, Italy and Germany. LBT Corporation partners are: The University of Arizona on behalf of the Arizona university system; Istituto Nazionale di Astrofisica, Italy;
LBT Beteiligungsgesellschaft, Germany, representing the Max-Planck Society,
the Astrophysical Institute Potsdam, and Heidelberg University; The Ohio State
University, and The Research Corporation, on behalf of The University of Notre Dame,
University of Minnesota, and University of Virginia. SC acknowledges support by the European Research Council No. 740120 `INTERSTELLAR'. MP is supported by the Programa Atracci\'on de Talento de la Comunidad de Madrid via grant 2018-T2/TIC-11715. RM acknowledges supports by the Science and Technology Facilities Council (STFC) and from ERC Advanced Grant 695671 ``QUENCH''.
\end{acknowledgements}

%
%
\bibliographystyle{bibtex/aa}
\bibliography{bibtex/MyBib}
\begin{appendix} 

\section{Integrated spectra: line properties and derived quantities}
\label{sect:spectra_derived_quantities}
From the best-fit of the integrated spectra (see Sec.~\ref{ssec:integrated_spectrum}), we directly retrieved the line FWHM, and the velocity-integrated flux of the line ($F_{line}$). Then, we also inferred line luminosity ($L_{line}$), [CII] mass ($M_{\rm{[CII]}}$), total gas mass ($M_{gas}$), and the [CII]-based star formation rate ($\rm{SFR_{[CII]}}$). The line luminosities were computed following \citet{Solomon2005}:
\eq{L_{line}\,[L_{\astrosun}] =1.04\times10^{-3}F_{line}\nu_{\it rest}(1+z)^{-1}D_L^2,}
where $F_{line}$ is in unit of $\si{Jy\, km\,s^{-1}}$; $\nu_{\it rest}$ in $\si{GHz,}$ and $D_{L}$ in $\si{Mpc}$.

Then, by analogy with \citet{Venemans+2017b}, assuming optically thin [CII] emission and local thermodynamical equilibrium (LTE) of the carbon line, we estimated the mass of singly ionised carbon in galaxies as:
\begin{align}\nonumber M_{\rm [CII]}\,[M_{\astrosun}]&= Cm_{C}\frac{8\pi k_{B}\nu_{\it rest}^2}{hc^3A_{ul}}Q(T_{ex})\frac{1}{4}e^{91.2/T_{ex}}L'_{\rm [CII]}\\
&=2.92\times10^{-4}Q(T_{ex})\frac{1}{4}e^{91.2/T_{ex}}L'_{\rm [CII]},\end{align}
where $C$ is the conversion factor between $\si{pc^2}$ and $\si{cm^2}$, $m_C$ the mass of a carbon atom, $A_{ul}=2.29\times10^{-6}\si{s^{-1}}$ the Einstein coefficient, $Q(T_{ex})=2+4e^{-91.2/T_{ex}}$ the CII partition function, and $T_{ex}$ the excitation temperature that we set equal to $T_{ex}=100\,\si{K}$ \citep[see][]{Venemans+2017b}. Then, assuming that all carbon atoms are singly ionised, we also derived a lower limit on the total gas mass ($M_{gas}$) using the carbon abundance relative to hydrogen atom \citep{Asplund+2009} $M_{C}/M_{H} = 3.54\times10^{-3}$. Finally, we estimated the SFRs using the ${\rm{SFR}}-L_{\rm{[CII]}}$ relation for high-redshift ($z>0.5$) galaxies from \citet{DeLooze+2014}:
\eq{{\rm{SFR_{[CII]}}}\,[M_{\astrosun}\,{\rm{yr}}^{-1}] = 3.0\times 10^{-9}\tonda{L_{\rm{[CII]}}/L_{\astrosun}}^{1.18,}} with a systematic uncertainty of a factor of $\sim2.5$. In Table~\ref{tbl:line_fit_results}, we listed the results of spectral fits and the derived quantities for those sources with dynamical mass constrained. The reported quantities are consistent within $\sim 2\sigma$ to the estimates published in other works \citep[e.g.][]{Banerji+2017,Wang+2013,Decarli+2017,Decarli+2018,Trakhtenbrot+2017,Venemans+2017c}.

\setcounter{magicrownumbers}{0}
\begin{table*}[!htbp]
\setcellgapes{1.2pt}\makegapedcells
\caption{Key parameters and derived quantities estimated from the fits of integrated spectra.}             
\label{tbl:line_fit_results}      
\centering          
\begin{tabular}{l l  c c c c c c}
\hline\hline       
No.$^{a}$               &Object ID                                              & FWHM$_{\rm{[CII]}}$     &$F_{\rm{[CII]}}$               &$L_{\rm{[CII]}}$                       &$M_{\rm{[CII]}}$                       &$M_{gas}$                                      &$\textrm{SFR}_{\rm{[CII]}}$\\          
                &                                                                       & ($\si{km\,s^{-1}}$)             &($\si{Jy\,km\,s^{-1}}$)        &($10^9\si{L_{\astrosun}}$)     &($10^6\si{M_{\astrosun}}$)     &($10^{9}\si{M_{\astrosun}}$)           &($\si{M_{\astrosun}\,yr^{-1}}$)\\
\hline
2               &\object{VHS J2101-5943}$^{\dagger}$                             &$222^{+19}_{-17}$              &$0.48^{+0.04}_{-0.04}$ &$0.019^{+0.020}_{-0.017}$\\

6               &\object{SDSS J1328-0224}                                                &$277^{+24}_{-22}$              &$2.43^{+0.17}_{-0.16}$ &$1.63^{+0.11}_{-0.11}$         &$4.3^{+0.3}_{-0.3}$                    &$1.22^{+0.08}_{-0.08}$                 &$223^{+18}_{-17}$\\
                
7               &\object{SDSS J0923+0247}                                                &$328^{+13}_{-12}$              &$5.10^{+0.18}_{-0.18}$ &$3.44^{+0.12}_{-0.12}$         &$9.1^{+0.3}_{-0.3}$                    &$2.58^{+0.09}_{-0.09}$                 &$537^{+22}_{-22}$\\

9               &\object{SDSS J0129-0035}                                                &$189^{+9}_{-9}$                &$2.14^{+0.08}_{-0.09}$ &$2.01^{+0.08}_{-0.08}$         &$5.3^{+0.2}_{-0.2}$                    &$1.50^{+0.06}_{-0.06}$                 &$284^{+13}_{-13}$\\

10              &\object{SDSS J1044-0125}                                                &$422^{+69}_{-60}$              &$1.31^{+0.19}_{-0.19}$ &$1.23^{+0.18}_{-0.18}$         &$3.3^{+0.5}_{-0.5}$                    &$0.92^{+0.14}_{-0.13}$                 &$159^{+28}_{-27}$\\

11              &\object{SDSS J1306+0356}                                                &$232^{+28}_{-24}$              &$2.4^{+0.2}_{-0.2}$            &$2.4^{+0.2}_{-0.2}$                    &$6.3^{+0.6}_{-0.6}$                    &$1.77^{+0.16}_{-0.16}$                 &$346^{+37}_{-36}$\\    

12              &\object{SDSS J2310+1855}                                                &$381^{+13}_{-13}$              &$8.0^{+0.2}_{-0.2}$            &$7.9^{+0.2}_{-0.2}$                    &$21.0^{+0.6}_{-0.6}$           &$5.92^{+0.18}_{-0.18}$                 &$1434^{+52}_{-50}$\\

14              &\object{SDSS J2054-0005}                                                &$243^{+10}_{-10}$              &$3.36^{+0.12}_{-0.12}$ &$3.36^{+0.12}_{-0.12}$         &$8.9^{+0.3}_{-0.3}$                    &$2.51^{+0.09}_{-0.09}$                 &$522^{+23}_{-22}$\\    

19              &\object{ULAS J1319+0950}                                                &$484^{+33}_{-30}$              &$2.60^{+0.16}_{-0.15}$ &$2.65^{+0.16}_{-0.16}$         &$7.0^{+0.4}_{-0.4}$                    &$1.99^{+0.12}_{-0.12}$                 &$395^{+29}_{-28}$\\    

22              &\object{PSO J308-21}                                                    &$403^{+35}_{-33}$              &$1.03^{+0.08}_{-0.08}$ &$1.08^{+0.08}_{-0.08}$         &$2.7^{+0.2}_{-0.2}$                    &$0.81^{+0.06}_{-0.06}$                 &$137^{+13}_{-13}$\\    

25              &\object{PSO J183+05}                                            &$382^{+19}_{-17}$              &$5.1^{+0.2}_{-0.2}$            &$5.6^{+0.2}_{-0.2}$                    &$14.8^{+0.6}_{-0.6}$           &$4.19^{+0.17}_{-0.17}$                 &$954^{+47}_{-44}$\\    

26              &\object{PSO J167-13}                                                    &$499^{+17}_{-17}$              &$3.20^{+0.10}_{-0.10}$ &$3.57^{+0.11}_{-0.11}$         &$9.5^{+0.3}_{-0.3}$                    &$2.68^{+0.08}_{-0.08}$                 &$562^{+20}_{-20}$\\    

28              &\object{VIKING J0305-3150}                                      &$245^{+15}_{-14}$              &$3.69^{+0.19}_{-0.18}$ &$4.2^{+0.2}_{-0.2}$                    &$11.2^{+0.6}_{-0.5}$           &$3.15^{+0.16}_{-0.15}$                 &$682^{+41}_{-38}$\\    

\hline                  
\end{tabular}
\tablefoot{$^{a}$Source identification numbers in agreement with those listed in the first column of Table~\ref{tbl:sample}. $^{\dagger}$This source is observed in CO(3-2). In this case, FWHM and flux refer to this line. We thus do not derive the [CII]-based quantities for that.}
\end{table*}

\section{Mass support from random motions}
\label{sect:mass_sigma}
We investigate here the turbulent pressure support term, which arises from non-rotational motions, on the  total dynamical mass of our QSO host galaxies \citep[see e.g.][]{Epinat+2009, Taylor+2010}. This term is not taken into account in our disc model since the gas is circularly rotating in a thin disc. 

Following \citet{Epinat+2009}, we quantify the mass supported by random motions inside the galaxy through the virial theorem:
\eq{M_{\sigma} = C\frac{\sigma_0^2 R_D}{G},\label{eq:m-sigma}}
where $R_D$ is the scale radius of the exponential brightness profile (see Sect.~\ref{sect:kinematical_model}); $\sigma_0$ an estimate of constant velocity dispersion throughout the whole galaxy, and $C$ is a parameter depending on the mass distribution and geometry. Here, we assume $C=2.25,$ which is the average value of known galactic mass distribution models \citep{BinneyTremaine2008}.

\begin{figure*}[!t]
        \centering
        \includegraphics[width=\textwidth]{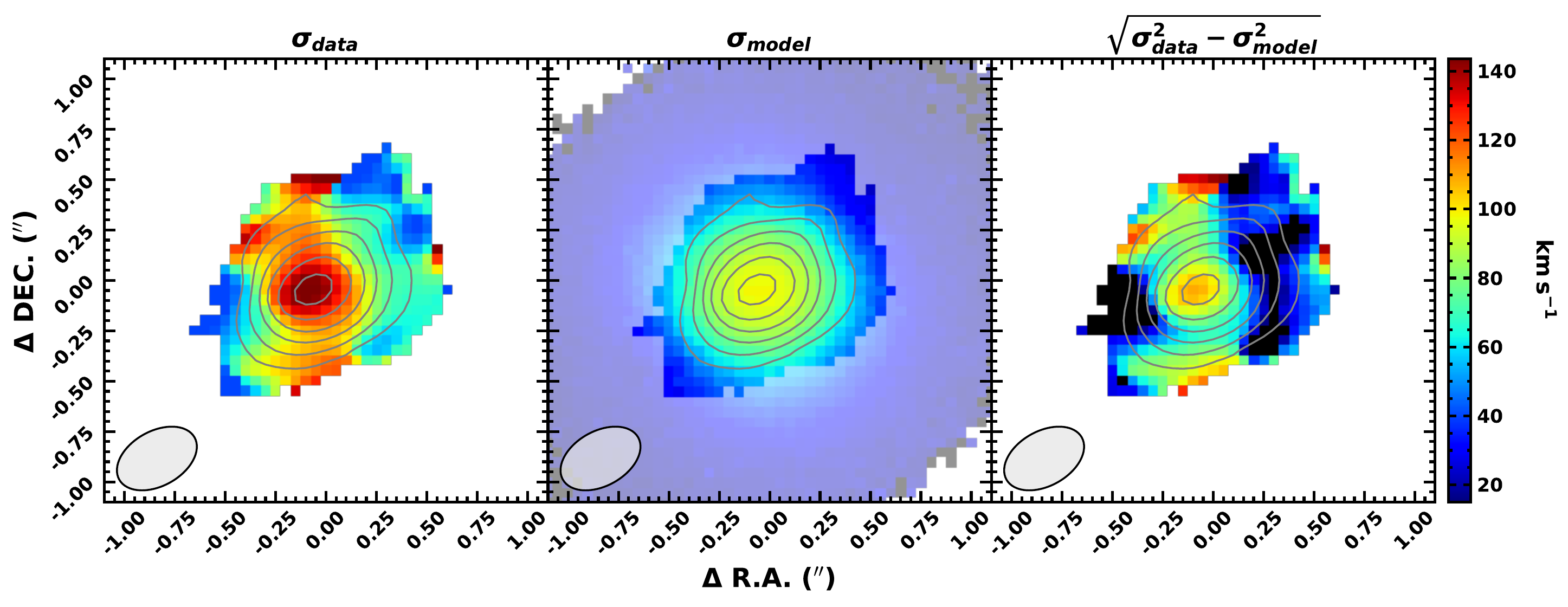}
         \caption{Estimation of local velocity dispersion in J0923+0247. \textit{Left panel:} observed velocity-dispersion field. \textit{Central panel:} simulated flux-weighted velocity-dispersion map along the line of sight, corresponding to the best-fit model of the velocity field. \textit{Right panel:} quadratic residuals representing the local velocity dispersion along the line of sight. The contours show the line-velocity integrated map. The contours correspond to $[0.25I_0,0.5I_0,I_0/e,0.68I_0,0.9I_0]$, where $I_0$ is the maximum value of the observed flux. In the bottom-left corner of each panel, the FHWM of the synthesised beam is shown.}
         \label{fig:disp_res_map}
\end{figure*}

The intrinsic velocity dispersion can be estimated from the observed velocity-dispersion map after taking into account the angular and spectral resolution of observations. As a representative example, we choose SDSS J0923+0247. The velocity field of the latter shows a clear velocity gradient, and we estimated the disc inclination of $\sim29\,\rm{deg}$ and dynamical mass of $\sim6.0\times10^{11} M_{\astrosun}$ (see Fig.~\ref{fig:fitted_maps}). Although our rotating thin disc model reproduces the observed velocity map very well, the observed velocity dispersion is slightly ($\sim1.4\times$) higher than what was expected by our best-fit (see Fig.~\ref{fig:disp_res_map}). Indeed, the best-fitting velocity-dispersion field includes only the effect of the unresolved velocity gradients and enlargement of the emission line profile due to the beam smearing and the instrumental line-spread function (set to $\sigma_{\rm LSF}=15\,{\rm km\,s^{-1}}$). As previously mentioned, our model does not include random motions due to the physics of the gas. The intrinsic velocity dispersion of the gas can be estimated by quadratically subtracting the model from the measured velocity-dispersion map. The model and quadratic residuals are shown in Fig.~\ref{fig:disp_res_map}. Then, we computed the '1/errors'-weighted mean velocity dispersion ($\sigma_0$), where the aforementioned errors are associated with the observed velocity-dispersion values that are estimated from the spaxel-by-spaxel line fit of the data cube. The resulting value is $\sigma_0\sim 54\,{\rm km\,s^{-1}}$. Using this value in Eq.~\ref{eq:m-sigma}, we obtain $M_{\sigma} \sim 1.3\times10^9\,M_{\astrosun}$. This mass budget accounts for only $\sim2\%$ of the estimated dynamical mass obtained with kinematical modelling assuming rotating thin disc geometry. By comparing this value with the uncertainties of the dynamical mass estimate ($\sim 0.5\,{\rm dex}$), we conclude that the mass support from random motions is negligible in this system. We adopt the same technique for all the sources for which we obtained a dynamical mass estimate and verified that the eventual mass budget arising from random motions is always included within the dynamical mass uncertainties. Therefore, we conclude that the rotating gas disc model provides a good description for the gas kinematics of our QSO host galaxies. 

We stress that, for systems that show complex or perturbed kinematics (e.g. due to the presence of outflow or a companion in the close environment of quasar, see Sect~\ref{sect:comparison}), the assumption of rotating disc geometry is undoubtedly less meaningful than for relaxed systems. This is the case of, for example, SDSS J0129-0356 and ULAS J1319+0950, for which the best fits are consistent with face-on discs. On the other hand, the observed velocity dispersion is still high ($\sim100-200\,{\rm km\,s^{-1}}$, but consistent within $2-3\sigma$ to the model), which is not expected for a face-on disc. Although this evidence supports the fact that, at least in these extreme cases, the hypothesis of a thin rotating disc is too simplified, and the dynamical mass estimates should be considered tentative; deeper observations are needed in order to properly describe the kinematics in these complex systems, rather than a more accurate kinematical modelling of the current observations. However, our fitting method enables us to homogeneously study the whole sample.

\section{Comparison between dynamical masses and the spectroastrometric mass estimates}
\label{sect:spec-comparison}
\setcounter{magicrownumbers}{0}
\begin{table*}
\setcellgapes{1.2pt}\makegapedcells
\caption{Parameters estimated from a 2D Gaussian fit of the flux maps, virial mass estimates, and the results of the spectroastrometry method.}             
\label{tbl:results_data_spec}      
\centering          
\begin{tabular}{l l c c c c c c }
\hline\hline       
No.$^{a}$       &Object ID              &       FWHM$_{maj}$$^{b}$&              FWHM$_{min}$$^{c}$&     $R_{em}$&                       $M_{vir}$&                              $r_{spec}$&             $\mu_{spec}$                            \\
                &                               &       ($\si{mas}$)                            &($\si{mas}$)                    &$(\si{kpc})$                   &$(10^{10}M_{\astrosun})$         &$(\si{kpc})$           &$(10^{10}M_{\astrosun})$       \\
\hline
2               &\object{VHS J2101-5943} &       \multicolumn{2}{c}{\textsc{point-like source}}&                       $<8.34$&                                $<8.0$&                                 $<2.3$&                 $<2.6$\\        

6               &\object{SDSS J1328-0224}&       $339\pm50$&                             $189\pm35$&                     $1.13^{+0.17}_{-0.17}$& $1.7^{+0.5}_{-0.4}$&                    $0.3\pm0.2$&            $0.5\pm0.3$\\   

7               &\object{SDSS J0923+0247}&       $385\pm50$&                             $340 \pm 58$&                        $1.28^{+0.17}_{-0.17}$& $5.7^{+8.7}_{-2.1}$&                    $0.7\pm0.1$&            $1.8\pm0.4$\\   

9               &\object{SDSS J0129-0035}        &       $303\pm51$&                             $271\pm59$&                     $0.91^{+0.15}_{-0.15}$& $1.2^{+1.9}_{-0.5}$&                    $0.37\pm0.06$&  $0.30\pm0.06$\\ 

10              &\object{SDSS J1044-0125}        &       \multicolumn{2}{c}{\textsc{point-like source}}&                       $<1.38$&                                $<4.8$&                                 $0.4\pm0.3$&            $1.8\pm1.5$     \\      

11              &\object{SDSS J1306+0356}        &       $1290\pm430$&                   $410\pm240$&                    $3.8^{+1.3}_{-1.2}$&            $3.2^{+1.4}_{-1.0}$&                    $1.1\pm0.7$&            $1.4\pm0.9$\\   

12              &\object{SDSS J2310+1855}        &       $512\pm61$&                             $287\pm135$ &               $1.50^{+0.18}_{-0.18}$& $4.1^{+3.1}_{-1.0}$&                    $0.41\pm0.07$&  $1.4\pm0.3$\\   

14              &\object{SDSS J2054-0005}        &       $352\pm72$&                             $293 \pm 97$&                        $1.0^{+0.2}_{-0.2}$&            $1.7^{+2.4}_{-0.6}$&                    $0.33\pm0.08$&  $0.45\pm0.12$\\ 

19              &\object{ULAS J1319+0950}        &       $536\pm95$&                             $328 \pm 78$&                        $1.6^{+0.3}_{-0.3}$&    $7.9^{+3.9}_{-1.9}$&                    $0.64\pm0.12$&  $3.5\pm1.0$\\   

25              &\object{PSO J183+05} &          $604\pm63$&                             $477\pm52$&                     $1.70^{+0.18}_{-0.18}$& $8.5^{+6.6}_{-2.2}$&                    $0.44\pm0.12$&  $1.5\pm0.5$\\   

26              &\object{PSO J167-13}&           $1068\pm74$&                            $467\pm53$&                     $3.0^{+0.2}_{-0.2}$&            $12.1^{+1.2}_{-1.1}$&           $1.16\pm0.12$&  $6.7\pm1.0$\\
\hline                  
\end{tabular}
\tablefoot{$^{a}$Source identification numbers in agreement with those listed in the first column of Table~\ref{tbl:sample}. $^{b,\,c}$Major and minor FWHM of the 2D best-fit Gaussian of the flux maps deconvolved from the beam. Point-like sources are explicitly indicated. In these cases, minor FWHM of the synthesised beam is taken as upper limit on the angular dimension of the emitting region. Uncertainties on quantities are statistical errors ignoring any possible systematic biases.}
\end{table*}
\begin{figure*}[!t]
        \centering
        \includegraphics[width=\textwidth]{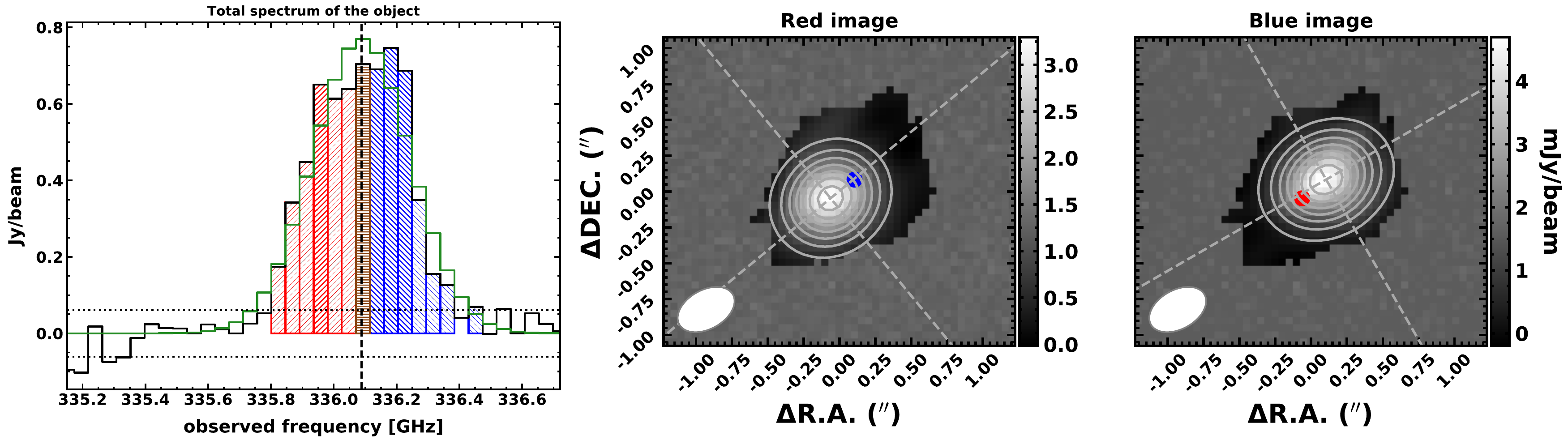}
         \caption{Integrated spectrum and red/blue maps of SDSS J0923+0247. \textit{Left panel}: the data and the best-fit model is shown with black and green lines, respectively. The red dashed vertical line indicates the central frequency; the rms of the residuals is indicated by the dotted horizontal lines. Channels used to create the red and blue maps (show on the right panels) are filled with their respective colours. The central brown bin is added to the red and blue side of the collapsed images with a weight given by the fraction of red and blue bins. \textit{Right panels}: the best 2D Gaussian model is shown with white contours. The blue and red circles indicate the centroid positions of the counterpart map. Here, pixels not defined in the maps are replaced with simulated noise in order to avoid numerical drawbacks in the fitting process.}
         \label{fig:spectroimages}
\end{figure*}

Since the dynamical masses estimated in this work are measured through a full kinematical modelling of the velocity field, they can be considered reliable mass estimates. Comparing them with mass measurements obtained with other methods enables us~to test the consistency of the results. In this section, we compare the host galaxy's dynamical mass measurements listed in Table~\ref{tbl:results_data} with the mass estimates obtained through the spectroastrometry method by \citealt{Gnerucci+2011} and the virial mass estimates obtained in Sect.~\ref{sect:virial_mass}.

Spectroastrometry is a technique that combines spatial and spectral resolution to probe spatial scales smaller than the angular resolution of the observations. We applied it to the case of our high-$z$ quasar sample to estimate the product $\mu=M_{dyn}\sin^2\beta$. Following \citet{Gnerucci+2011}, we measured the FWHM and the central frequency of the line from the integrated spectra, (e.g. see Fig.\ref{fig:tot_flux}), then we collapsed the redshifted and blueshifted channels obtaining 'red' and 'blue' maps, respectively (see Fig.~\ref{fig:spectroimages}). If the galaxy disc is at least marginally resolved, the latter two maps are spatially shifted due to the rotation of the gas. Then, we performed a 2D fit of collapsed maps using an elliptical Gaussian function, and we determined the position of the two centroids. Thus, we computed the spectroastrometric radius ($r_{spec}$), as the half distance between 'red' and 'blue' centroids. Finally, we used the FWHM and $r_{spec}$ measurements, and we estimated the spectroastrometric mass (see Eq.~2 in \citealt{Gnerucci+2011}):
\eq{M_{spec}\sin^2\beta = f_{spec}\,\mu_{spec}=f_{spec}\frac{\text{FWHM}_{line}^2r_{spec}}{G,}
\label{eq:spectroastrometric_mass}}
where $f_{spec}$ is the calibrator factor. Here, we used the value of \citet{Gnerucci+2011}; $f_{spec}=1.0\pm0.1$. The $r_{spec}$ values and $\mu_{spec}$ are listed in Table~\ref{tbl:results_data_spec}.

\begin{figure}[!t]
        \centering
        \includegraphics[width=0.77\hsize]{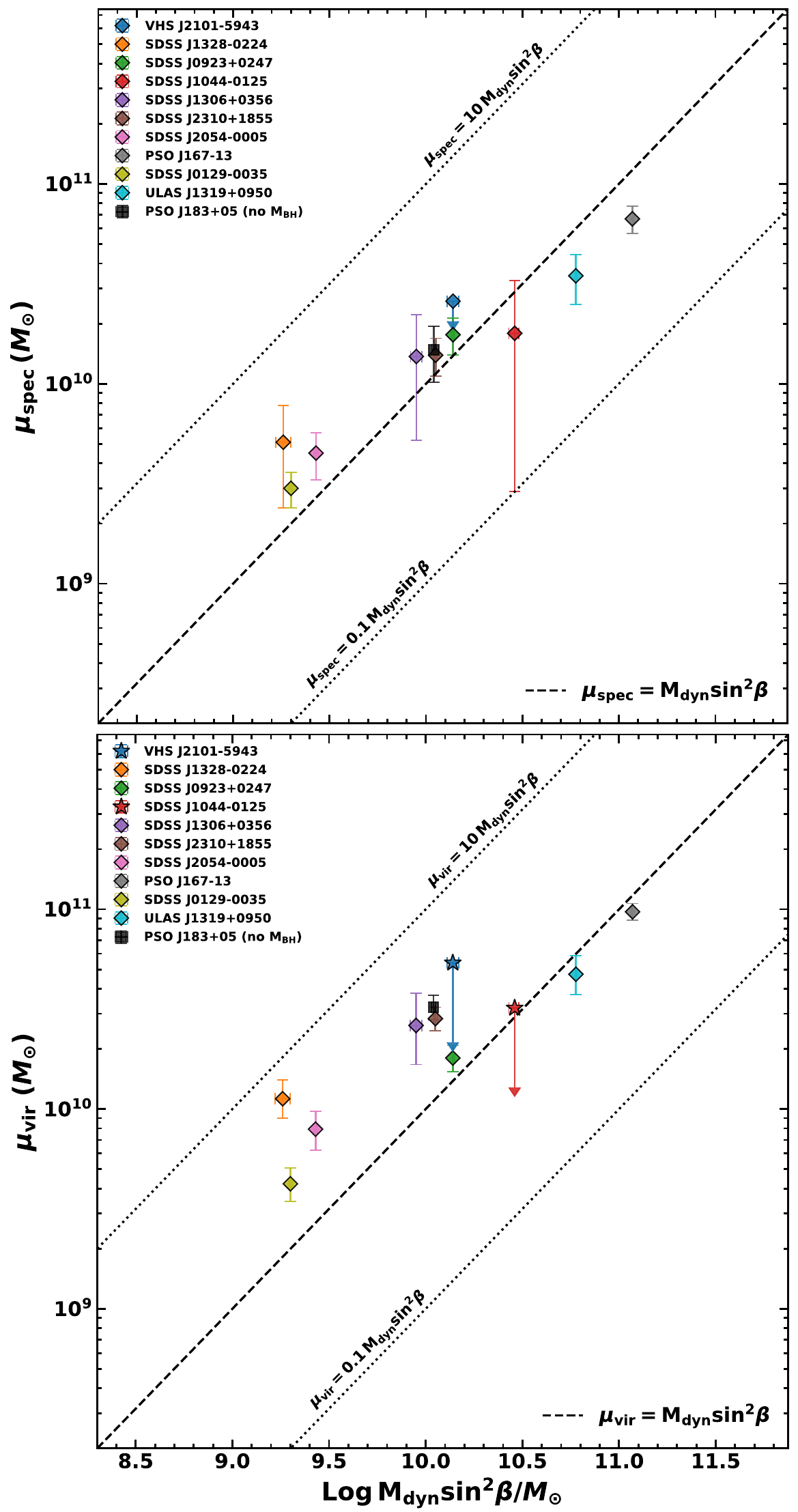}
         \caption{Comparison of galaxy mass factors obtained with spectroastrometry (\textit{upper panel}) and virial estimates (\textit{bottom panel}) with $M_{dyn}\sin^2\beta$ from full kinematical modelling of the host galaxy's gas velocity field. The dashed black line represents the 1:1 relation. The object symbols are the same as in Fig.~\ref{fig:mvir_mdyn}.}
         \label{fig:spectroastrometry}
\end{figure}

In Fig.~\ref{fig:spectroastrometry} (upper panel), we compare $\mu_{spec}$ from spectroastrometry with $M_{dyn}\sin^2\beta$ measured from the full kinematical modelling (see Table~\ref{tbl:results_data}). We conclude that the two estimators are consistent within uncertainties. We also compare $\mu_{vir} = M_{vir}\sin^2\beta$ from virial estimates (see Sect.~\ref{sect:virial_mass}) with $M_{dyn}\sin^2\beta$ (bottom panel of Fig.~\ref{fig:spectroastrometry}). The results show that the virial mass factor $\mu_{vir}$ and $M_{dyn}\sin^2\beta$ have a non-linear relation with large dispersion. Therefore, we conclude that the classical virial method does not provide a reliable prediction of $M_{dyn}\sin^2\beta$. On the other hand, although the spectroastrometry method also suffers from large uncertainties and biases, we conclude that $\mu_{spec}$ are in better agreement with dynamical mass factor obtained with the full kinematical modelling of galaxy discs presented in this work. In fact, as discussed by \citet{Gnerucci+2011}, the classical virial mass estimate can be biased by systematic errors mostly associated with the measurement of galaxy dimensions.
This result confirms the reliability and usefulness of the spectroastrometry method, especially in the typical case of both poor spatial resolution and S/N ratio of the majority of the current available observations of high-$z$ galaxies.

\section{LBT observations and NIR spectra}
\label{sect:LBT_observations}
The observations of the five quasars (SDSS J0129-0035, SDSS J2054-0005 and SDSS J2310+1855, PSO J308-21, and PSO J138+05) were executed between 2018 September and 2019 June (PI: G. Cresci) with LUCI in seeing-limited conditions using the standard strategy for near-infrared long-slit spectroscopic observations: we dithered the objects along the $1\si{\arcsecond}$ slit following an ABAB cycle in order to subtract the sky. We made use of the low-resolution grating (G200, $\lambda/\Delta\lambda \approx 2000$) and the N1.8 camera (pixel size $\sim0.25\si{\arcsecond/pix})$ to maximise the signal-to-noise ratio. In order to obtain an accurate flux calibration of the spectra, that is required to estimate $L_{\lambda}$, we also obtained images of the QSOs with the $J$ or $K$ filter using the N3.75 camera (pixel size $\sim0.12\si{\arcsecond/pix}$). The total exposure time for spectroscopy is $\sim 2$ hours per target, and $\sim100,200$ or $800$ seconds for the imaging of the targets, depending on their apparent magnitude. The data were reduced and delivered by the LBT Imaging Data Center using the dedicated pipelines.

The requested time was derived assuming for the MgII line flux a typical value of $5\times10^{-16}\,\si{erg^{-1}\,s^{-1}\,cm^{-2}}$ at $z\sim 6$ \citep{Mazzucchelli+2017}, and for the CIV line a flux  $F_{\rm CIV}\sim3\times F_{\rm MgII}$ \citep{Shen+2011b}, and a line width of $4000\,\si{km\,s^{-1}}$.
Unfortunately, we did not achieve the requested sensitivities due to bad weather conditions, and we detected BLR emission only in J2310+1855. We therefore present here the $\Mbh$ mass of J2310+1855 derived from the observations of CIV BLR line. We also note that both \citet{Feruglio+2018} and \citet{Shen+2019} obtained independent NIR spectra of this target with different facilities. In particular, \citet{Shen+2019} published NIR spectra of a large sample $z\sim5.7$ QSOs, also providing an $\Mbh$ measurement for the J2310+1855 through virial relations based on CIV and MgII broad emission line.

Before modelling the CIV line in our LBT spectrum, we subtracted the continuum emission, fitting a power law at both sides of the ionised carbon line (in the two windows at $1450\,\AA$ and $1700\,\AA$). Then, we used a single Gaussian model to reproduce the CIV BLR emission profile. In fact, the low SNR does not allow us to constrain the possible contribution from iron emission in the region around the CIV, which is expected to be negligible \citep[see e.g.][]{Shen+2008,Shen+2011b}, nor the possible emission from the CIV NLR line \citep[e.g.][]{Shen+2011b}.

From the best fit of the CIV line, we derived a ${\rm FWHM}\sim12500 \,\si{km\,s^{-1}}$, and from the extrapolated continuum at $1350\,\AA$, a flux of $F_{1350} \sim1.1\times10^{-17}\si{erg\,s^{-1}cm^{-2}\AA^{-1}}$, and a luminosity of $\sim10^{45.8}\,\si{erg\,s^{-1}}$. Using the \citet{Vestergaard+2006} relation, we obtained %
$\log\Mbh\sim9.8$, consistent with \citet{Shen+2019}. The CIV line is blueshifted with respect to the [CII]$_{158\mu m}$ systemic of $\Delta v\sim -7200\,\si{km\,s^{-1}}$, strongly suggesting the presence of outflows in this source. We note that the values of ${\rm FWHM}$ and $\Delta v$ of CIV broad line in the J2310+1855 spectrum %
are consistent with the typical values estimated in high-$z$ QSOs \citep[see e.g.][]{Vietri+2018}. The CIV-based $\Mbh$ estimate can be therefore strongly biased; by adopting the different calibrations introduced to correct for the outflow contribution in CIV lines \citep[see e.g.][and references therein]{Vignali+2018}, we obtained mass estimates in the range $\log\Mbh = 8.9 - 9.4$, which is in agreement with the estimate by \citet{Feruglio+2018}. The uncertainties on these measurements are dominated by the intrinsic scatter ($\approx 0.3$ dex; see e.g. \citealt{Vestergaard+2006,Denney2012,Park+2017}) in the single-epoch calibrations, which are much larger than the typical uncertainties ascribed to the measurements of the line widths and fluxes. 

\begin{figure}[!t]
        \centering
        \includegraphics[width=0.99\hsize]{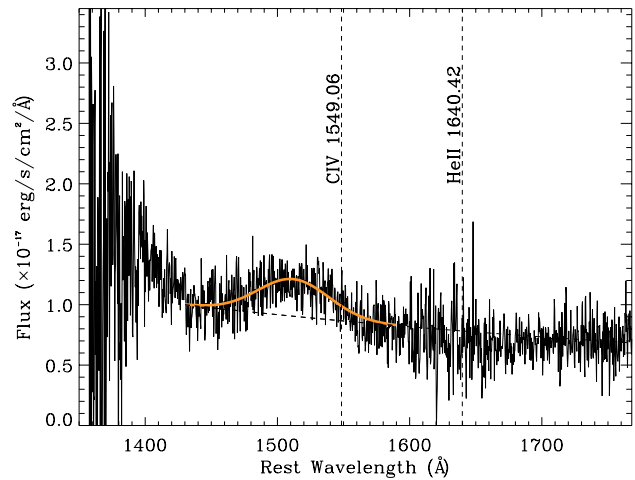}
         \caption{Portion of LBT/LUCI $zJ$ spectrum of SDSS J2310+1855 around CIV line for which the expected wavelength is indicated in the figure according to [CII]-based redshift. The orange and black dashed curves indicate the spectrum's best fit (line $+$ continuum and continuum, respectively).
         }
         \label{fig:LBTJ2310spectrum}
\end{figure}

The latter values are consistent with the MgII-based $\Mbh$ reported in \citet{Shen+2019}, which is used in this paper to study the $\Mbh-M_{dyn}$ relation. In Fig.~\ref{fig:LBTJ2310spectrum}, we show the NIR spectrum of J2310+1855 with the best fit of the CIV broad line. 

\section{Maps, integrated spectra, and the results of the kinematical modelling} 
\label{sect:all_maps}
\begin{figure*}[!htbp]
        \centering
        \includegraphics[width=\textwidth]{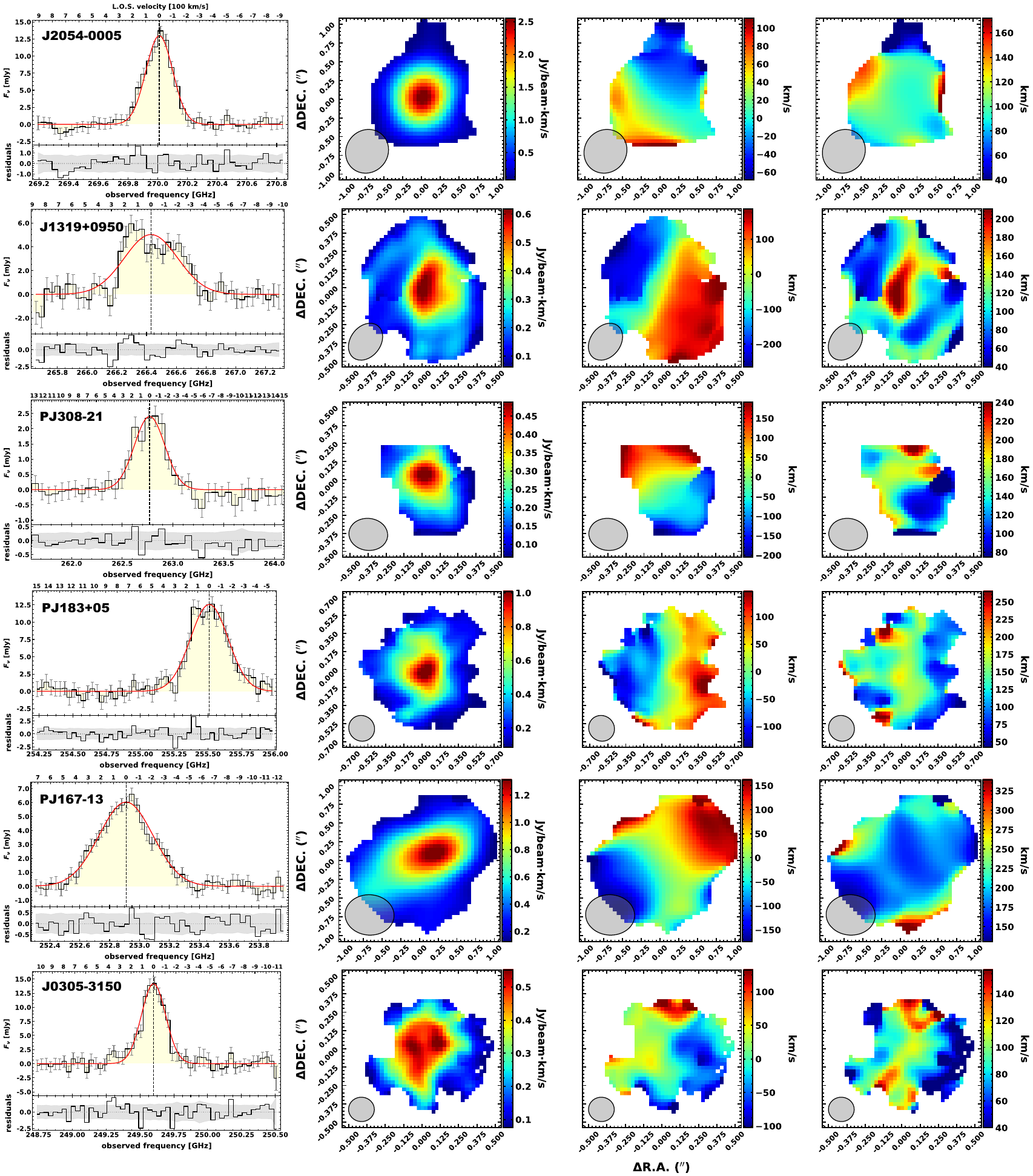}
        \caption{Line integrated spectra, flux, velocity, and velocity-dispersion maps along the line of sight for objects listed in Table~\ref{tbl:results_data}. See Fig.~\ref{fig:tot_flux} and Fig.~\ref{fig:3_maps} for the description of each panels.}
        \label{fig:second-spectromaps}
\end{figure*}

\begin{figure*}[!htbp]
        \centering
        \includegraphics[width=0.95\textwidth,trim={0 2cm 0 1cm},clip]{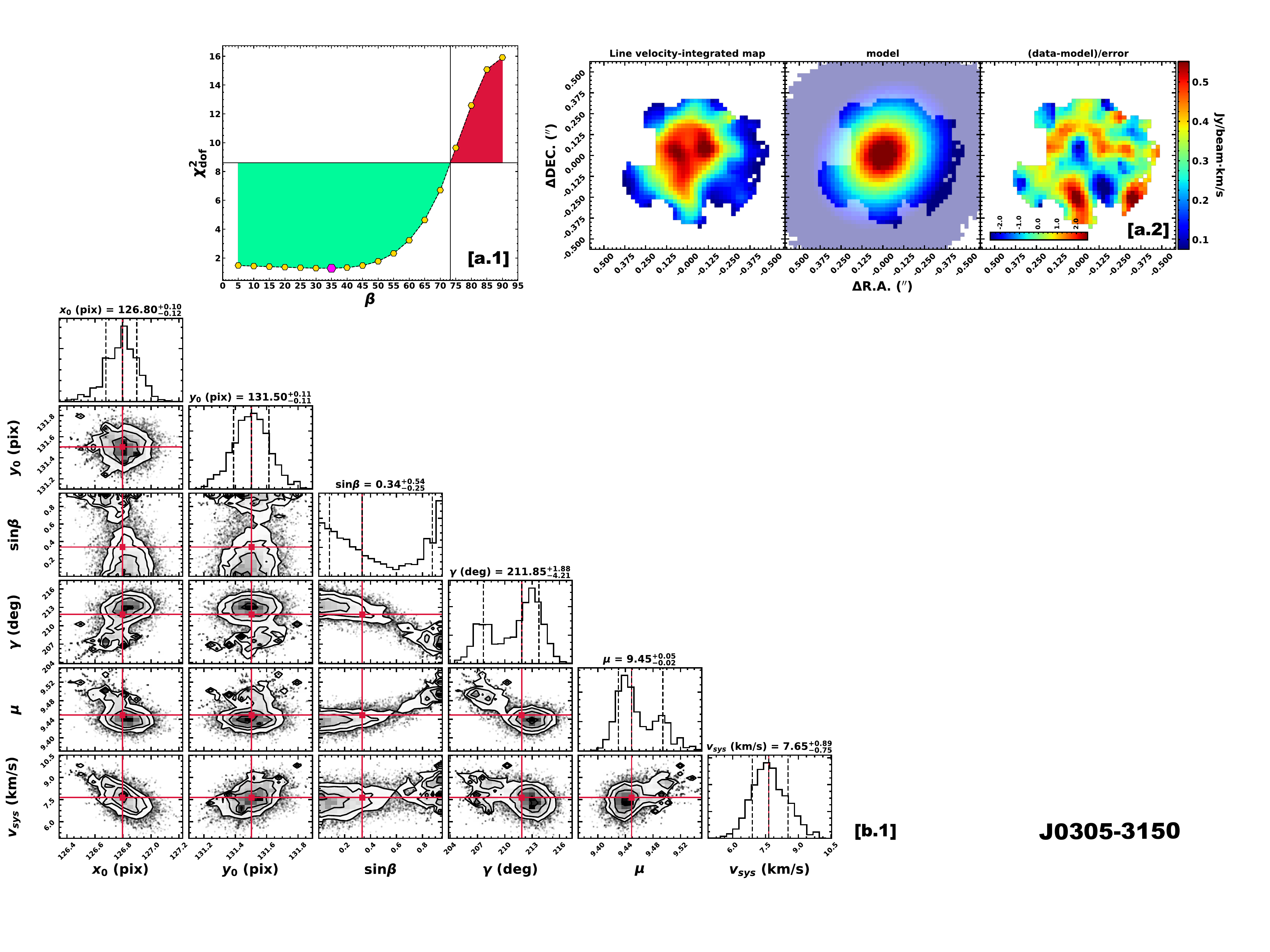}
        \caption{Results of 2D kinematical modelling for objects listed in Table~\ref{tbl:results_data}. We refer to Fig.~\ref{fig:fitted_maps} for descriptions of individual panels.}
         \label{fig:last-modelling}
\end{figure*}

Here, we report the integrated spectra, flux, velocity, and velocity-dispersion maps for objects in Table~\ref{tbl:results_data} (Fig.~\ref{fig:second-spectromaps}). See Fig.~\ref{fig:tot_flux} and Fig.~\ref{fig:3_maps} for the descriptions of each panel. We also report the 2D best fit of the flux and velocity maps (Fig.~\ref{fig:last-modelling}). The different panels are labelled as they are in Fig.~\ref{fig:fitted_maps}; we refer to the latter for a description of the figures.
\begin{figure*}[!htbp]
        \centering
        \includegraphics[width=\textwidth]{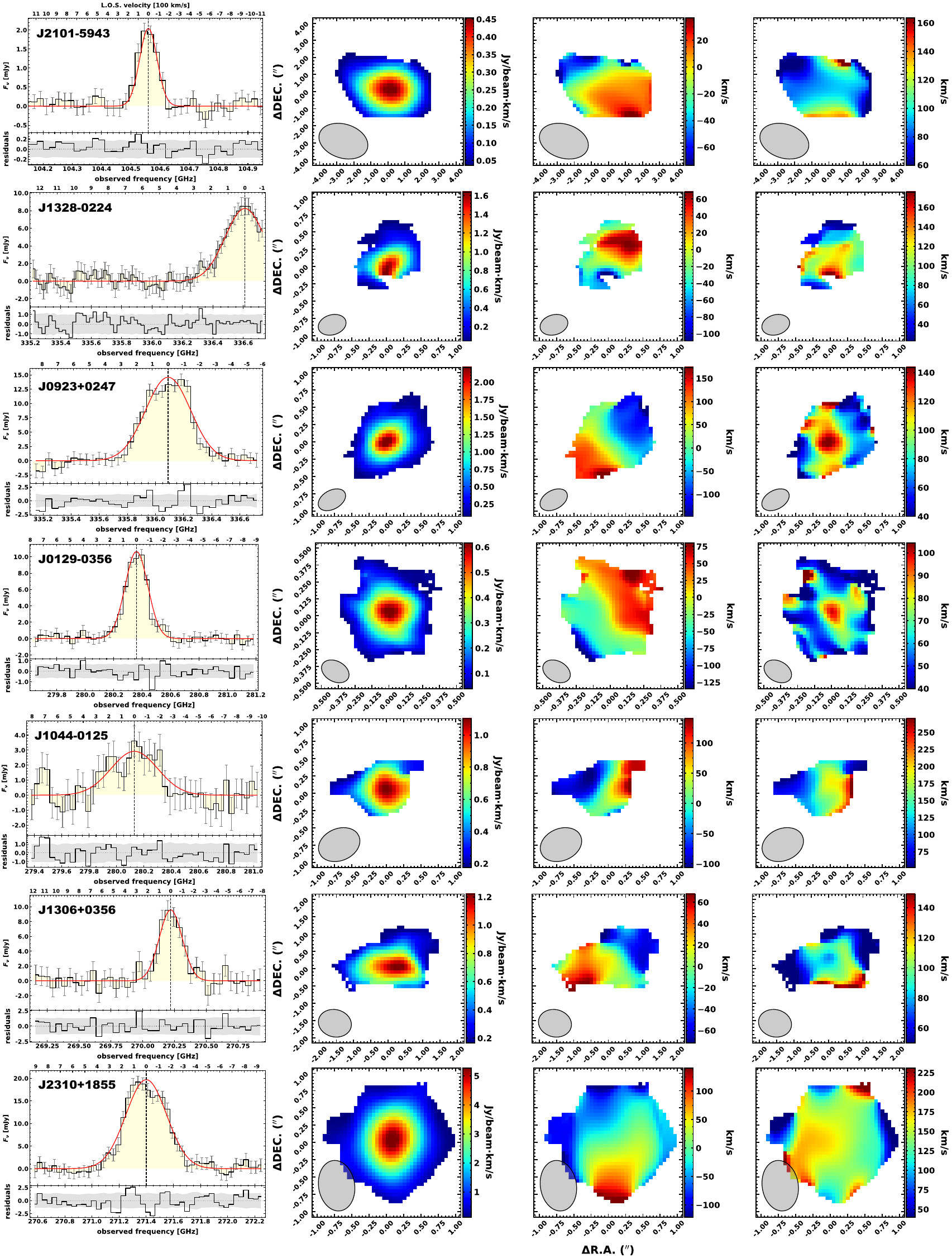}
         \label{fig:first-spectromaps}
\end{figure*}

\begin{figure*}[!htbp]
        \centering
        \includegraphics[width=0.95\textwidth,trim={0 2cm 0 1cm},clip]{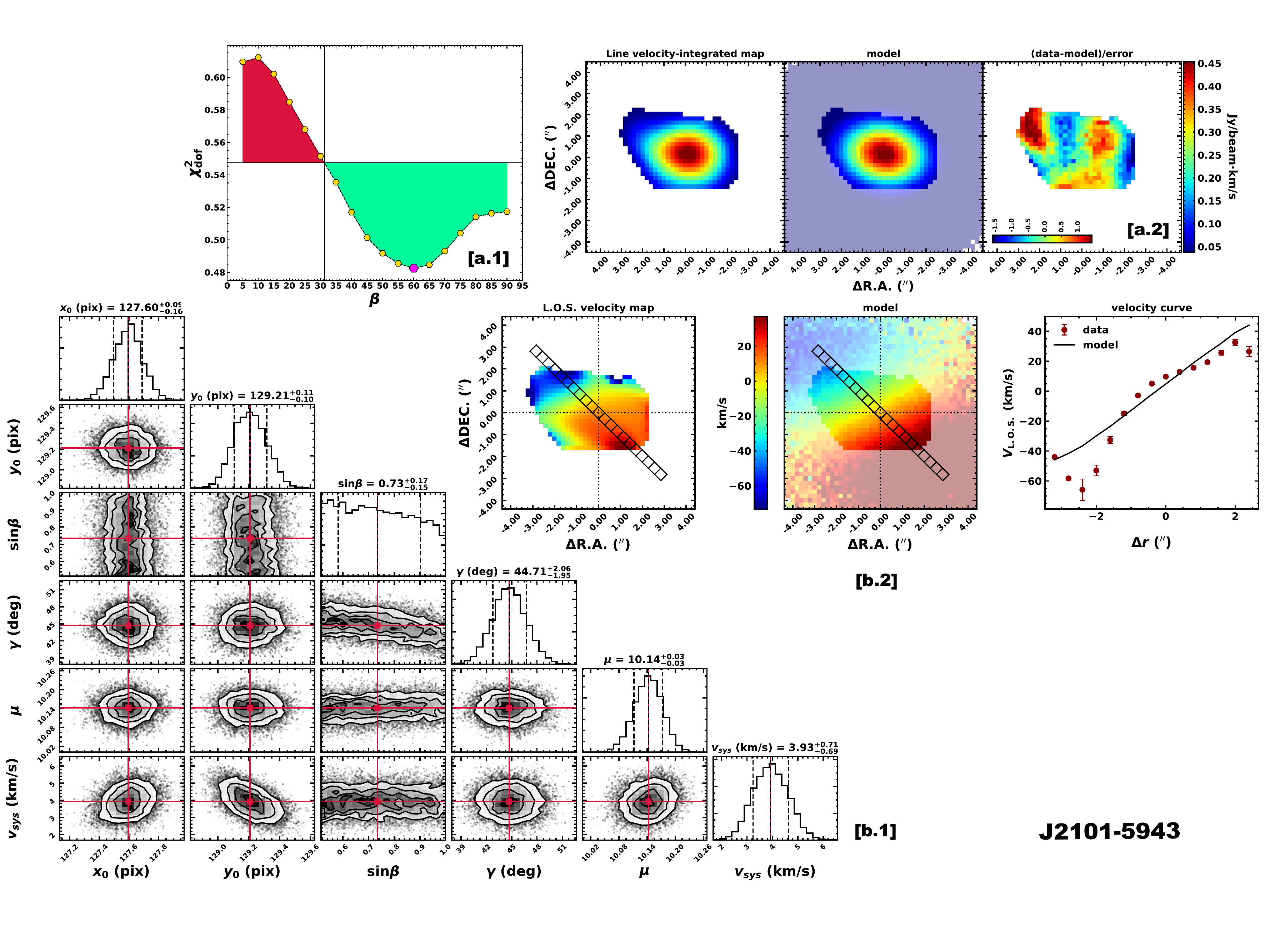}
\end{figure*}
\begin{figure*}[!htbp]
        \centering
        \includegraphics[width=0.95\textwidth,trim={0 2cm 0 1cm},clip]{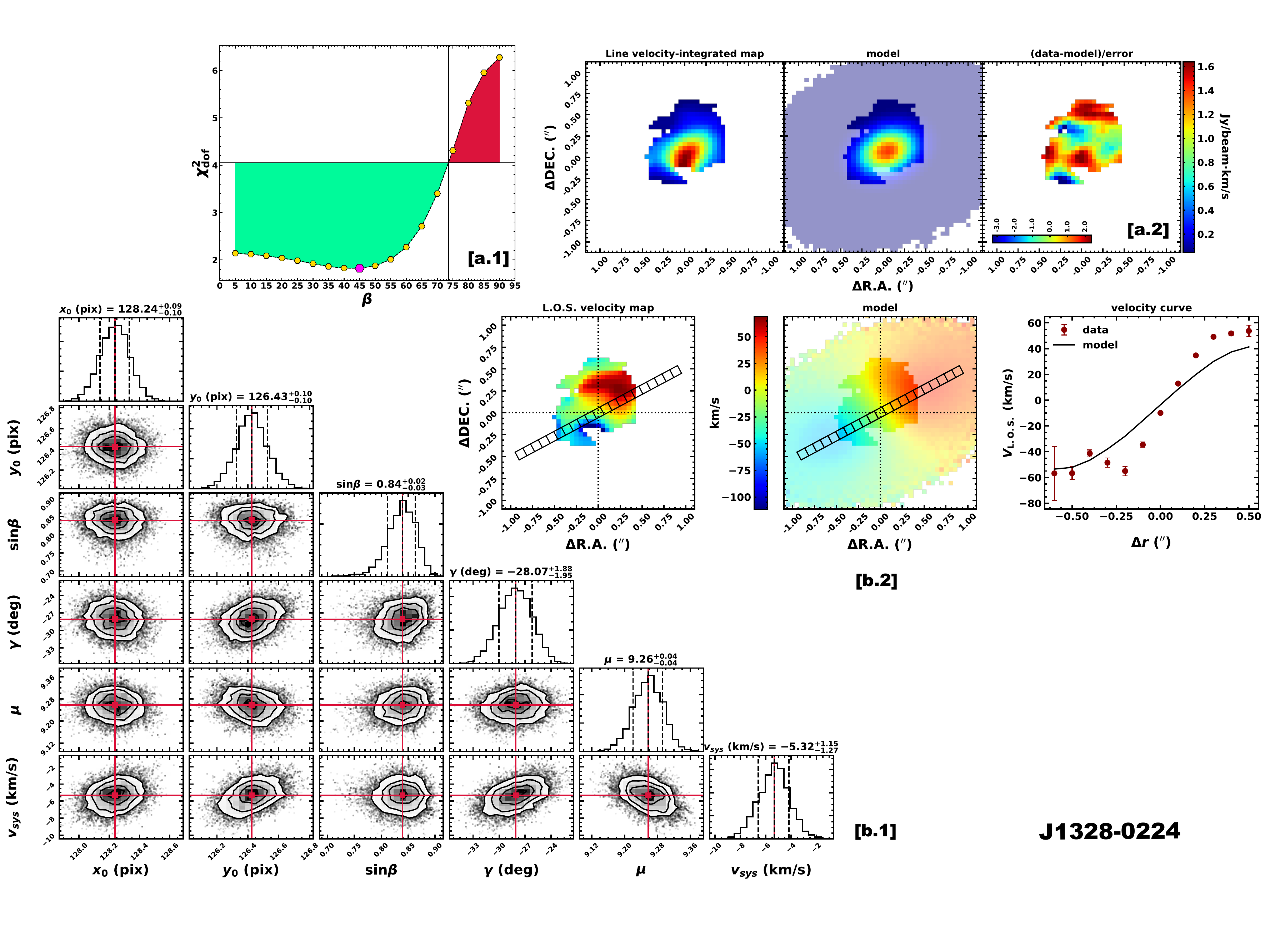}
\end{figure*}
\begin{figure*}[!htbp]
        \centering
        \includegraphics[width=0.95\textwidth,trim={0 2cm 0 1cm},clip]{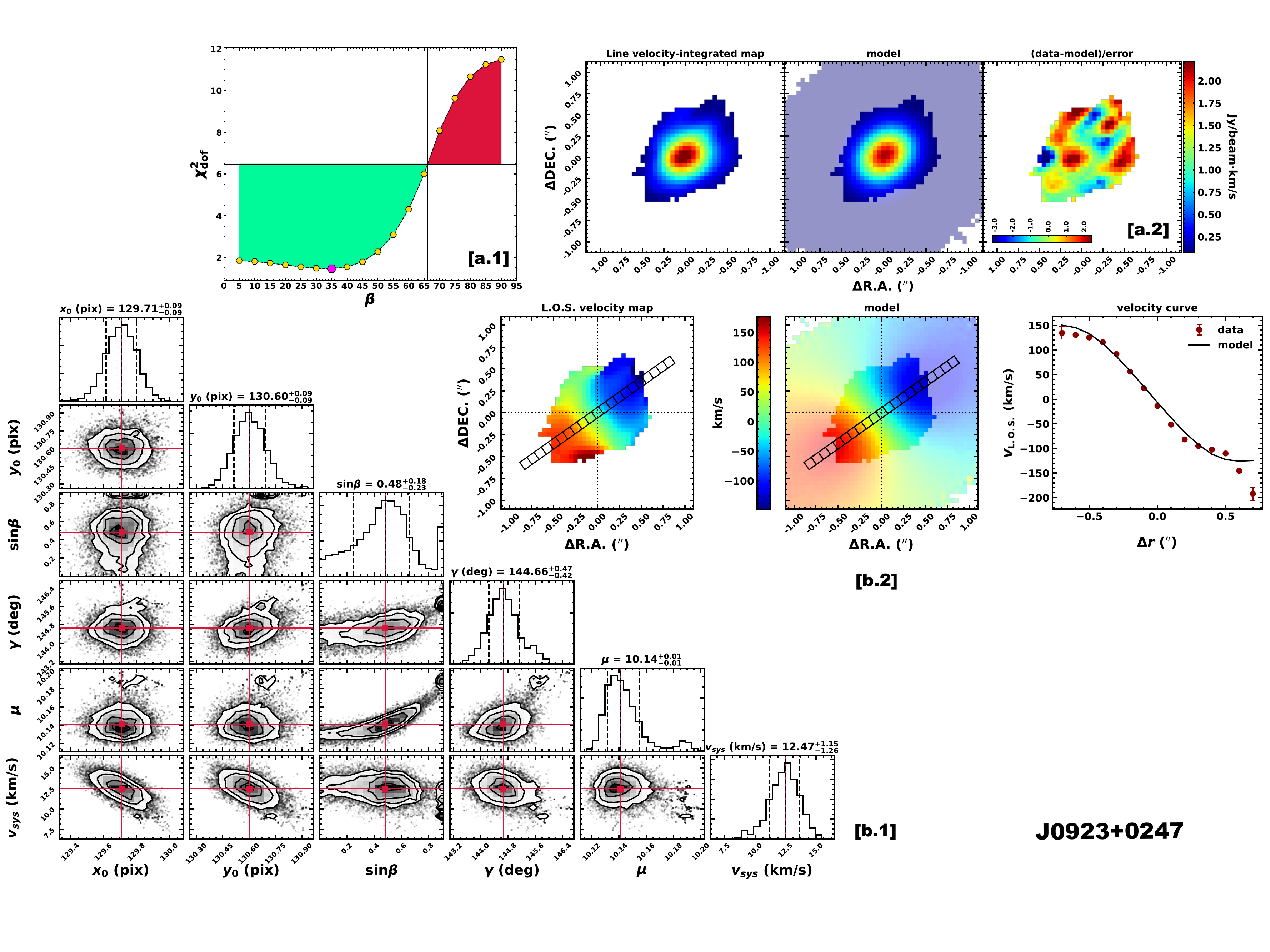}
\end{figure*}
\begin{figure*}[!htbp]
        \centering
        \includegraphics[width=0.95\textwidth,trim={0 2cm 0 1cm},clip]{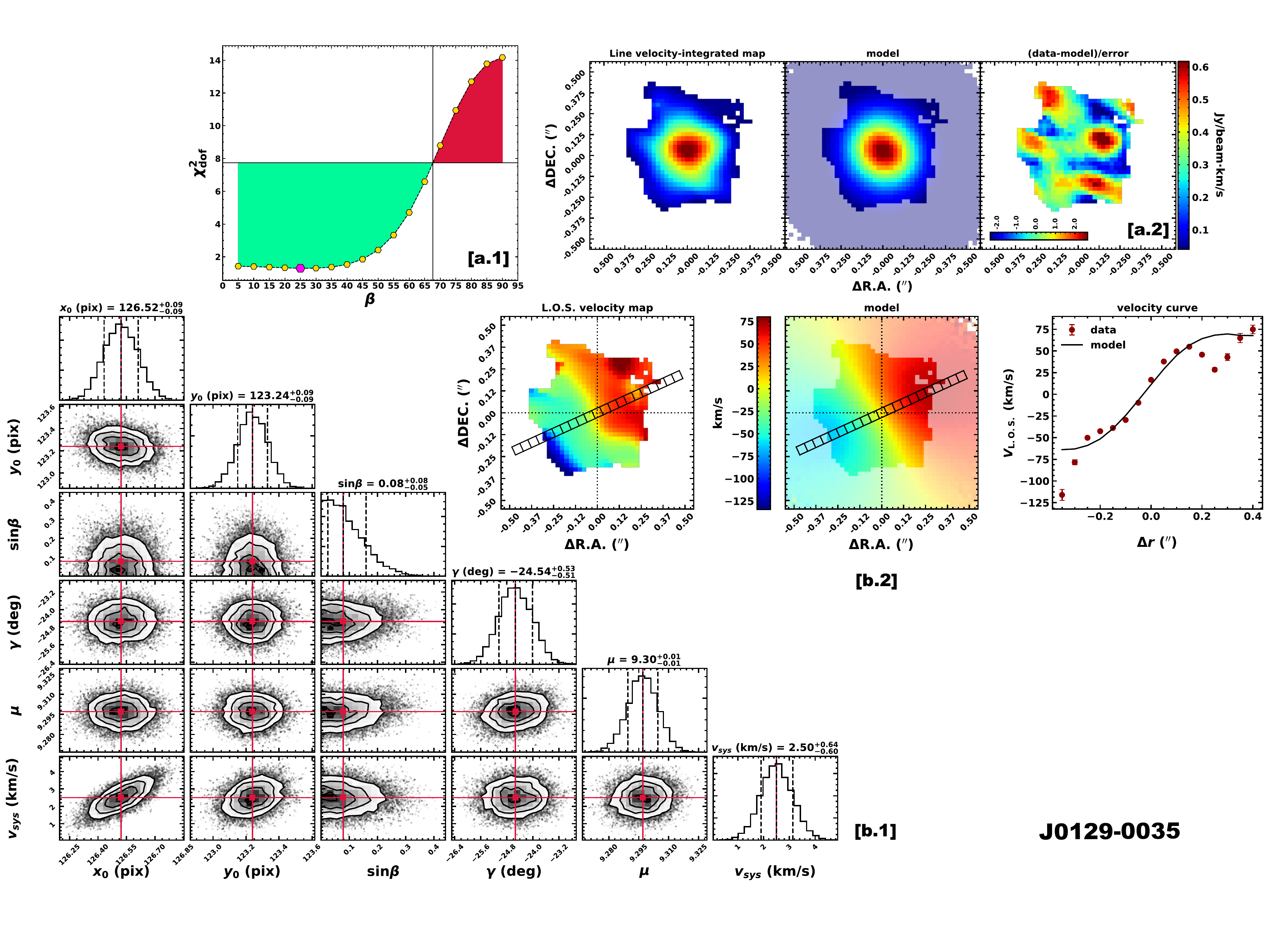}
\end{figure*}
\begin{figure*}[!htbp]
        \centering
        \includegraphics[width=0.95\textwidth,trim={0 2cm 0 1cm},clip]{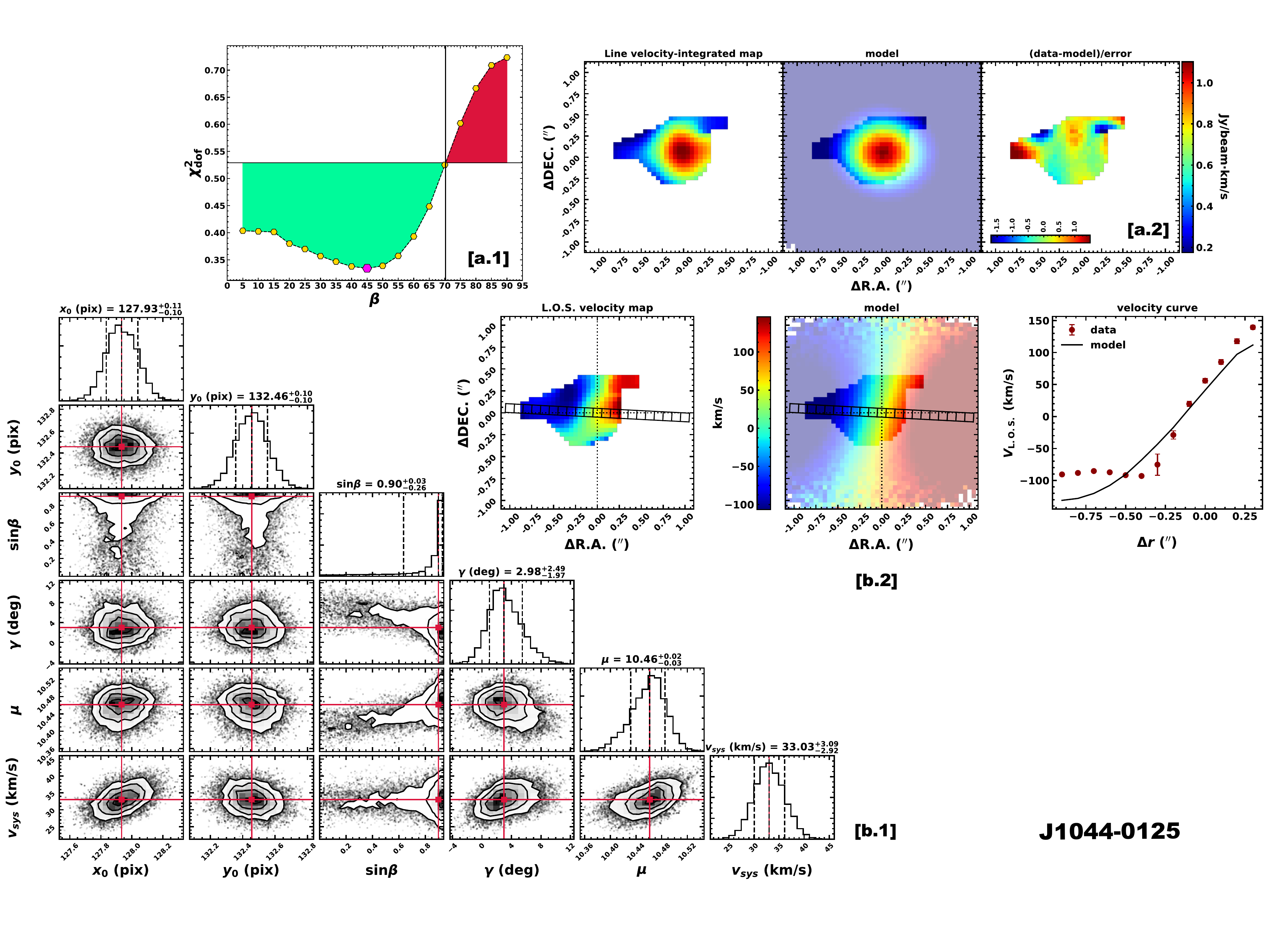}
\end{figure*}
\begin{figure*}[!htbp]
        \centering
        \includegraphics[width=0.95\textwidth,trim={0 2cm 0 1cm},clip]{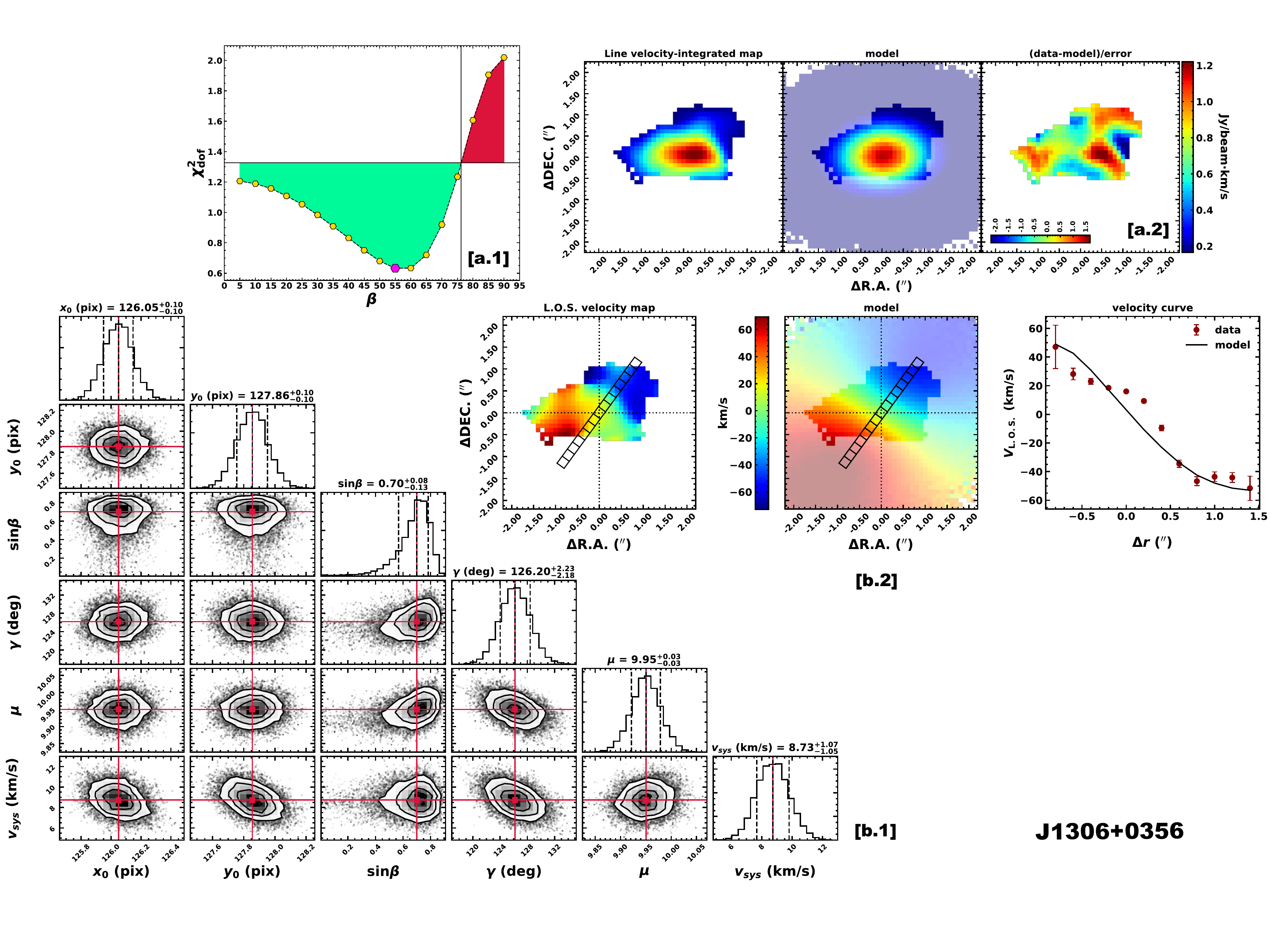}
\end{figure*}
\begin{figure*}[!htbp]
        \centering
        \includegraphics[width=0.95\textwidth,trim={0 2cm 0 1cm},clip]{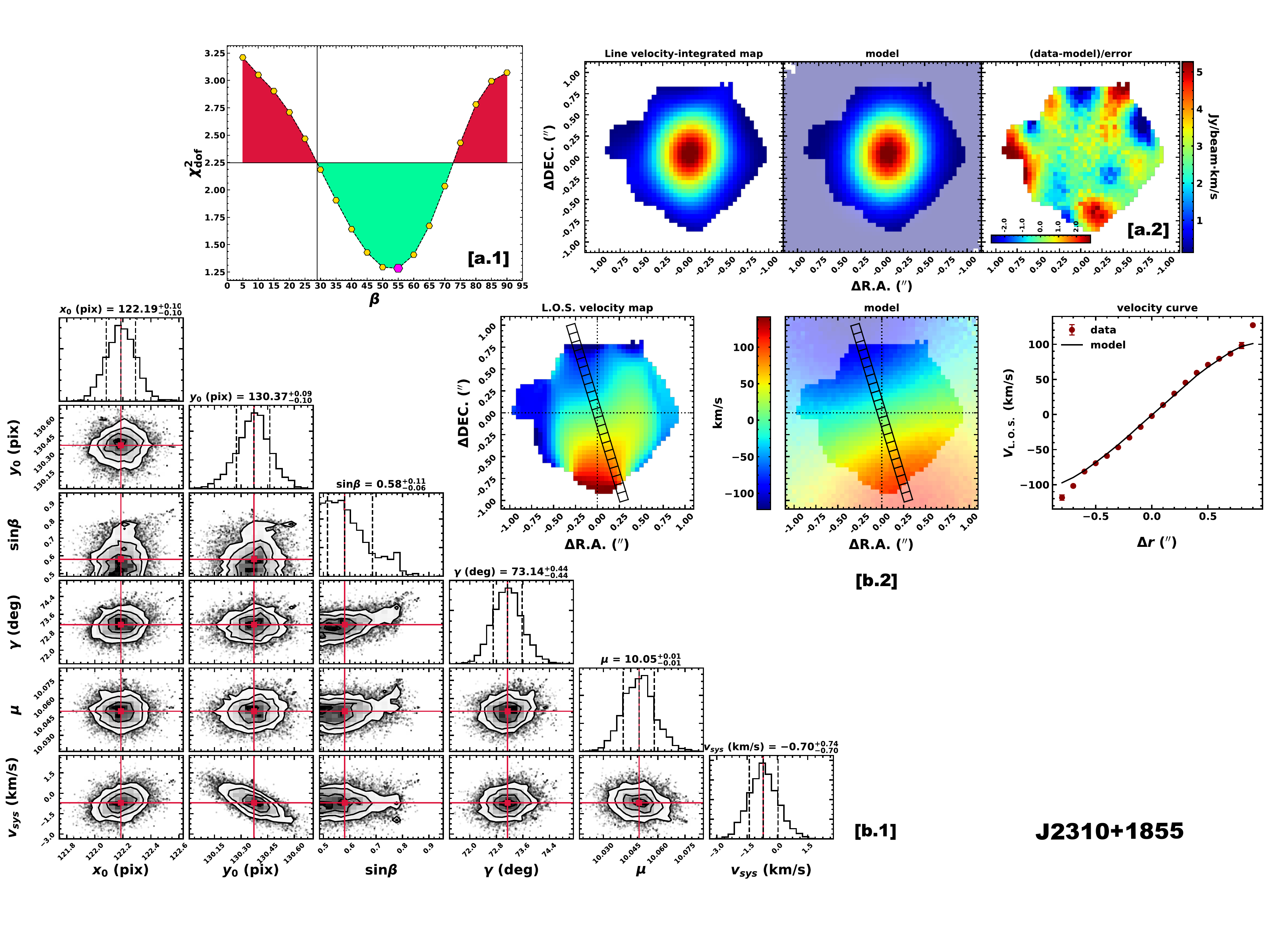}
\end{figure*}
\begin{figure*}[!htbp]
        \centering
        \includegraphics[width=0.95\textwidth,trim={0 2cm 0 1cm},clip]{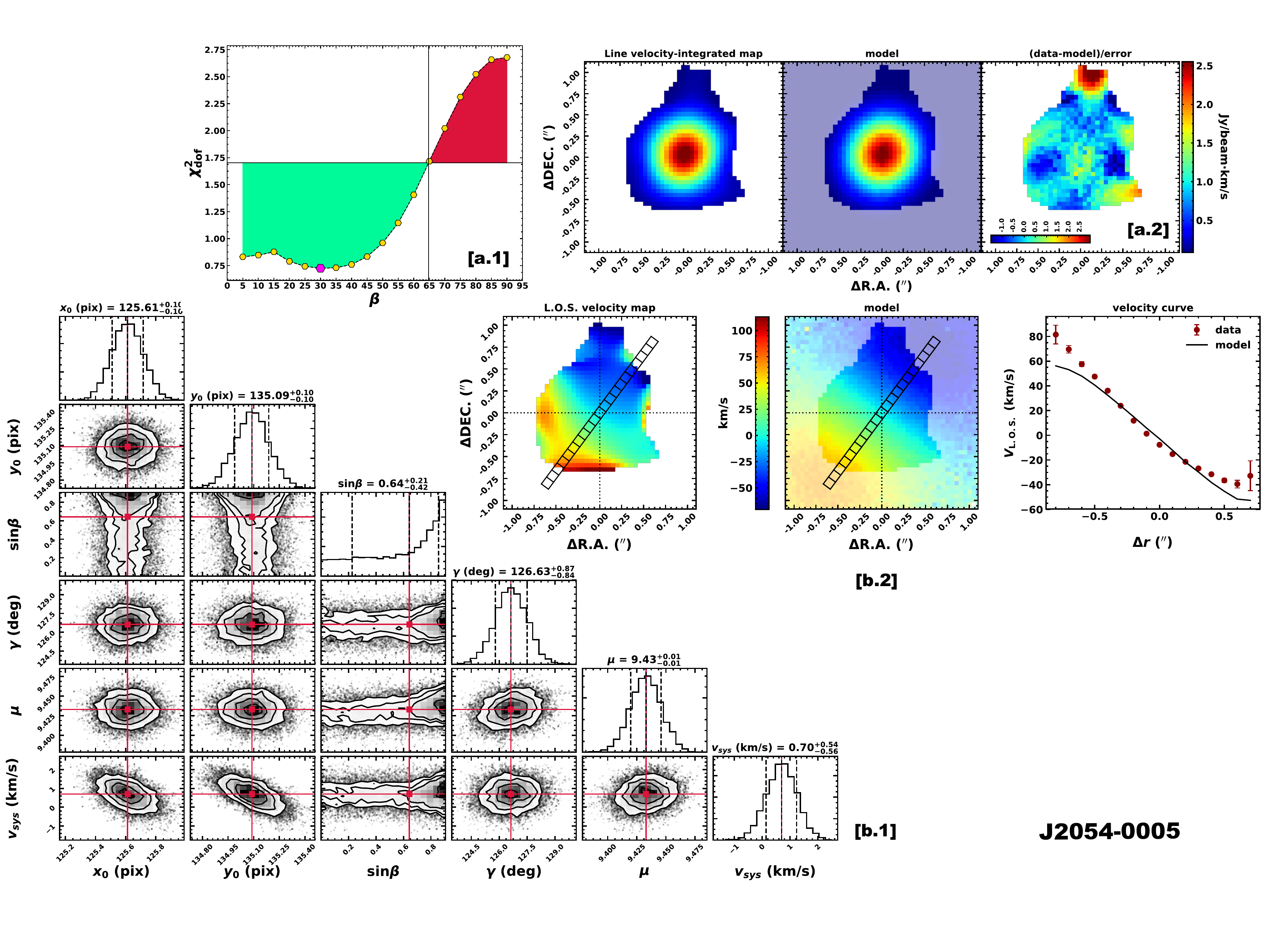}
\end{figure*}
\begin{figure*}[!htbp]
        \centering
        \includegraphics[width=0.95\textwidth,trim={0 2cm 0 1cm},clip]{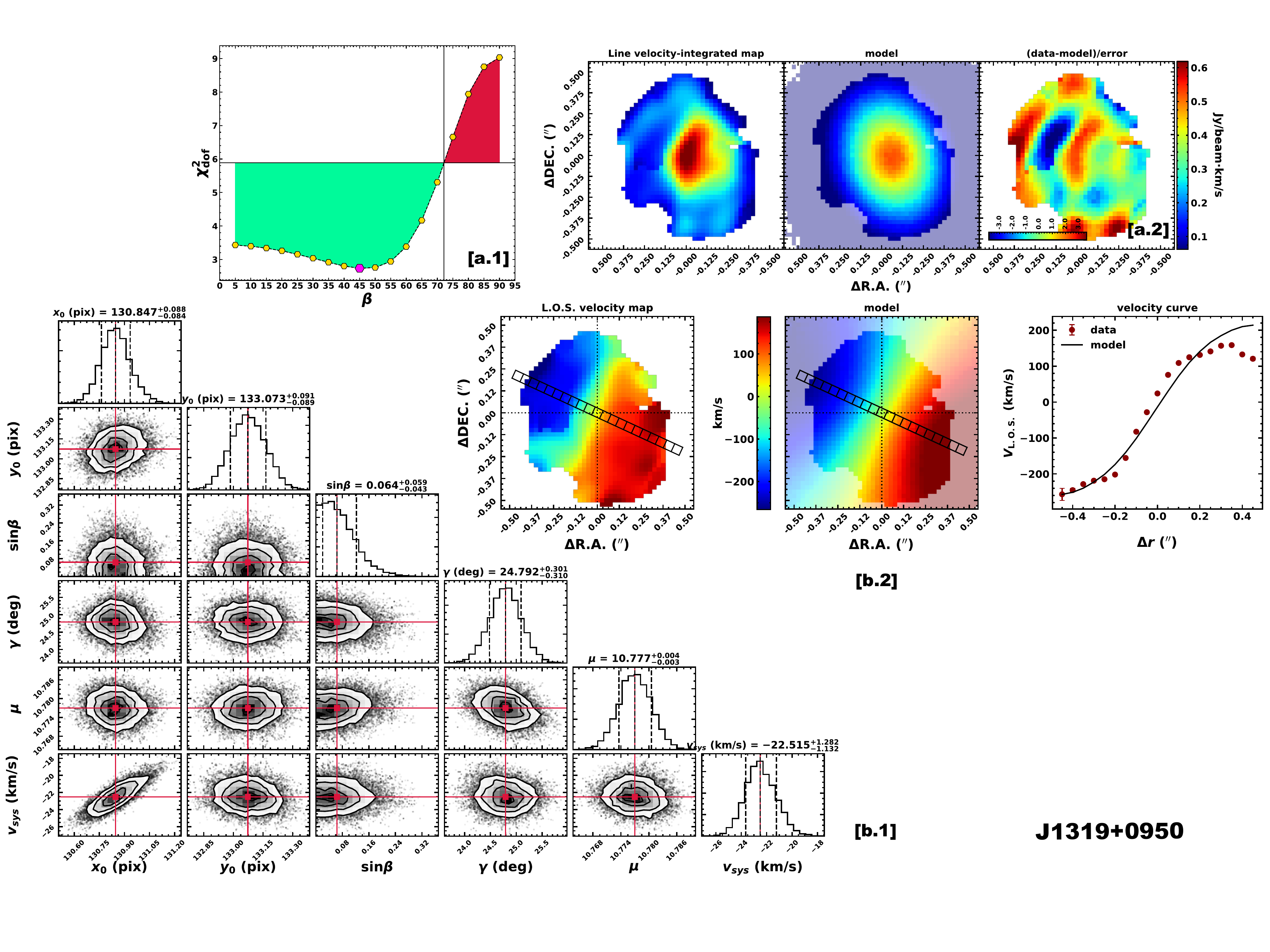}
\end{figure*}
\begin{figure*}[!htbp]
        \centering
        \includegraphics[width=0.95\textwidth,trim={0 2cm 0 1cm},clip]{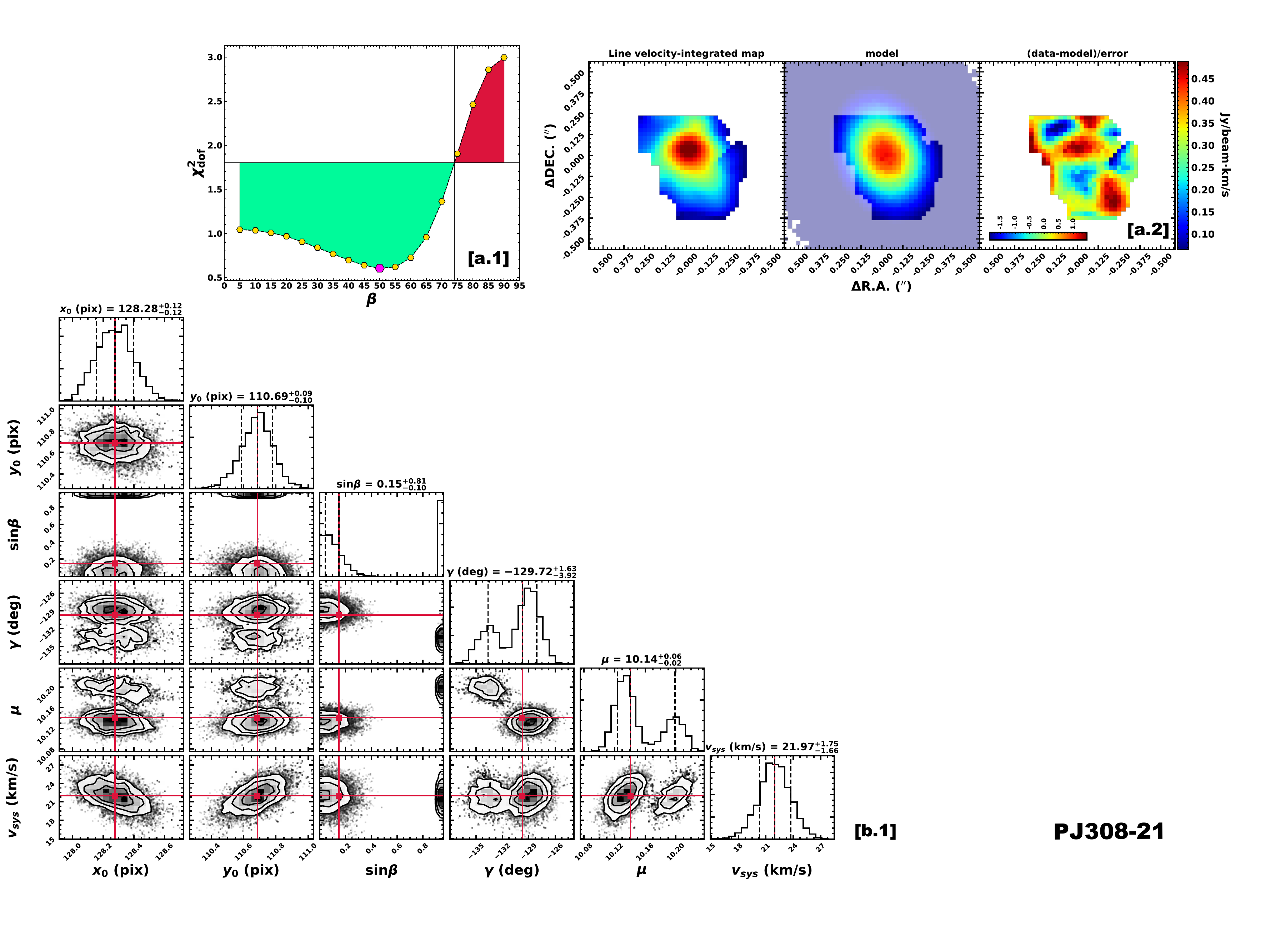}
\end{figure*}
\begin{figure*}[!htbp]
        \centering
        \includegraphics[width=0.95\textwidth,trim={0 2cm 0 1cm},clip]{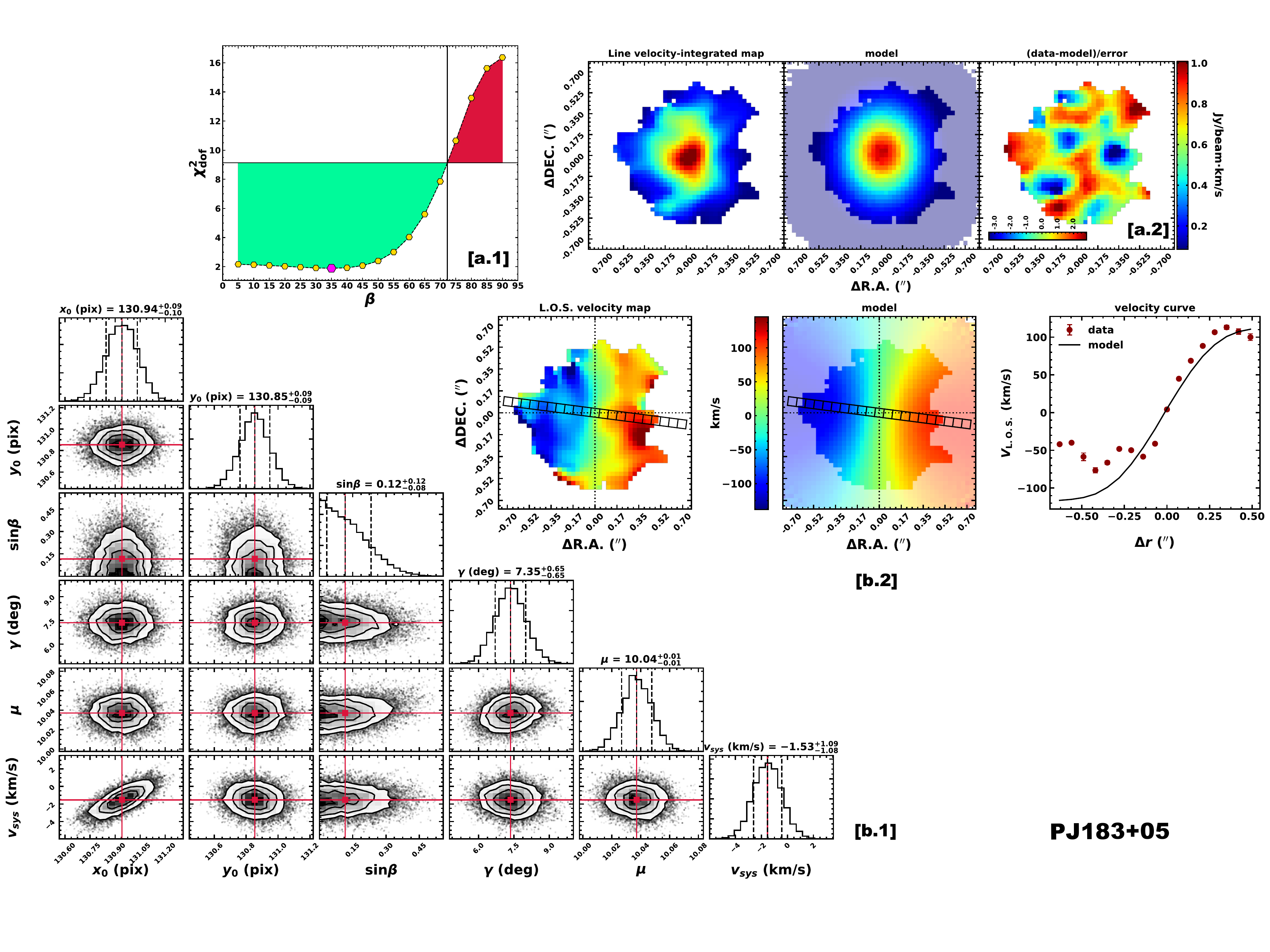}
\end{figure*}
\begin{figure*}[!htbp]
        \centering
        \includegraphics[width=0.95\textwidth,trim={0 2cm 0 1cm},clip]{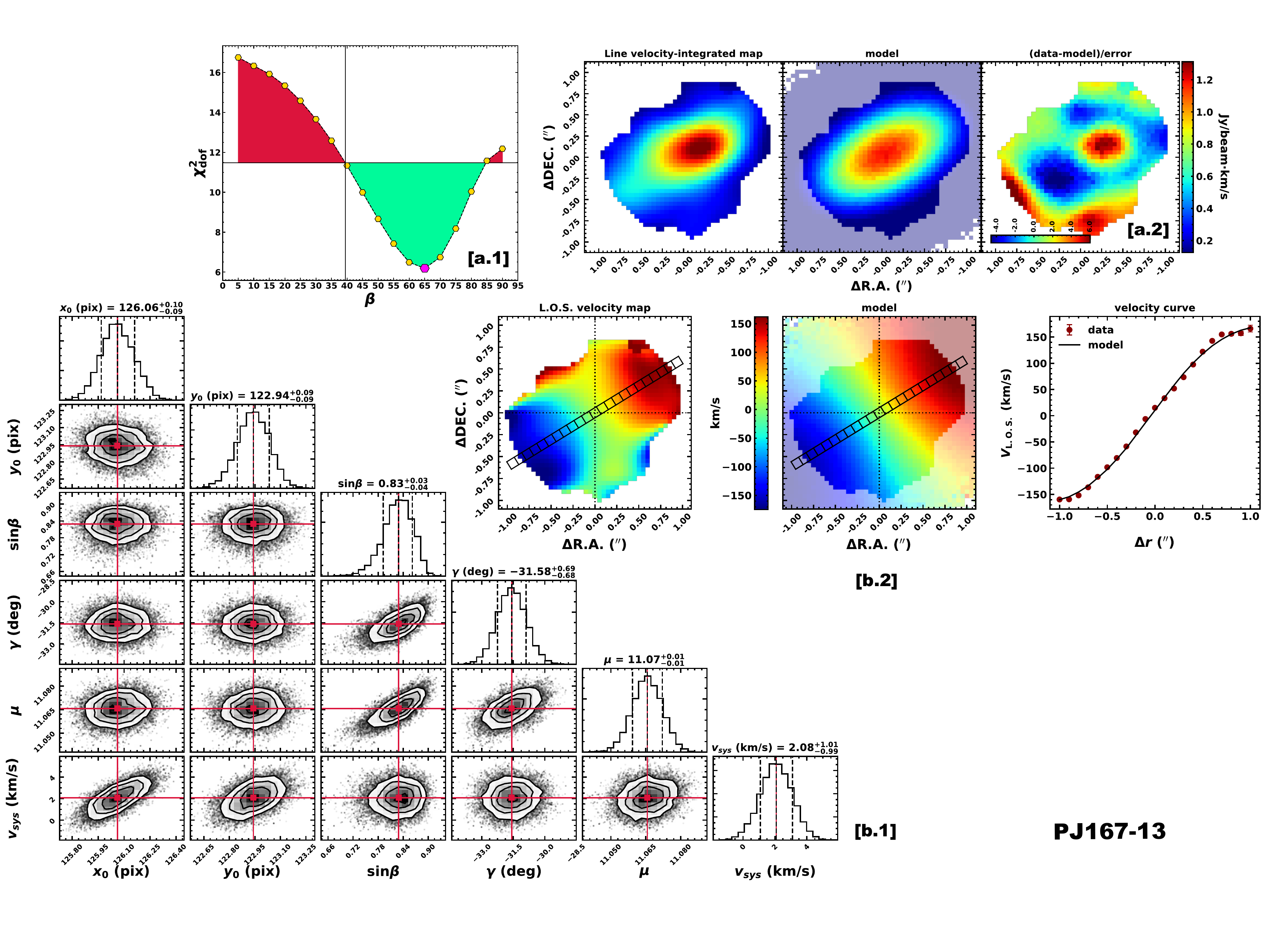}
\end{figure*}

\end{appendix}

\end{document}